%% file: ANA-HMBS-2024-05-PAPER.tex
\documentclass[PAPER, atlasdraft=false, UKenglish,orcidlogo,cernpreprint, texmf, orcidlogo]{atlasdoc}
\usepackage{atlaspackage}
\usepackage[block=space]{atlasbiblatex}

\usepackage[BSM]{atlasphysics}
\graphicspath{{logos/}{figures/}}

\usepackage{ANA-HMBS-2024-05-PAPER-defs}

\AtlasTitle{Search for squarks and gluinos in \textit{pp} collisions at $\sqrt{s} = 13$~TeV and $13.6$~TeV in events with $\tau$-leptons, jets and missing transverse momentum using the ATLAS detector}

\AtlasAbstract{%
A search for R-parity-conserving supersymmetry in events with large missing transverse momentum, jets and
at least one hadronically decaying $\tau$-lepton is presented. Both gluino and squark pair production are considered, with the cascade decay of each gluino or squark producing either a $\tau$-slepton or a $\tau$-sneutrino.
Three channels are examined, requiring either exactly one hadronically decaying $\tau$-lepton and no other leptons, exactly one hadronically decaying $\tau$-lepton and at least one other lepton, or two or more hadronically decaying $\tau$-leptons.
Analyses in the three channels are optimised independently and combined statistically.
Two separate analysis strategies, either a cut-and-count or machine-learning approach, are used. The search uses 140~$\mathrm{fb}^{-1}$ and 51.8~$\mathrm{fb}^{-1}$
of  \textit{pp} collision data recorded by the ATLAS detector at the Large Hadron Collider during 2015--2018 at $\sqrt{s} = 13$~TeV and 2022--2023 at $\sqrt{s} = 13.6$~TeV, respectively. Gluino masses below $2.25$~TeV and squark masses up to $1.7$~TeV are excluded.

}

\AtlasRefCode{HMBS-2024-05}

\PreprintIdNumber{CERN-EP-2025-132}

\AtlasJournalRef{Eur.Phys.J.C 85 (2025) 12, 1437}
\AtlasDOI{10.1140/epjc/s10052-025-14957-6}

\hypersetup{pdftitle={ATLAS document},pdfauthor={The ATLAS Collaboration}}

\begin{document}

\maketitle

\tableofcontents

\FloatBarrier
\section{Introduction}
%

The Standard Model (\SM) of particle physics is a powerful and experimentally well-validated theory. The \SM describes excitations and interactions of quantum fields, representing elementary particles, and can be used to give extremely precise predictions of the results of elastic and inelastic particle collisions.
The latest triumph of the \SM is its prediction of the Higgs boson and the discovery in 2012 of a compatible particle by the ATLAS~\cite{HIGG-2012-27} and CMS~\cite{CMS-HIG-12-028} experiments in the proton--proton ($\pp$) collisions produced by CERN's Large Hadron Collider (LHC)~\cite{Evans:2008zzb}.
Although the precision of the \SM and its impact on particle physics is hard to underestimate, it is fundamentally an incomplete, effective theory. The usual list of limitations of the \SM includes its inability to explain the large difference between the electroweak scale and the Planck scale of the gravitational interactions (known as the \enquote{hierarchy problem}), the associated fine-tuning of the Higgs potential, and its lack of viable dark matter candidates (whose existence is corroborated by multiple astronomical observations~\cite{Clowe_2004,Bertone:2004pz,Zyla:2020zbs}).

When trying to describe the symmetries of special relativity and the internal gauge symmetries of quantum fields by a common mechanism in an attempt to extend the \SM, it can be shown that the only way they can be combined into a common symmetry Lie group is as a direct product~\cite{Coleman:1967ad}, meaning that no interaction between them is possible.
One of the only ways to overcome this restriction is through the introduction of graded Lie superalgebras~\cite{Haag:1974qh}, which gives rise to supersymmetry (SUSY) ~\cite{Golfand:1971iw_fix,Volkov:1973ix,Wess:1974tw,Wess:1974jb,Ferrara:1974pu,Salam:1974ig}, an extension of the \SM which postulates the existence of an additional (super)symmetry between bosons and fermions. A direct consequence of SUSY is that every known \SM particle has at least one superpartner with the same quantum numbers, but a spin that differs by one half, resulting in a plethora of new particles. This \enquote{fascinating mathematical structure}~\cite{Witten:1983mna} leads to additional loop corrections to the Higgs boson mass which perfectly cancel out with those from the \SM~\cite{Sakai:1981gr,Dimopoulos:1981yj,Ibanez:1981yh,Dimopoulos:1981zb}, thus solving the hierarchy problem.

To accommodate experimental observations, especially the stability of the proton, a conserved quantum number $R= (-1)^{3(B-L) + 2s}$~\cite{Farrar:1978xj} is often postulated, where $B$ and $L$ are the baryon and total lepton numbers, and $s$ is the spin of the particle.
Phenomenologically, this implies that SUSY particles are pair-produced at collider experiments, and that the lightest supersymmetric particle (LSP) is stable and can be a viable dark matter candidate~\cite{Goldberg:1983nd,Ellis:1983ew}. If supersymmetry was preserved, the superpartners would have the same mass as the \SM particles, but this has not been observed. Therefore, supersymmetry must be broken, and this introduces a wide range of possible SUSY mass spectra~\cite{Martin:1997ns}.
A common feature of SUSY models is the mixing of binos, winos and higgsinos (the superpartners of the electroweak gauge bosons and the Higgs boson, respectively). Their mass eigenstates are known as charginos ($\tilde{\chi}^\pm_{1,2}$) and neutralinos ($\tilde{\chi}^0_{1,2,3,4}$), where the subscript denotes their ordering in increasing mass.
Third-generation squarks and sleptons are more likely to have lower masses than their first- and second-generation counterparts and be next-to-lightest supersymmetric particle (NLSP) candidates~\cite{PhysRevD.84.095015,Alves:2011wf,Alwall:2008ag}, and are of special interest for experimental searches because they provide distinct signatures.
The third-generation sneutrino is another option for the NLSP~\cite{Covi:2007xj,Ellis:2008as,Figy:2010hu,Katz:2009qx,Katz:2010xg,Bhattacharya:2011zn,Chala:2017jgg,Alvestad:2021sje} that can produce similar signatures.
The ATLAS Collaboration has an extensive programme of searches for SUSY-inspired signatures~\cite{SUSY-2023-10_fix,SUSY-2014-06}.


\FloatBarrier
\section{Analysis targets}
%

\begin{figure}[htbp!]
\centering
\subfloat[]{\includegraphics[width = 0.45\textwidth]{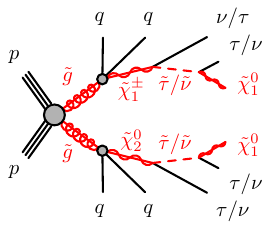}\label{fig:gluinoFeynmanDiagram}}~~
\subfloat[]{\includegraphics[width = 0.45\textwidth]{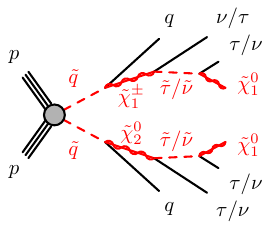}\label{fig:squarkFeynmanDiagram}}
\caption{Diagrams showing the production and decay of \protect\subref{fig:gluinoFeynmanDiagram} a gluino pair and \protect\subref{fig:squarkFeynmanDiagram} a squark pair. In the gluino case, the two-step decay via an off-shell squark $\tilde{g} \to \tilde{q}^* q \to q q \ninotwo/\chinoonepm$ is symbolised by a four-body vertex. }%
\label{fig:FeynmanDiagram}
\end{figure}

This \paper describes a search for pair production of gluinos or first- or second-generation left-handed squarks (superpartners of gluons and quarks, respectively), with $\tau$-sleptons (\tauslepton) and $\tau$-sneutrinos (\tausneut) as co-NLSPs, targeting final states with $\tau$-leptons, jets, and missing transverse momentum.
Gluinos are allowed to decay into an off-shell squark and a quark, a common decay mode in the minimal supersymmetric model (MSSM).
Squarks are assumed to be left-handed and from the first or second generation, and to decay via the second-lightest neutralino \ninotwo or the lightest chargino \chinoonepm, as these are the common decay modes for the left-handed squarks.
Both the \chinoonepm and \ninotwo are assumed to couple to \tauslepton and \tausneut in association with a \SM $\tau$-lepton or $\nu_\tau$, with the branching ratios set to $0.5$. The \tauslepton and \tausneut decay into the LSP \ninoone and a \SM $\tau$-lepton or $\nu_\tau$.
Such cascade chains arise naturally in the MSSM when the \tauslepton or \tausneut are light.
The simplified SUSY models~\cite{Alwall:2008ve,Alwall:2008ag,Alves:2011wf} considered in this \paper are shown in Figure~\ref{fig:FeynmanDiagram}. According to the pMSSM scan~\cite{SUSY-2014-08}, these final states are uniquely sensitive to a subset of SUSY models. The ATLAS Collaboration previously searched for such signals using data gathered during 2015--2016~\cite{SUSY-2016-01,SUSY-2016-30}, excluding gluino masses below $2$~\TeV and LSP masses below $1$~\TeV, and CMS considered similar signatures with data gathered during 2011~\cite{CMS-SUS-12-004}, excluding gluino masses below $1.15$~\TeV. The present search uses 140~$\mathrm{fb}^{-1}$ and 51.8~$\mathrm{fb}^{-1}$ of $\pp$ collision data recorded by the ATLAS detector~\cite{PERF-2007-01} during 2015--2018 at centre-of-mass energies of $\sqrt{s} = 13$~\tev and 2022--2023 at $\sqrt{s} = 13.6$~\tev.

The kinematic properties of the final-state particles depend heavily on the masses of SUSY particles in the decay chains. The simplified models are parameterised in terms of the gluino or squark mass \susymass and the LSP mass $m(\ninoone)$. The other SUSY particles in the decay chain, namely \ninotwo and \chinoonepm, and \tauslepton and \tausneut, are considered to be pairwise mass-degenerate. Their masses are determined from the two free parameters of the model as:
\begin{equation}
m(\tilde\chi_1^\pm) = m(\tilde\chi_2^0)= \frac{1}{2}\left[\susymass + m(\tilde\chi_1^0)\right] \; , \quad
m(\tilde\tau)=m(\tilde\nu_\tau)=\frac{1}{2}\left[m(\tilde\chi_1^\pm) + m(\tilde\chi_1^0)\right].
\label{eq:eq1}
\end{equation}
This parameterisation follows the conventions used in ATLAS searches for gluinos and squarks with cascade decays~\cite{SUSY-2018-17,SUSY-2018-22} used to simplify the model parameter phase space.
The sensitivity of the analysis does not depend strongly on the assumptions of Eq.~(\ref{eq:eq1}) as long as the mass differences between SUSY particles $\Delta m(\tilde{g} / \tilde{q}, \tilde\chi_1^\pm / \tilde\chi_2^0)$ and $\Delta m(\tilde\chi_1^\pm / \tilde\chi_2^0, \tilde\chi_1^0)$ are larger than a few~\GeV.
The \ninoone is set to be bino-like, while the \ninotwo and \chinoonepm are wino-like, to facilitate the decays considered.
The coupling to all other SUSY particles is set to 0. The narrow-width approximation is utilised for all decays. The simplified models considered are primarily inspired by the MSSM, but other SUSY models such as the NUHM~\cite{Matalliotakis:1994ft,Ellis:2002iu}, mSUGRA~\cite{Chamseddine:1982jx}, or GMSB~\cite{Giudice:1998bp} can also produce similar final states.

These simplified models result in a range of $\tau$-lepton-rich signatures. The search is split into three orthogonal channels, which are combined in a simultaneous fit, and which contain events with:
\begin{itemize}
\item exactly one hadronically decaying $\tau$-lepton and no electron or muon (\nolep);
\item exactly one hadronically decaying $\tau$-lepton and at least one electron or muon (\onelep);
\item at least two hadronically decaying $\tau$-leptons (\twotau).
\end{itemize}
The additional kinematic information in the \onelep and \twotau channels is used to separate signal from background more efficiently. In principle, a \textsc{0tau1lep} channel would also be sensitive to the considered signal models, but the higher branching ratio and competitive reconstruction and identification efficiency for hadronically decaying $\tau$-leptons make \nolep a more promising channel. The \textsc{0tau1lep} channel would also overlap heavily with the dedicated searches for SUSY with exactly one electron or muon in the final state~\cite{SUSY-2018-10}, while \nolep is complementary to it.

The analysis uses two independent approaches. In the first one, machine learning (ML) classifiers are used to split the whole available phase space into various analysis regions. This is the most inclusive way to analyse the data because it uses every event (with a minor exception discussed below), and no additional kinematic requirement is applied. The second is based on the more traditional cut-and-count approach used in previous search iterations~\cite{SUSY-2016-01,SUSY-2016-30}. The event selection is tailored to target different signal topologies and suppress the main \SM backgrounds.
The use of kinematic variables makes the interpretation simpler than in the ML approach. The combination of improved analysis techniques, a larger data sample, and enhanced calibration and identification algorithms for jets and $b$-jets, light leptons (electrons and muons, denoted by $\ell$), and $\tau$-leptons, significantly improves the sensitivity of the search to new physics.

The \paper is structured as follows. In Section~\ref{sec:detector}, a brief description of the ATLAS detector is given. Section~\ref{sec:samples} offers an overview of the Monte Carlo (MC) samples used by the analysis. Section~\ref{sec:reconstruction} describes the reconstruction, identification, and calibration procedures for final-state objects.
The machine-learning and cut-and-count approaches are presented in Section~\ref{sec:analysis_strategy}. Section~\ref{sec:faketau} discusses the data-driven estimations of events with jets misidentified as $\tau$-leptons (fake taus), common to both approaches. The statistical procedures and systematic uncertainties affecting the analysis are gathered in Section~\ref{sec:systematics}. Finally, the results are presented in Section~\ref{sec:result}.


\FloatBarrier
\section{ATLAS detector}
\label{sec:detector}

%

The ATLAS experiment~\cite{PERF-2007-01,GENR-2019-02} is a multipurpose cylindrical particle detector at the LHC with nearly \(4\pi\) coverage in
solid angle.\footnote{ATLAS uses a right-handed coordinate system with its origin at the nominal interaction point (IP)
in the centre of the detector and the \(z\)-axis along the beam pipe.
The \(x\)-axis points from the IP to the centre of the LHC ring,
and the \(y\)-axis points upwards.
Cylindrical coordinates \((r,\phi)\) are used in the transverse plane,
\(\phi\) being the azimuthal angle around the \(z\)-axis.
The pseudorapidity is defined in terms of the polar angle \(\theta\) as \(\eta = -\ln \tan(\theta/2)\).
Angular distance is measured in units of \(\Delta R \equiv \sqrt{(\Delta\eta)^{2} + (\Delta\phi)^{2}}\).
}
Located at the heart of the ATLAS detector, the inner tracking detector begins $33$~mm from the beam axis and is surrounded by a thin superconducting solenoid
providing a \SI{2}{\tesla} axial magnetic field. It consists of a silicon pixel detector (starting with the \enquote{insertable B-layer}~\cite{ATLAS-TDR-2010-19,PIX-2018-001} added before \RunTwo), a silicon microstrip tracker, and a transition radiation tracker, and covers the pseudorapidity range \(|\eta| < 2.5\).
Beyond the solenoid, lead/liquid-argon (LAr) sampling calorimeters measure
the energy and position of electromagnetic showers with high granularity.
In the central region \(|\eta| < 1.7\) the steel/scintillator-tile calorimeter provides
hadronic energy measurements. In the endcap and
forward regions, LAr calorimeters with copper or tungsten absorbers
are used for both electromagnetic and hadronic shower measurements up
to \(|\eta| = 4.9\). The outermost layer of the ATLAS detector is the muon
spectrometer. It consists of separate trigger and high-precision
tracking chambers, operating in a magnetic field generated by three large
air-core toroidal superconducting magnets with eight coils each.

A two-level trigger system is used to select and record events of interest. %
The level-1 trigger is implemented in custom hardware. It uses information from the calorimeters and the muon spectrometer
to accept events at a rate below \SI{100}{\kHz}. This is followed by the high-level trigger, which is fully software-based~\cite{TRIG-2016-01,TRIG-2022-01}. It is used to reduce the average accepted event rate to \qty{1.25}{\kHz} in \RunTwo and \qty{3}{\kHz} in \RunThr.

An extensive software suite~\cite{ATL-SOFT-PUB-2021-001} is used in data simulation, in the reconstruction and analysis of real and simulated data, in detector operations, and in the trigger and data acquisition systems of the experiment.


\FloatBarrier
\section{Data and simulated event samples}
\label{sec:samples}

The analysis is performed on two data samples collected with the ATLAS detector during proton--proton collisions at $\sqrt{s} = 13$~\TeV\ and $\sqrt{s} = 13.6$~\TeV
with a bunch-crossing separation of 25~ns. The $13$~\TeV\ data sample corresponds to an integrated luminosity of $140~\mathrm{fb}^{-1}$ collected during 2015--2018 (\RunTwo of the LHC). The $13.6$~\TeV\ data sample was recorded during 2022--2023 (early \RunThr) and corresponds to $51.8~\mathrm{fb}^{-1}$. The luminosity is measured mainly by the LUCID-2 detector~\cite{LUCID2}, which records Cherenkov light produced in the quartz windows of photomultipliers located close to the beampipe. A set of data quality requirements is applied to the data to ensure that the LHC beam conditions were stable and the ATLAS detector was fully functional~\cite{DAPR-2018-01}. The quoted integrated luminosity is computed after the rejection of luminosity blocks failing these requirements.

All events used by the analysis are required to pass a missing transverse momentum (\met) trigger~\cite{TRIG-2016-01,TRIG-2022-01,TRIG-2019-01} requirement. During data-taking, the selection thresholds increased gradually from $70$ to $115$~\GeV, and the computation algorithms were modified, e.g. by including information on charged-particle tracks~\cite{TRIG-2022-01}.
These triggers are able to select signal events with an efficiency of $80$\%--$90\%$ for most of the models considered, and around $40\%$ for the most compressed scenarios (i.e. the $\masssplit \lessapprox 100$~\GeV region).
The analysis also relies on single-lepton~\cite{TRIG-2018-01,TRIG-2018-05} and single-jet~\cite{TRIG-2012-01} triggers in some low-\met regions used for auxiliary measurements to estimate how many jets are misidentified as $\tau$-leptons (further described in Section~\ref{sec:faketau}).

The analysis makes use of MC simulations to model the \SM backgrounds and SUSY signals.
The event generators and the cross-section computation accuracies are summarised in Table~\ref{tab:samples_desc_run2}.
The ATLAS simulation framework~\cite{SOFT-2010-01,SOFT-2022-02} was used to produce all MC samples.
Its detailed simulation of the detector response based on \GEANT~\cite{Agostinelli:2002hh} ensures that the same reconstruction and identification algorithms can be applied to data and MC events.
The effect of multiple interactions in the same and neighbouring bunch crossings (\pileup) was modelled by overlaying each simulated hard-scattering event with inelastic $pp$ collision events.
The \pileup model combines collision events from \AtlasMC[2.0.1.4]{EPOS}~\cite{Porteboeuf:2010um}, using the LHC tune, and  \PYTHIA[8]~\cite{Sjostrand:2007gs}, using the A3 tune~\cite{ATL-PHYS-PUB-2016-017} and the \NNPDF[2.3lo] set of parton distribution functions (PDFs)~\cite{Ball:2012cx}.

The two main \SM background categories consist of events with jets misidentified as $\tau$-leptons (fake taus, modelled inclusively for all production processes with a data-driven approach described in Section~\ref{sec:faketau}), and \ttbar and single-top-quark events, collectively referred to as \enquote{Top}. The $W(\tau\nu)$+jets event contribution is significant in the \nolep channel, while diboson and $Z(\tau\tau)$+jets events become more relevant in the \onelep and \twotau channels. Events from all other \SM processes (\Wjets and \Zjets without $\tau$-leptons, and various \ttV and Higgs boson production processes) are grouped into the \enquote{Other} category.

Independent SUSY signal event samples were simulated for different simplified models.  The range of gluino (squark) masses considered is $0.5$--$2.6~(0.4$--$2.0)$~\TeV, and the corresponding range of \ninoone masses is $0.1$--$1.7~(0.1$--$1.3)$~\TeV, with a granularity of $200$~\GeV for \susymass and $100$~\GeV for $m(\ninoone)$ (with additional mass points in the compressed region). The matrix element (ME) calculation is performed at tree level and includes the emission of up to two additional partons. The ME--parton-shower matching uses the CKKW-L prescription~\cite{Lonnblad:2001iq,Lonnblad:2011xx}, with the merging scale set to one quarter of the mass of pair-produced particles.
Signal cross-sections are calculated to approximate next-to-next-to-leading order in the strong coupling constant, adding the resummation of soft gluon emission at next-to-next-to-leading-logarithm accuracy (giving \enquote{approximate NNLO+NNLL} accuracy)~\cite{Beenakker:2024nlf,Beenakker:2016lwe,Beenakker:2014sma,Beenakker:2013mva,Beenakker:2011sf,Beenakker:2009ha,Kulesza:2009kq,Kulesza:2008jb,Beenakker:1996ch}. The nominal cross-section and its uncertainty are derived using the \PDFforLHC[21\_40\_pdfas] PDF set, following the recommendations of Ref.~\cite{Ball:2022oua}.

\begin{table}[!h]
\centering
\caption{
Summary of the generators used to simulate the signal and background processes.
The \enquote{dedicated} tune is the set of tuned parton-shower parameters developed by the \SHERPA authors. The symbol \enquote{$V$} stands for a vector boson.}
\label{tab:samples_desc_run2}
\begin{adjustbox}{width=1\textwidth}
\begin{tabular}{llllllll}
\toprule
Physics process     & $\sqrt{s}$~[\TeV] & Generator 	& Parton Shower 	&  Accuracy & Tune 			& PDF (generator) & PDF (shower)\\
\midrule
$Z(\ell\ell )$+jets    & $13$ &  \SHERPA[2.2.11]~\cite{Bothmann:2019yzt}                        & \SHERPA[2.2.11]~\cite{Schumann:2007mg}         & NNLO~\cite{Anastasiou:2003ds}                 & Dedicated       & \NNPDF[3.0nnlo]~\cite{Ball:2014uwa,Carrazza:2015aoa} & \NNPDF[3.0nnlo] \\
& $13.6$ &  \SHERPA[2.2.14]  & \SHERPA[2.2.14]   & NNLO   & Dedicated       & \NNPDF[3.0nnlo] & \NNPDF[3.0nnlo] \\
$Z(\tau\tau )$+jets & $13/13.6$   & \SHERPA[2.2.14]  & \SHERPA[2.2.14] & NNLO  & Dedicated       & \NNPDF[3.0nnlo] & \NNPDF[3.0nnlo] \\
$W(\ell\nu )$+jets & $13$   & \SHERPA[2.2.11] & \SHERPA[2.2.11] & NNLO   & Dedicated     & \NNPDF[3.0nnlo] & \NNPDF[3.0nnlo] \\
& $13.6$   & \SHERPA[2.2.14] & \SHERPA[2.2.14] & NNLO   & Dedicated     & \NNPDF[3.0nnlo] & \NNPDF[3.0nnlo] \\
$VV$, $VVV$  & $13$  &  \SHERPA[2.2.1],\,2.2.2,\,2.2.14  &   \SHERPA[2.2.1],\,2.2.2,\,2.2.14 & NLO~\cite{Hoeche:2011fd,Hoeche:2012yf,Catani:2001cc,Hoeche:2009rj}                  &  Dedicated  &\NNPDF[3.0nnlo] & \NNPDF[3.0nnlo] \\
& $13.6$  &  \SHERPA[2.2.14] &   \SHERPA[2.2.14] & NLO  &  Dedicated  &\NNPDF[3.0nnlo] & \NNPDF[3.0nnlo] \\
$t\bar{t}$ & $13/13.6$ & \POWHEGBOX[v2]~\cite{Frixione:2007nw,Nason:2004rx,Frixione:2007vw,Alioli:2010xd}       		& \PYTHIA[8]~\cite{Sjostrand:2014zea}  		& NNLO+NNLL~\cite{Beneke:2011mq,Cacciari:2011hy,Baernreuther:2012ws,Czakon:2012zr,Czakon:2012pz,Czakon:2013goa,Czakon:2011xx} 		& A14~\cite{ATL-PHYS-PUB-2014-021} & \NNPDF[3.0nlo]& \NNPDF[2.3lo]~\cite{Ball:2012cx}\\
Single-top: \\
$~~~~Wt$  & $13/13.6$  & \POWHEGBOX[v2] 	& \PYTHIA[8]	&NLO+NNLL~\cite{Kidonakis:2010ux,Kidonakis:2013zqa}	& A14	& \NNPDF[3.0nlo] & \NNPDF[2.3lo]    \\
~~~~$s$- and $t$-channels & $13/13.6$     & \POWHEGBOX[v2] & \PYTHIA[8] &NLO~\cite{Aliev:2010zk,Kant:2014oha}  & A14   & \NNPDF[3.0nlo]  & \NNPDF[2.3lo]      \\
$t\bar{t}{+}V$ & $13/13.6$    & \MGNLO[2.3.3]~\cite{Alwall:2014hca}      & \PYTHIA[8] & NLO~\cite{deFlorian:2016spz_fix}      & A14 & \NNPDF[3.0nlo] & \NNPDF[2.3lo] \\
$t\bar{t}{+}H$ & $13/13.6$ & \POWHEGBOX[v2]~\cite{Hartanto:2015uka}  & \PYTHIA[8]& NLO	~\cite{Alwall:2014hca} & A14 & \NNPDF[3.0nlo] & \NNPDF[2.3lo]\\

$V H (\tau\tau)$ & $13/13.6$ &\POWHEGBOX[v2]~\cite{Alioli:2010xd,Hamilton:2013fea,Hamilton:2015nsa} & \PYTHIA[8] &NNLO + NLO ~\cite{Ciccolini:2003jy,Brein:2003wg,Brein:2011vx,Altenkamp:2012sx,Denner:2014cla,Brein:2012ne,Harlander:2014wda}   &AZNLO~\cite{STDM-2012-23}      & \NNPDF[3.0nlo] & \NNPDF[2.3lo] \\
$ggH(\tau\tau)$ & $13$  & \POWHEGBOX v2 & \PYTHIA[8]    & NNLL+NNLO  & AZNLO & \NNPDF[3nlo] & \NNPDF[2.3lo]\\
& $13.6$  &  \POWHEGBOX[v2] & \PYTHIA[8] & NNLL+NNLO ~\cite{Hamilton:2012np,Campbell:2012am,Hamilton:2012rf,Catani:2007vq,Bozzi:2005wk,deFlorian:2011xf} & A14 & \NNPDF[3nlo]  & \NNPDF[2.3lo]\\
VBF $H(\tau\tau) $ & $13$ & \POWHEGBOX v2	& \PYTHIA[8] & NNLO+NLO	& AZNLO	& \NNPDF[3nlo]	& \NNPDF[2.3lo] \\
& $13.6$ & \POWHEGBOX[v2] & \PYTHIA[8]  & NNLO+NLO & A14 & \NNPDF[3nlo]   & \NNPDF[2.3lo] \\
SUSY signal & $13/13.6$ & \MADGRAPH[3.3.1] & \PYTHIA[8] &  NNLO+NNLL~\cite{Beenakker:2024nlf,Beenakker:2016lwe,Beenakker:2014sma,Beenakker:2013mva,Beenakker:2011sf,Beenakker:2009ha,Kulesza:2009kq,Kulesza:2008jb,Beenakker:1996ch} &  A14 & \NNPDF[3.0nlo]  &  \NNPDF[2.3lo]  \\
\bottomrule
\end{tabular}
\end{adjustbox}
\end{table}


\FloatBarrier
\section{Event reconstruction and kinematic variables}
\label{sec:reconstruction}

%

The reconstruction algorithms and selection criteria described in this section are harmonised across \RunTwo and \RunThr. If an event contains several objects of the same type, they are ordered in decreasing transverse momentum, $\pt$.

Primary vertices are reconstructed from inner-detector tracks with \(\pt>\SI{500}{\MeV}\)~\cite{PERF-2015-01,ATL-PHYS-PUB-2019-015}. The primary vertex with the largest sum of squared track \pt is chosen as the hard-scatter vertex, and serves as the origin for all the objects described below except for $\tau$-leptons, where a dedicated algorithm is used~\cite{ATL-PHYS-PUB-2022-044}.

Jets are reconstructed from particle-flow objects~\cite{PERF-2015-09} using the \antikt\ algorithm~\cite{Cacciari:2008gp,Fastjet} with a radius parameter of $R = 0.4$. Jets undergo a sequence of calibrations which include components derived both from simulation and in situ measurements to correct for the difference between data and MC simulations~\cite{JETM-2018-05}. Only jets with $\pt > 20$~\GeV\ and $|\eta| < 2.8$ are considered, this avoids objects whose energy is split between endcap and forward calorimeters. A neural-network-based jet vertex tagger (NNJVT)~\cite{PERF-2014-03} using the default (\enquote{FixedEffPt}) working point is applied to jets with $\pt < 60$~\GeV\ and $|\eta| < 2.5$ to suppress those from \pileup.
Jets originating from the hadronisation of $b$-quarks are tagged with the DL1dv01 tagger~\cite{ATL-PHYS-PUB-2022-047,ATL-PHYS-PUB-2020-014,FTAG-2019-07}. It uses a fixed $85\%$-efficiency working point corresponding to a selection that correctly identifies $85\%$ of the $b$-jets with $|\eta| < 2.5$, as determined in a sample of simulated \ttbar\ events,
and is applied only to jets in that $|\eta|$ range. The \(b\)-tagging efficiencies in simulated events are corrected to match those measured in data~\cite{FTAG-2018-01}.

Electron candidates are required to satisfy the \enquote{LooseAndBLayerLLH} identification criteria~\cite{EGAM-2018-01,EGAM-2021-02}, and have $\pt > 10$~\GeV\ and $|\eta| < 2.47$.
In addition, a $|z_0\sin\theta| < 0.5$~mm selection is applied to the longitudinal impact parameter $z_0$ to suppress non-prompt electrons.
Electron candidates have to survive the overlap removal procedure (OR, described below), to satisfy the \enquote{Tight\_VarRad} isolation criterion and the \enquote{TightLLH} electron identification~\cite{PERF-2017-01,EGAM-2018-01,EGAM-2021-02}, and to have a transverse impact parameter $d_{0}$ significance smaller than $5$ to be consider by the analysis.

Muon candidates are required to satisfy the \enquote{Medium} identification criteria~\cite{MUON-2018-03,MUON-2022-01}, and have $\pt > 10$~\GeV, $|\eta| < 2.7$ and $|z_0\sin\theta| < 0.5$~mm. If it survives the OR, a muon candidate is used by the analysis if it also satisfies the \enquote{PflowLoose\_VarRad} isolation criteria~\cite{MUON-2018-03,MUON-2022-01} and has a $d_{0}$ significance smaller than $3$.

Hadronically decaying $\tau$-leptons are seeded by $R = 0.4$ anti-$k_t$ jets reconstructed from topological cell clusters calibrated with the local hadronic cell-weighting scheme~\cite{PERF-2014-07,Barillari:1112035}. The production vertex of the visible part of a hadronic $\tau$-lepton decay (\tauhadvis) is then determined from tracks within $\Delta R = 0.2$ of the axis of the seed jet. The $\tau$-lepton candidate is built using tracks and calorimeter-cell clusters, and undergoes a sequence of calibrations that compose the \enquote{$\tauhadvis$ Energy Scale calibration}~\cite{ATL-PHYS-PUB-2015-045,ATL-PHYS-PUB-2022-044}. The $\tau$-lepton candidates are required to have $\pt > 20$~\GeV, $|\eta| < 2.5$, and one or three tracks associated with the \tauhadvis vertex (as identified by a dedicated classifier~\cite{ATL-PHYS-PUB-2022-044}) with the sum of their charges equal to $\pm1$. The region $|\eta|\in [1.37,1.52]$ is excluded, corresponding to the transition region between the barrel and endcap calorimeters. A classifier based on a recurrent neural network (RNN)~\cite{ATL-PHYS-PUB-2022-044} is used to separate $\tau$-leptons from background jets, using tracking and calorimeter measurement information, including information about individual tracks and clusters, as input. The analysis uses the \enquote{medium} $\tauhadvis$ identification (medium \tauhad ID) working point, corresponding to $75\%$ efficiency for $1$-track and $60\%$ efficiency for $3$-track $\tau$-leptons. To suppress electrons satisfying the medium \tauhad ID selection, the \enquote{Loose} working point of a dedicated electron-veto algorithm based on an RNN classifier is used.
Additionally, $\tau$-lepton candidates that fail to satisfy the medium \tauhad ID selection, but satisfy all other selections and a much looser RNN classifier score $>0.01$ selection, are used by the analysis to estimate the fake-tau background. These objects are referred to as \enquote{anti-ID taus}.

A single energy deposit in the calorimeter or a track in the inner detector can be used in the reconstruction of several analysis objects. To avoid double-counting and resolve the ambiguity, the OR procedure is performed for every event. It ensures that every detector signature is matched to at most one reconstructed object. The OR procedure used follows the one described in Ref.~\cite{HIGG-2019-09}, with the addition that if a $b$-jet is closer than $\Delta R = 0.2$ to an anti-ID tau, the anti-ID tau is discarded, similarly to Ref.~\cite{HDBS-2018-40}.

The missing transverse momentum \met is computed as the magnitude of \vecmet,   %
the negative vector sum of the transverse momenta of electrons, muons, $\tau$-leptons or anti-ID taus, jets, and a \enquote{soft term} composed of tracks from the primary vertex not associated with any reconstructed objects~\cite{JETM-2020-03}.
Only anti-ID taus that are \enquote{promoted} to $\tau$-leptons when modelling the fake-tau backgound (Section~\ref{sec:faketau}) are treated as $\tau$-leptons instead of jets in the computation.
The \enquote{tight} \met\ working point is used, defining the properties of the jets used for the calculation. Three additional variables related to \met are used in the analysis:
\begin{itemize}
\item The transverse mass \mt is defined as
$\mt^2(\vec{p}_{\text{T}}^{\,a}, \vec{p}_{\text{T}}^{\,b})
= 2\left(p_\text{T}^{a}p_\text{T}^{b} - \vec{p}_\text{T}^{\,a} \cdot \vec{p}_\text{T}^{\,b}\right)$,
where $a$ and $b$ represent any two massless particles. When the transverse mass of a singular object is used, e.g.\ \mttauone, the massless $\pt^\mathrm{miss}$ vector is used for the second object in the computation.

\item The stransverse mass \mttwo~\cite{Lester:1999tx,Lester:2014yga} is used to discriminate between processes where two leptons come from a resonant decay and processes where they come from non-resonant decays.
The \mttwo variable is computed as %
\begin{equation}\nonumber
\mttwo = \min_{\vec{p}_\text{T}^{a} + \vec{p}_\text{T}^{b} = \vecmet} \left(\max\left[\mt(\vec{p}_{\text{T}}^{\,\text{lep}_1},\vec{p}_\text{T}^{\,a}) , \mt(\vec{p}_{\text{T}}^{\,\text{lep}_2},\vec{p}_\text{T}^{\,b})   \right]\right),
\end{equation}
where $\vec{p}_\text{T}^{\,\text{lep}_1}$ and $\vec{p}_\text{T}^{\,\text{lep}_2}$ correspond to the transverse momenta of the two leading leptons, at least one of which is a $\tau$-lepton,
and $(a,b)$ refers to two invisible massless particles assumed to be produced with transverse momenta $\vec{p}_\text{T}^{\,a}$ and $\vec{p}_\text{T}^{\,b}$.

\item Object-based \met significance~\cite{ATLAS-CONF-2018-038} is a variable designed to separate events where the measured \met is consistent with being due to detector resolution smearing alone from events where \met is likely to be the result of neutrinos or SUSY particles not interacting with the detector material.
\end{itemize}
Another event-level variable used is $\HT$, the scalar sum of the $\pt$ of jets, $\tau$-leptons, electrons, and muons.


\FloatBarrier
\section{Analysis strategy}
\label{sec:analysis_strategy}

%

All events considered by the analysis are required to satisfy an inclusive preselection to ensure selection in the regime where the triggers operate at full efficiency. Since \met performance depends on the topology of the objects, the preselection affects not only the \met value, but also the jet selection. The events are required to have $\met > 200$~\GeV, at least one hadronically decaying $\tau$-lepton, and at least two jets, one of which has $\pt > 120$~\GeV. Additionally, to suppress contributions from QCD multijet processes, where the \met comes from mismeasurements of jet energies, the angular separation between the two leading jets and $\met$ is required to be larger than $0.4$ in the transverse plane.

Several signal regions (SRs) that are sensitive to the SUSY models with $\tau$-lepton\,+\,\met signatures are defined. The common strategy for the design of the SRs is the multi-bin approach, where not only the yields, but also the shape of the most sensitive variable is utilised. The same kinematic-variable selections are used for \RunTwo and \RunThr data samples, but the binning of each variable changes to ensure that at least five events are expected in each bin before any data-driven corrections are applied. All SRs are optimised using Asimov signal significance as the metric of merit~\cite{Cowan:2010js}.

Events with $\tau$-leptons that either originate from $\tau_\text{had}$ decays (true $\tau$-leptons) or from light leptons misidentified as $\tau$-leptons are modelled using MC simulation with data-driven corrections.
Dedicated control regions (CRs) enriched in major backgrounds are used to derive normalisation factors (NFs) in simultaneous fits of MC predictions to data (further described in Section~\ref{sec:systematics}). These affect only the overall normalisation of the MC simulation (i.e. the CRs are not binned in any variable), and only the events without jets misidentified as $\tau$-leptons. Validation regions (VRs) are defined to be kinematically closer to the SRs than the CRs. The results of the fit are propagated from the CRs to the VRs, to validate the extrapolation to the SRs. Events with jets misidentified as $\tau$-leptons are estimated in a data-driven way, as discussed in Section~\ref{sec:faketau}. These events are not normalised to the data, and no CRs are needed for them. The following subsections describe two different approaches used to define the SRs and CRs.

\subsection{Machine-learning approach}

The first approach is the ML-based analysis. The key ideas for this approach are that no kinematic variables are used to further define any of the regions except the classifier scores, and that the whole data sample is used inclusively.
This approach allows the classifier to fully evaluate and utilise the correlation of all variables it is trained on. A multi-class approach is chosen, with one category for signal and multiple categories for \SM processes. The different scores are normalised such that they are always positive and sum to $1$ for every event.

\begin{figure}[htb!]
\centering
\subfloat[\RunTwo \nolep]{\includegraphics[width=0.45\textwidth]{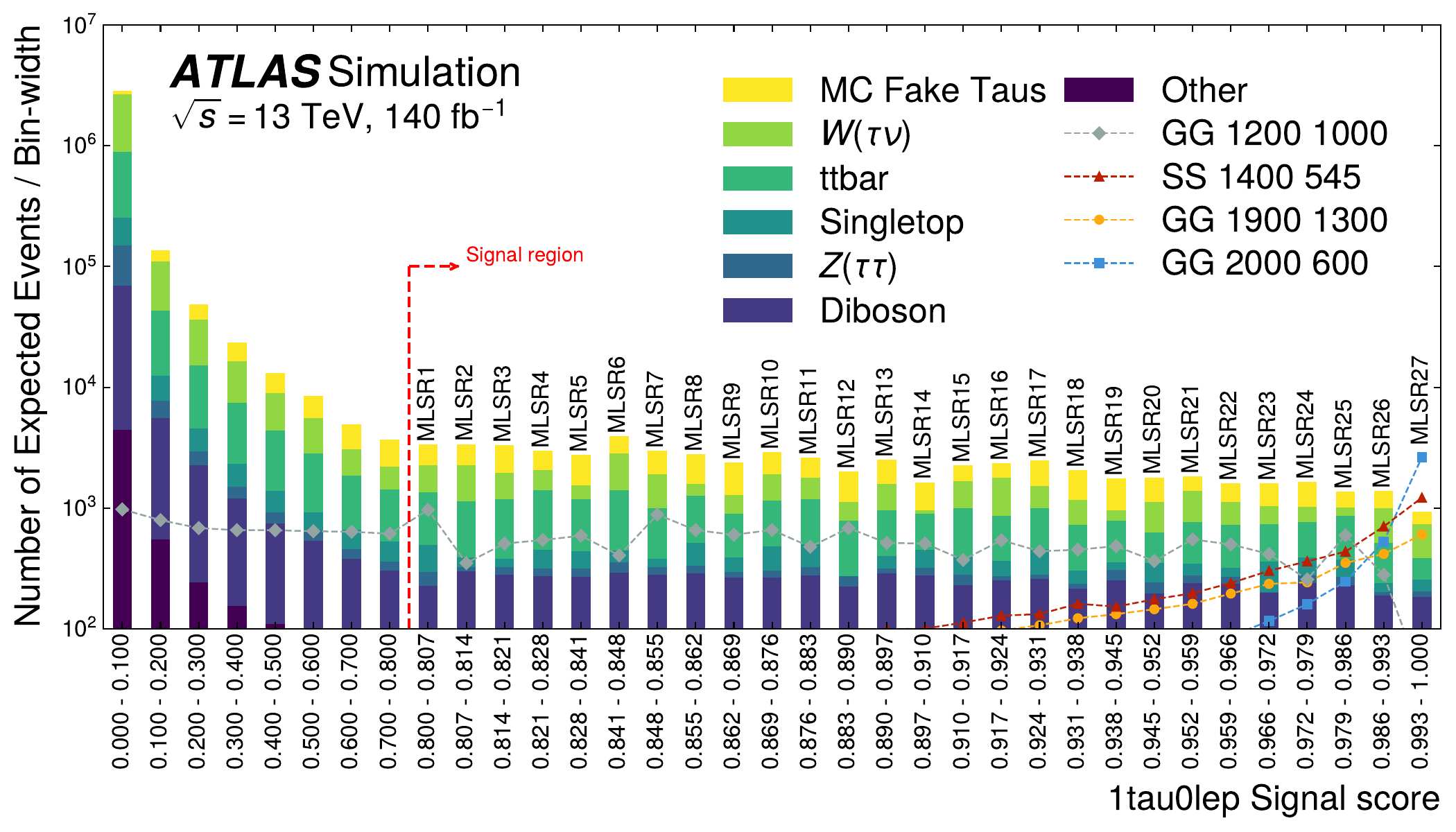}}
~~\subfloat[\RunThr \nolep]{\includegraphics[width=0.45\textwidth]{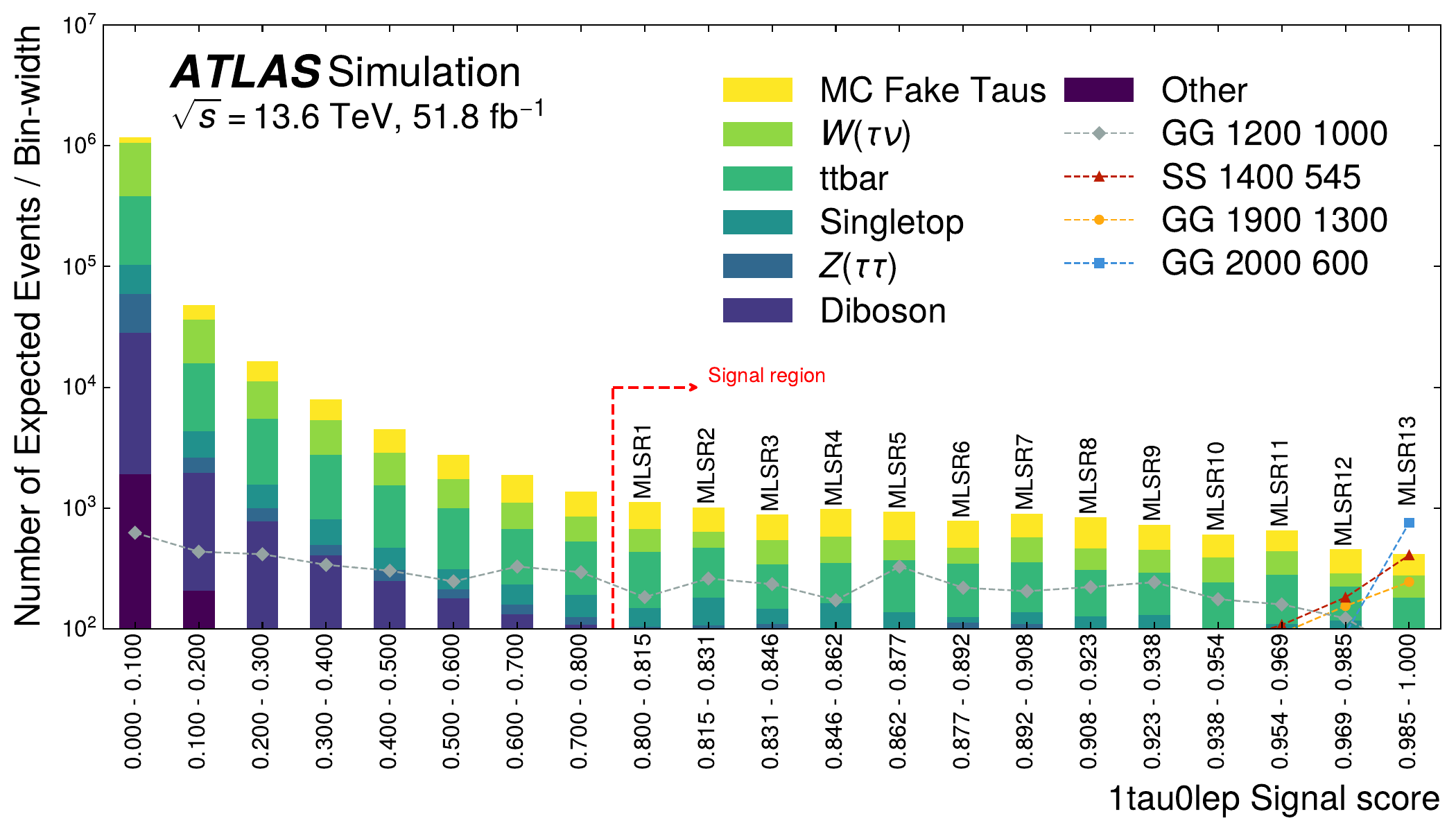}}\\
\subfloat[\RunTwo \onelep]{\includegraphics[width=0.45\textwidth]{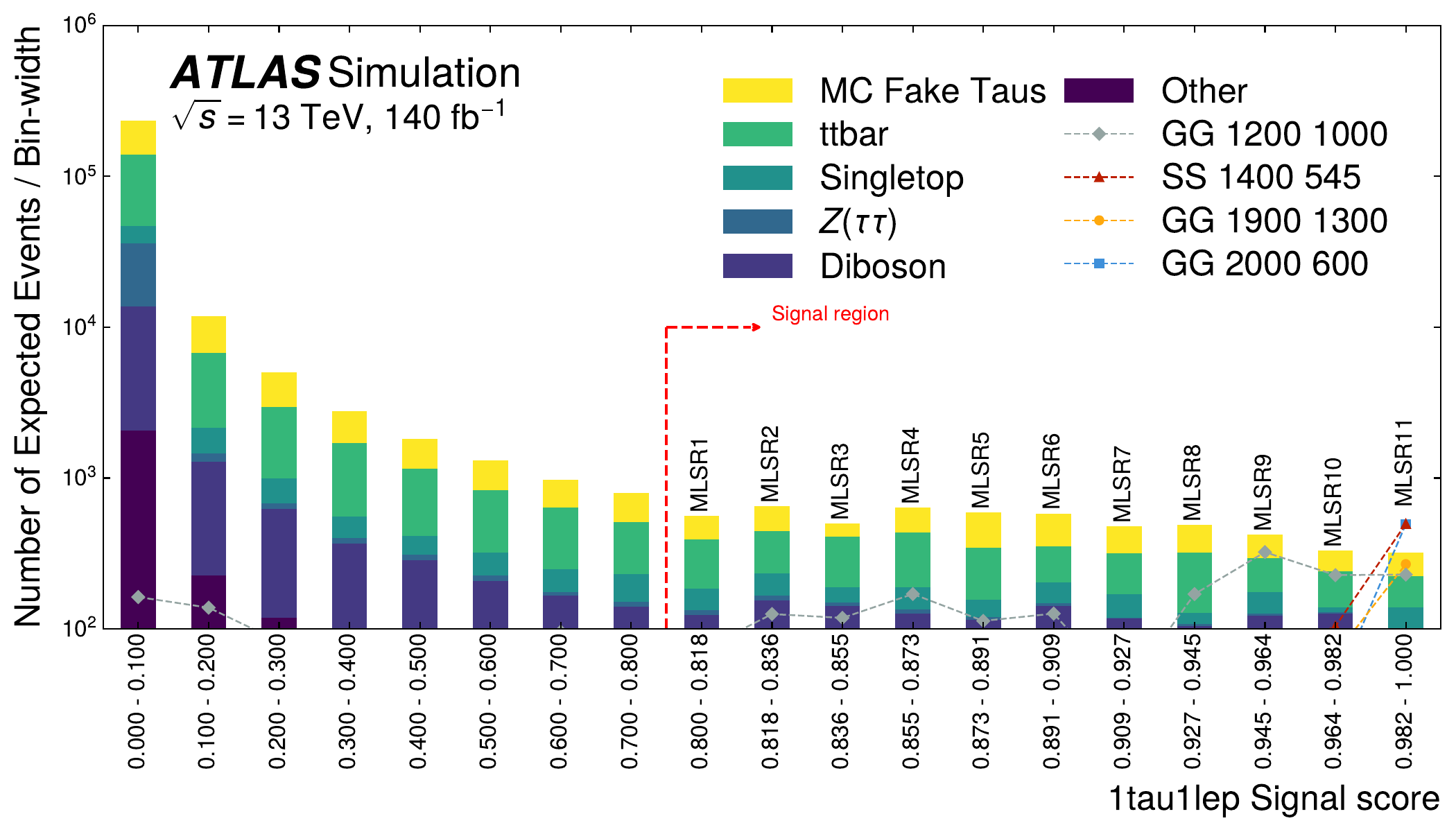}}
~~\subfloat[\RunThr \onelep]{\includegraphics[width=0.45\textwidth]{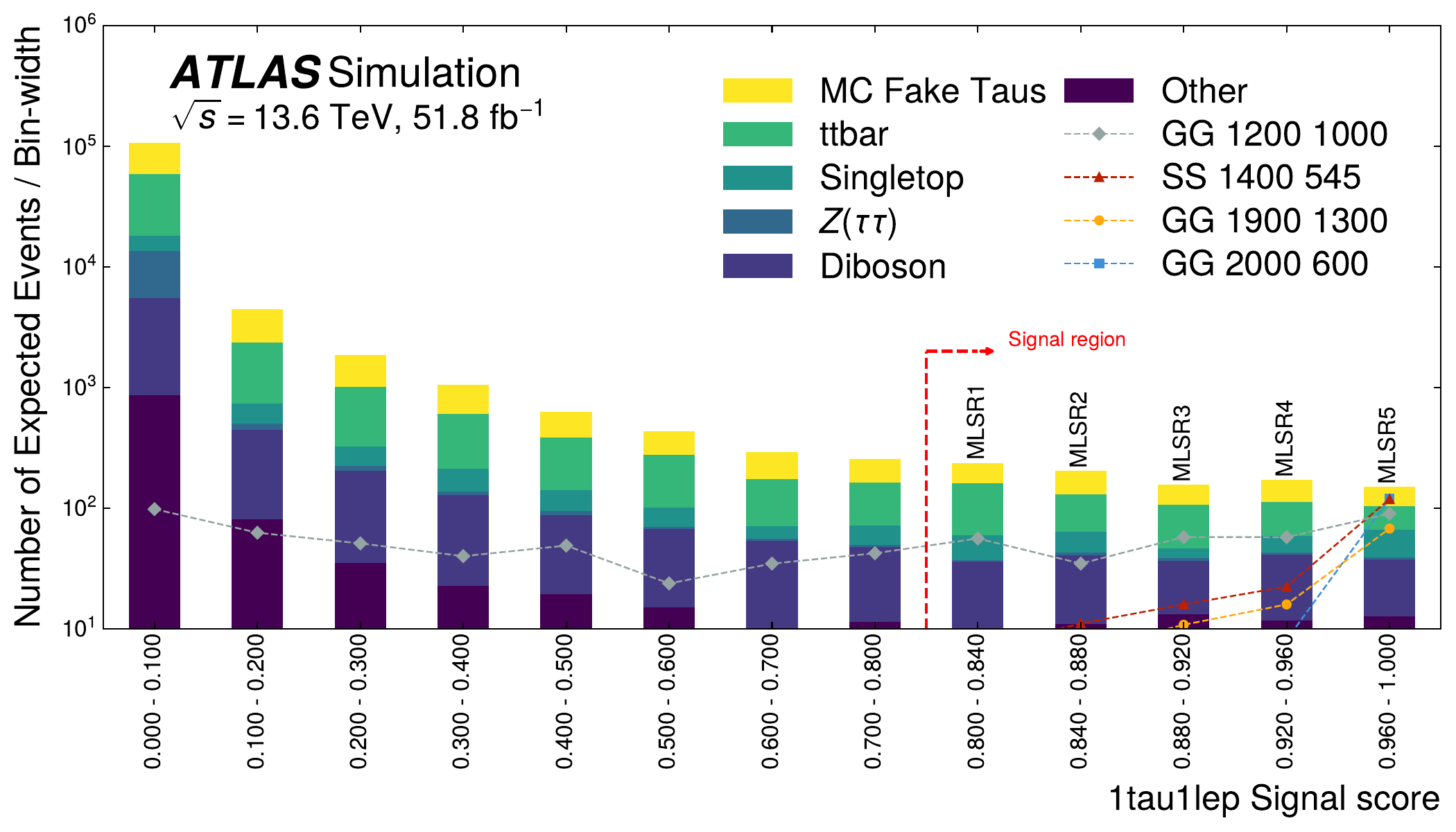}}\\
\subfloat[\RunTwo \twotau]{\includegraphics[width=0.45\textwidth]{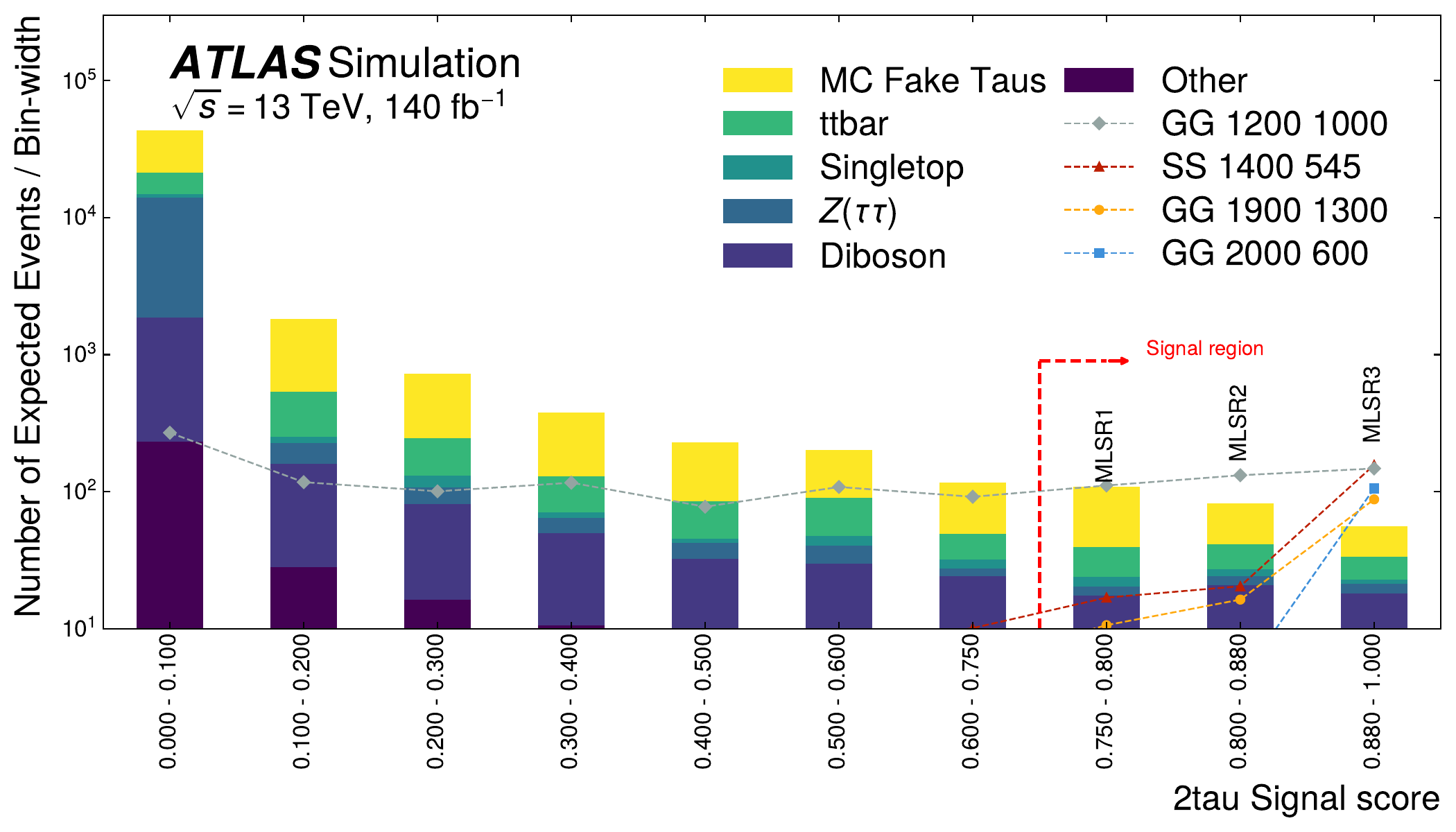}}
~~\subfloat[\RunThr \twotau]{\includegraphics[width=0.45\textwidth]{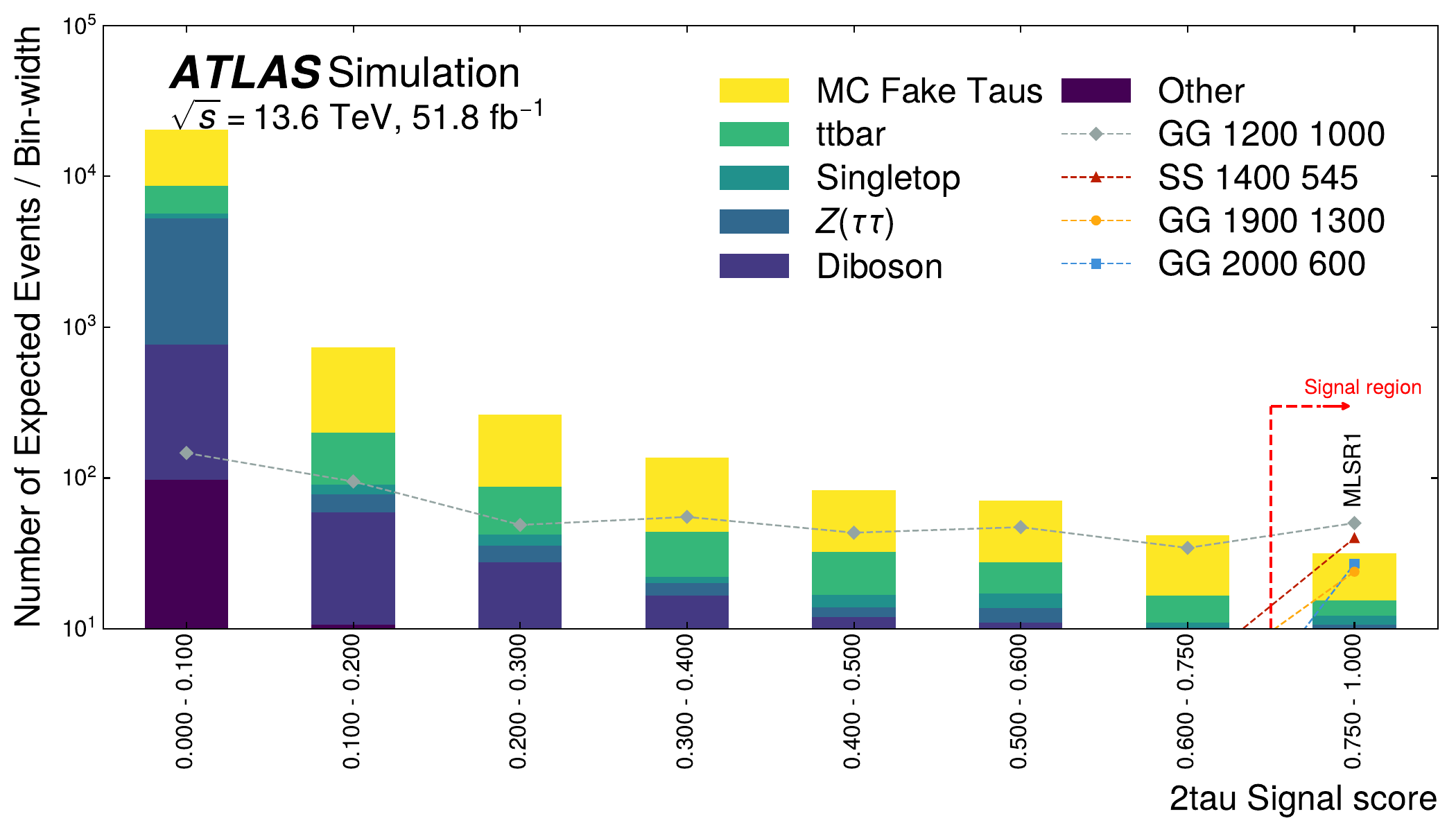}}
\caption{Expected background composition across the range of signal scores. The numbers of events shown are normalised to their bin-width. Benchmark signal points are overlayed. %
}
\label{fig:ml_expected_sr_comp}
\end{figure}

Three boosted decision trees (BDTs) are trained using the XGBoost library~\cite{Chen:2016:XST:2939672.2939785}, one for each of the three channels (\nolep, \onelep, \twotau).
Four-momentum and high-level kinematic variables associated with $\tau$-leptons and up to three leading jets are used for training, along with event-level variables such as \met, \HT, and \met significance. In particular, the addition of the transverse mass and angular separation of the $\tau$-lepton and \MET improves the performance significantly. Lepton kinematic variables, including \mttwo and the invariant mass of the leading lepton and $\tau$-lepton pair, are also used for training in the \onelep channel. Transverse masses of $\tau$-leptons, light leptons, and jets are some of the most important variables. The total number of leptons of each flavour and the stransverse mass \mttwo of the two $\tau$-leptons are used as additional input variables for the \twotau channel. The sum of the transverse masses of the $\tau$-leptons is also used, and is one of the most powerful variables.
Only simulated events are used for the training, including those with fake taus. A mixture of \RunTwo and \RunThr samples is used in the training of common classifiers; this approach outperforms separate training for \RunTwo and \RunThr because of the larger training samples. The training samples are rescaled with class weights such that the class distributions are balanced. All simplified SUSY model simulations are included as a common Signal class, ensuring that a wide range of kinematic signatures is covered.

\begin{figure}[htbp!]
\centering
\subfloat[\nolep]{\includegraphics[width=0.7\textwidth]{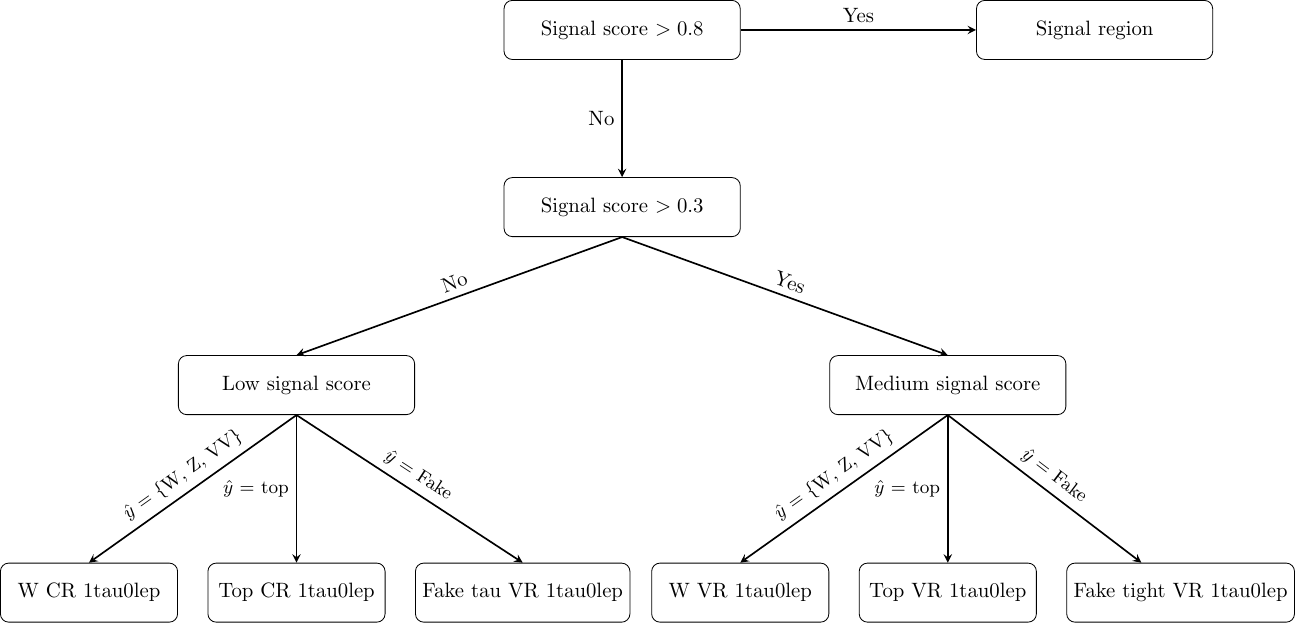}}\\
\subfloat[\onelep]{\includegraphics[width=0.7\textwidth]{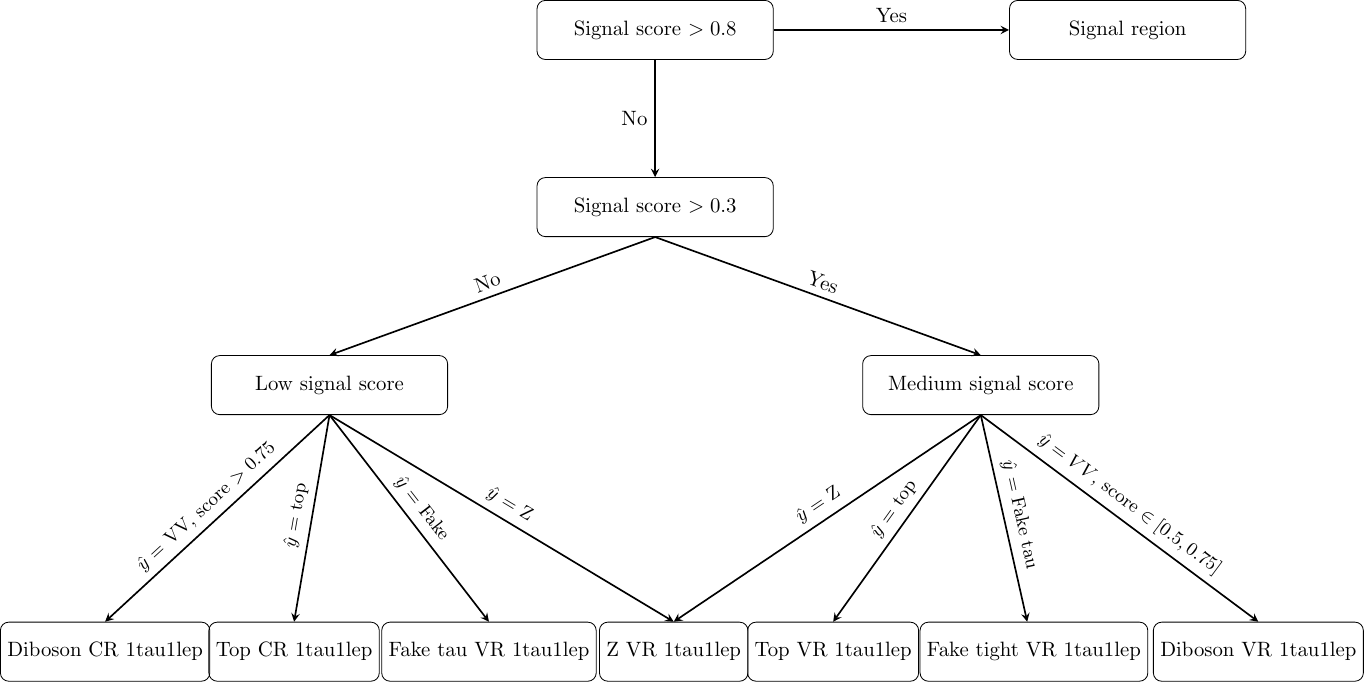}}\\
\subfloat[\twotau]{\includegraphics[width=0.5\textwidth]{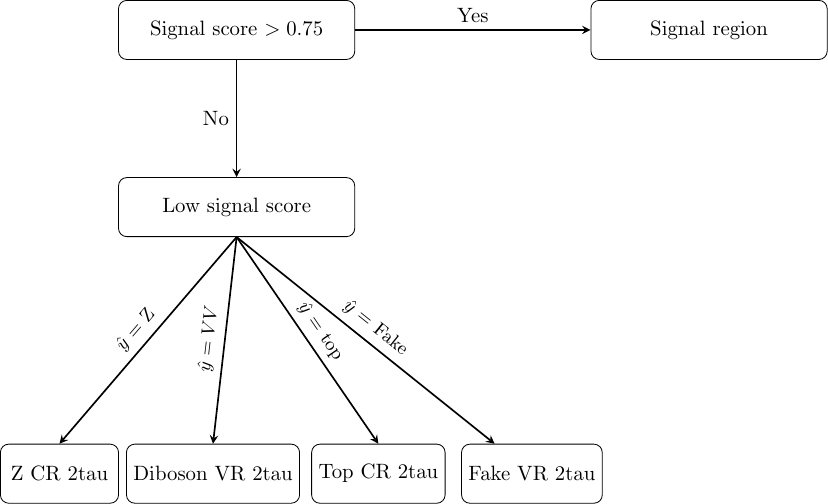}}
\caption{Schematic definitions of the ML-based SRs, CRs and VRs. The diboson VR in the \onelep channel additionally relaxes the signal score requirement to $>0.15~(0.05)$ for \RunTwo (\RunThr).}
\label{fig:ml_regions}
\end{figure}

Each BDT is trained using five-fold cross-validation~\cite{crossval} and early stopping is implemented, triggering whenever the validation loss does not improve after 100 iterations. The maximum number of trees in the ensemble, per training fold, is taken to be $10\,000$, but the early stopping mechanism always terminates the training before reaching this number. The typical number of iterations is around $15\,000$ in the \nolep channel, $5\,000$ in the \onelep channel, and below $2\,000$ in the \twotau channel.
The loss function is defined as the multi-class logloss~\cite{068cb646-8936-3df0-aa16-e97573eda53c}.
Hyperparameter optimisation was studied extensively, using both traditional grid searches and a modern optimiser using the Optuna software package~\cite{optuna_2019}.
A tree depth of five is found to maximise performance and minimise overtraining. For the regularisation parameters, the minimum information needed to decide on a node split ($\gamma$) is found to be the most important and is taken to be 0.1. The training is stable with respect to variations of other hyperparameters, so they are set to their default values. The training is optimised separately for the three search channels. However, utilising early stopping, the channels are found to converge to stable minima using the same set of hyperparameters with a learning rate of 0.01.

The final classification of an event is done in two steps. First, a selection based on the signal score is performed, separating events with high/medium/low signal scores into SRs/VRs/CRs. The high scores are selected to maximise the sensitivity to all the possible signal models considered, while the lower scores ensure that the signal contamination is below $1\%$ in the CRs. Then the CRs and VRs are split according to the the most likely classification if the \enquote{Signal} class is ignored, i.e.\ the argument of the maxima of all the non-signal scores $\hat{y} \in \{Z,W,VV,\text{Top},\text{fake tau}\}$. For example, $\hat{y} \in \{Z,VV\}$ refers to the set of events that enter the \enquote{$Z$} or \enquote{$VV$} class when the signal score is ignored.

In the \nolep channel, six different classes are defined -- \Wjets, \Zjets, diboson, Top (\ttbar + single-top-quark production), fake taus, and Signal. The SRs are defined by selecting events with a signal score larger than $0.8$. The events selected in this way are then binned in signal score. This approach utilises the information encoded in the shape of the classifier score distribution, increasing the sensitivity and constraining the detector-related systematic uncertainties. Since the \nolep channel has the largest yields, 27 (13) bins are defined in \RunTwo (\RunThr) \nolep SRs. The dominant backgrounds are due to \Wjets events, fake taus, and Top processes. %

The contributions of \Wjets events with true $\tau$-leptons to the \onelep and \twotau channels are strongly suppressed by the requirement of at least one additional lepton. Therefore, only five classes are used -- \Zjets, diboson, Top, fake taus, and Signal. In the \onelep channel, only events with a signal score larger than $0.8$ are selected, and they are then binned in signal score. In total, 11 (5) bins are defined in the \RunTwo (\RunThr) \onelep SRs. In the \twotau channel the signal score requirement is relaxed to $>0.75$ as the signal separation power of the classifier is much stronger; here, 3 bins are defined for \RunTwo, and 1 for \RunThr.

The expected \SM background distributions binned in the BDT scores for all channels are shown in Figure~\ref{fig:ml_expected_sr_comp}. The binning in the SRs corresponds to the one used for the statistical analysis. The fake-tau contribution is estimated from the MC simulation on which the classifiers are trained. Yields from several benchmark signal models are overlayed, corresponding to the low-\masssplit, high-\susymass, and high-$m(\ninoone)$ signal-parameter phase space. The convention for benchmark signal-model names is \enquote{GG/SS $\susymass$ $m(\ninoone)$} where GG/SS refers to the gluino/squark model, and the masses are given in~\GeV. The signal-model kinematic properties change with lower \masssplit values, becoming more background-like in the compressed region, which leads to a flatter classifier score distribution.

\begin{figure}[htb!]
\centering
\subfloat[\RunTwo, Top CR \nolep, $m_\mathrm{T}^{\tau_1}$]{\includegraphics[width=0.32\textwidth]{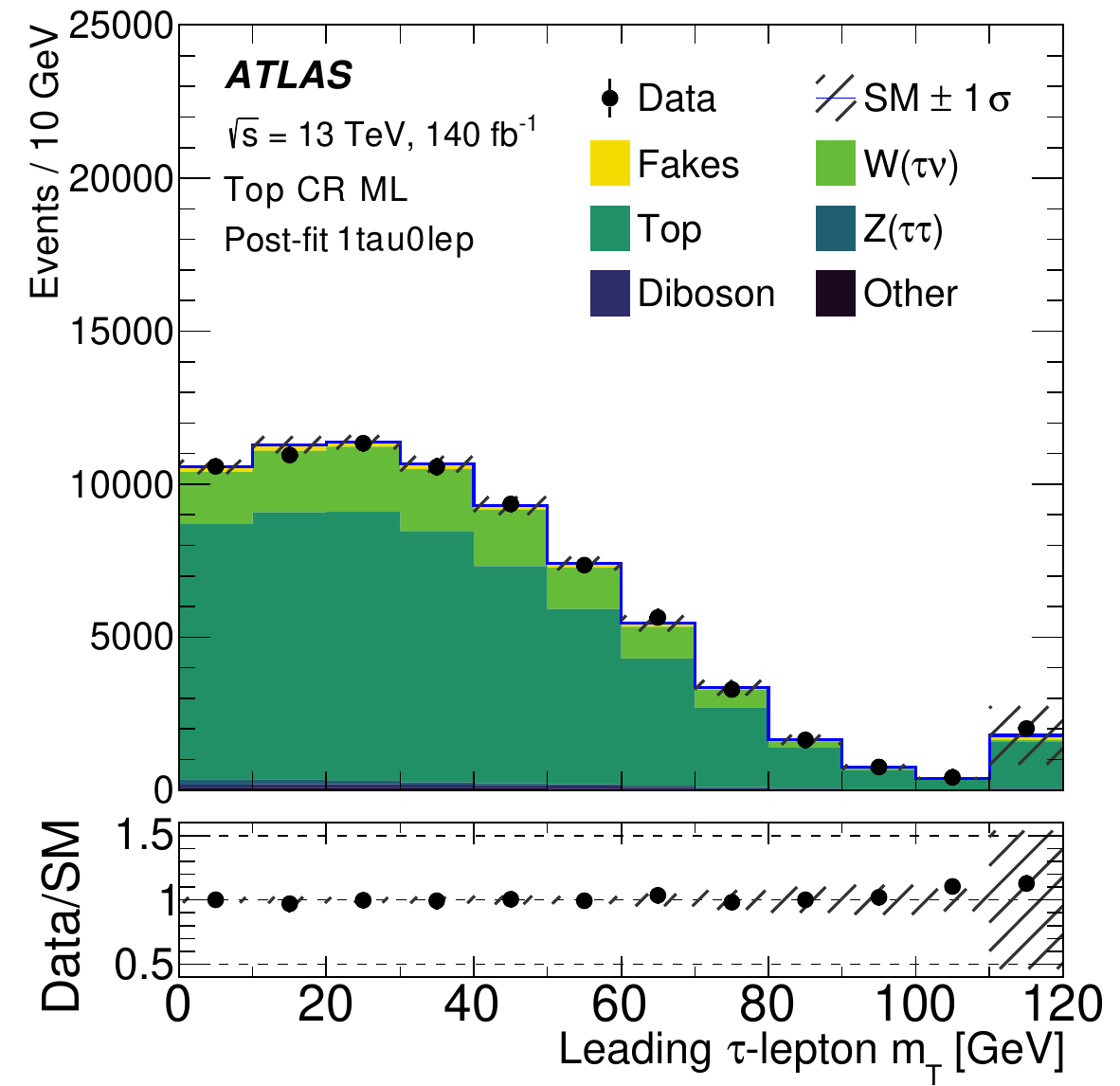}}
\subfloat[\RunTwo, $Z(\tau\tau)$ CR \twotau, $m_\mathrm{T}^{\tau_2}$]{\includegraphics[width=0.32\textwidth]{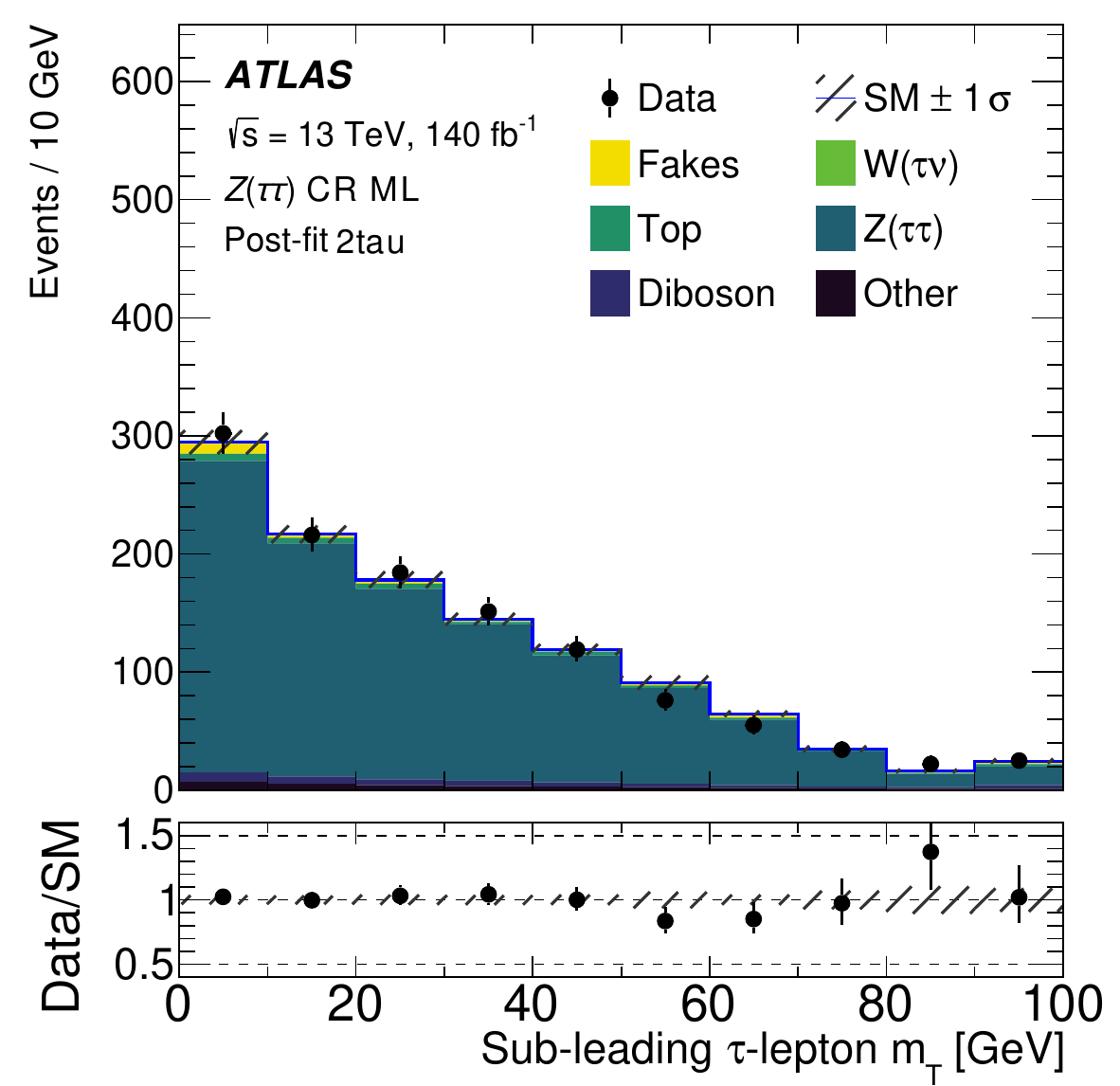}}
\subfloat[\RunTwo, $W(\tau\nu)$ CR, BDT score]{\includegraphics[width=0.32\textwidth]{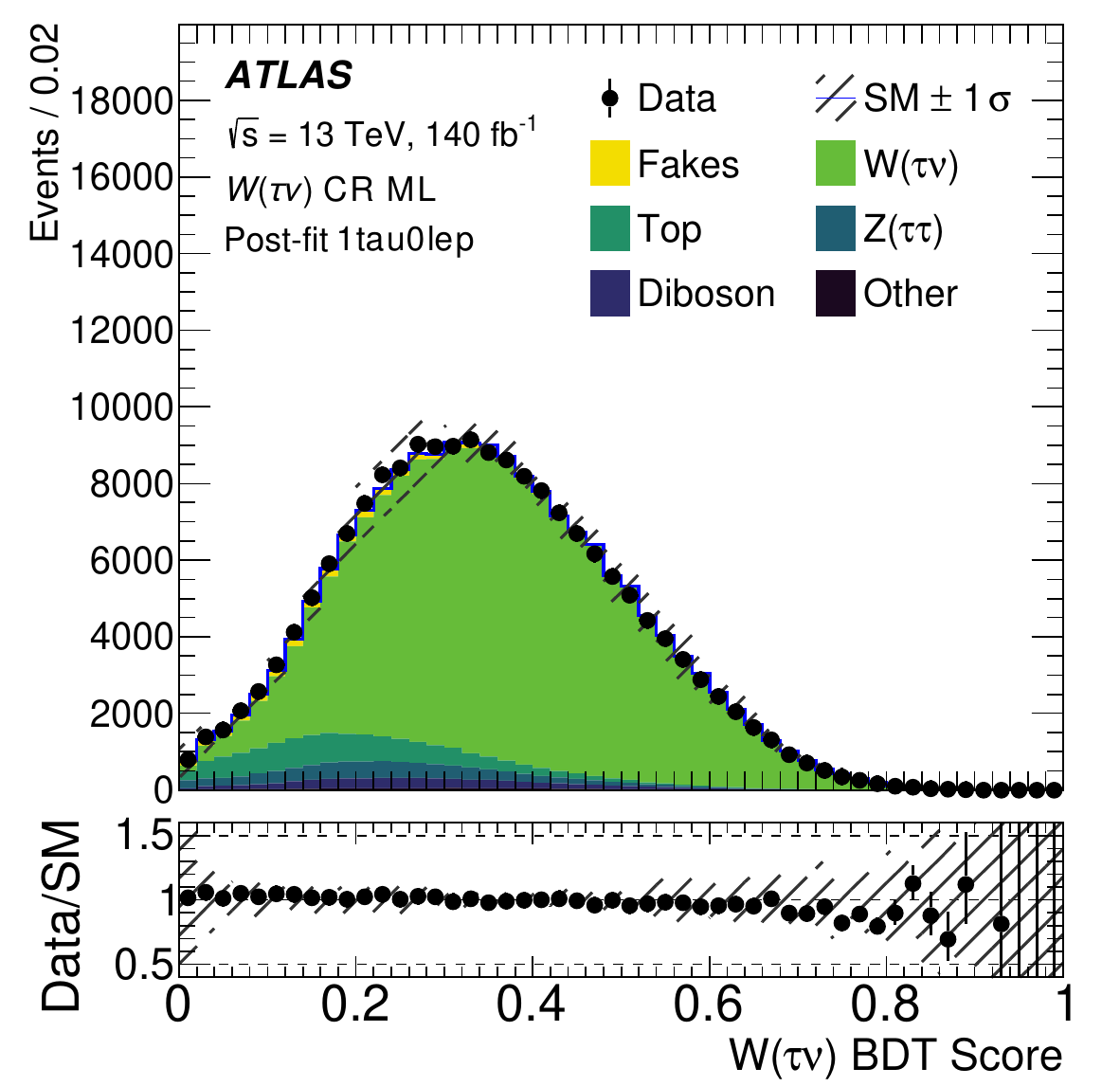}} \\
\subfloat[\RunThr, Top CR \nolep, $m_\mathrm{T}^{\tau_1}$]{\includegraphics[width=0.32\textwidth]{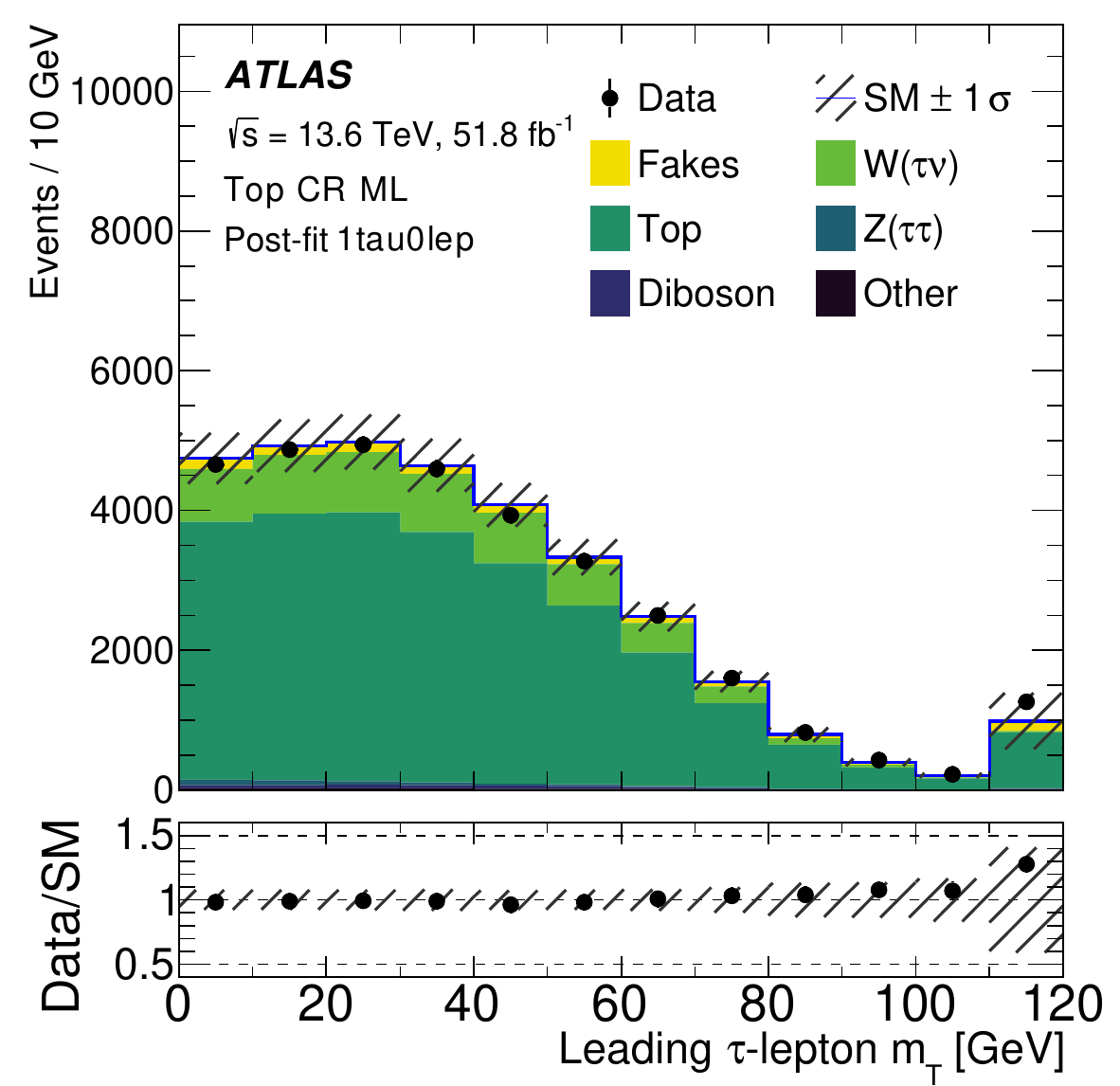}}
\subfloat[\RunThr, $Z(\tau\tau)$ CR \twotau, $m_\mathrm{T}^{\tau_2}$]{\includegraphics[width=0.32\textwidth]{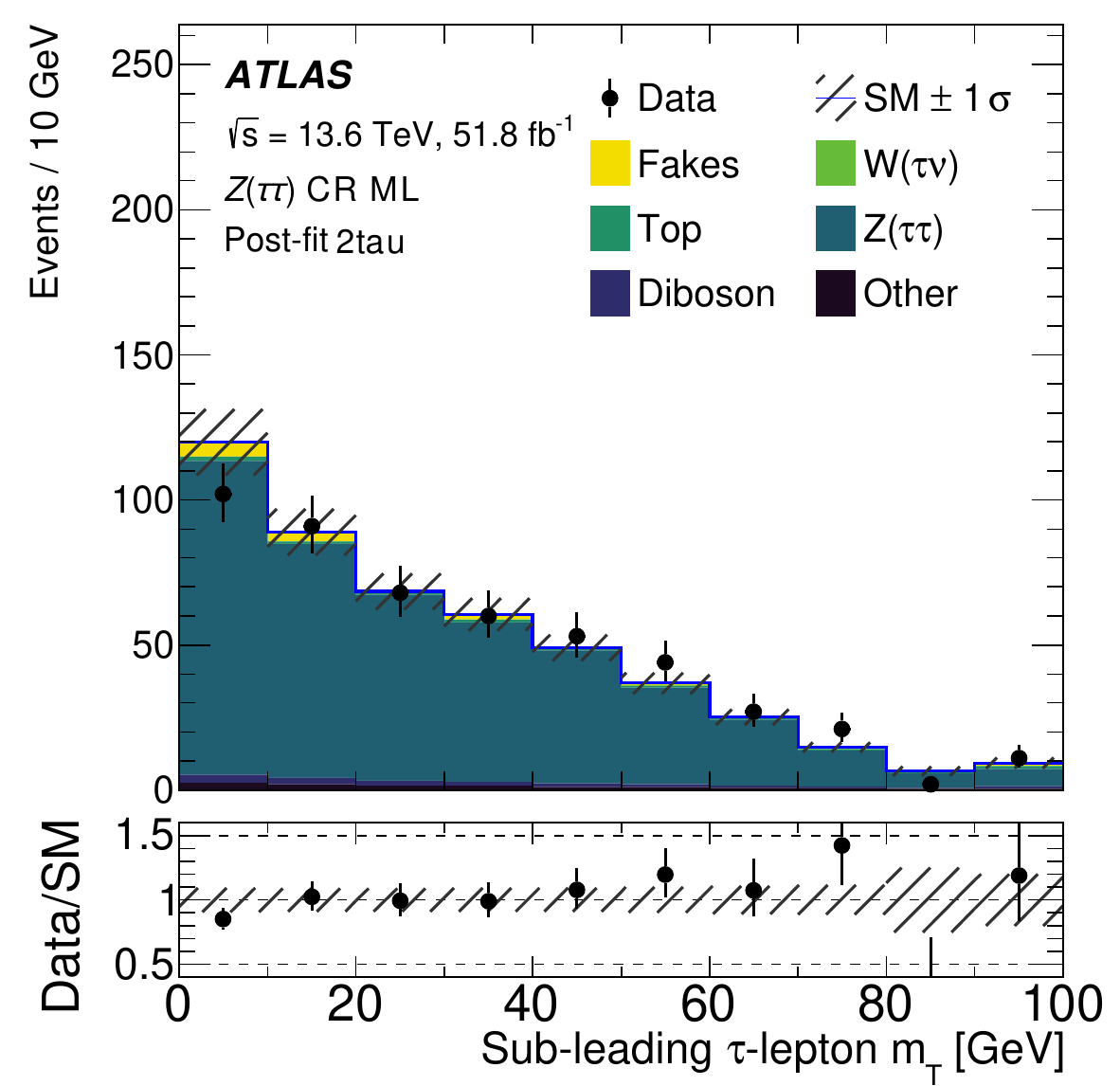}}
\subfloat[\RunThr, $W(\tau\nu)$ CR, BDT score]{\includegraphics[width=0.32\textwidth]{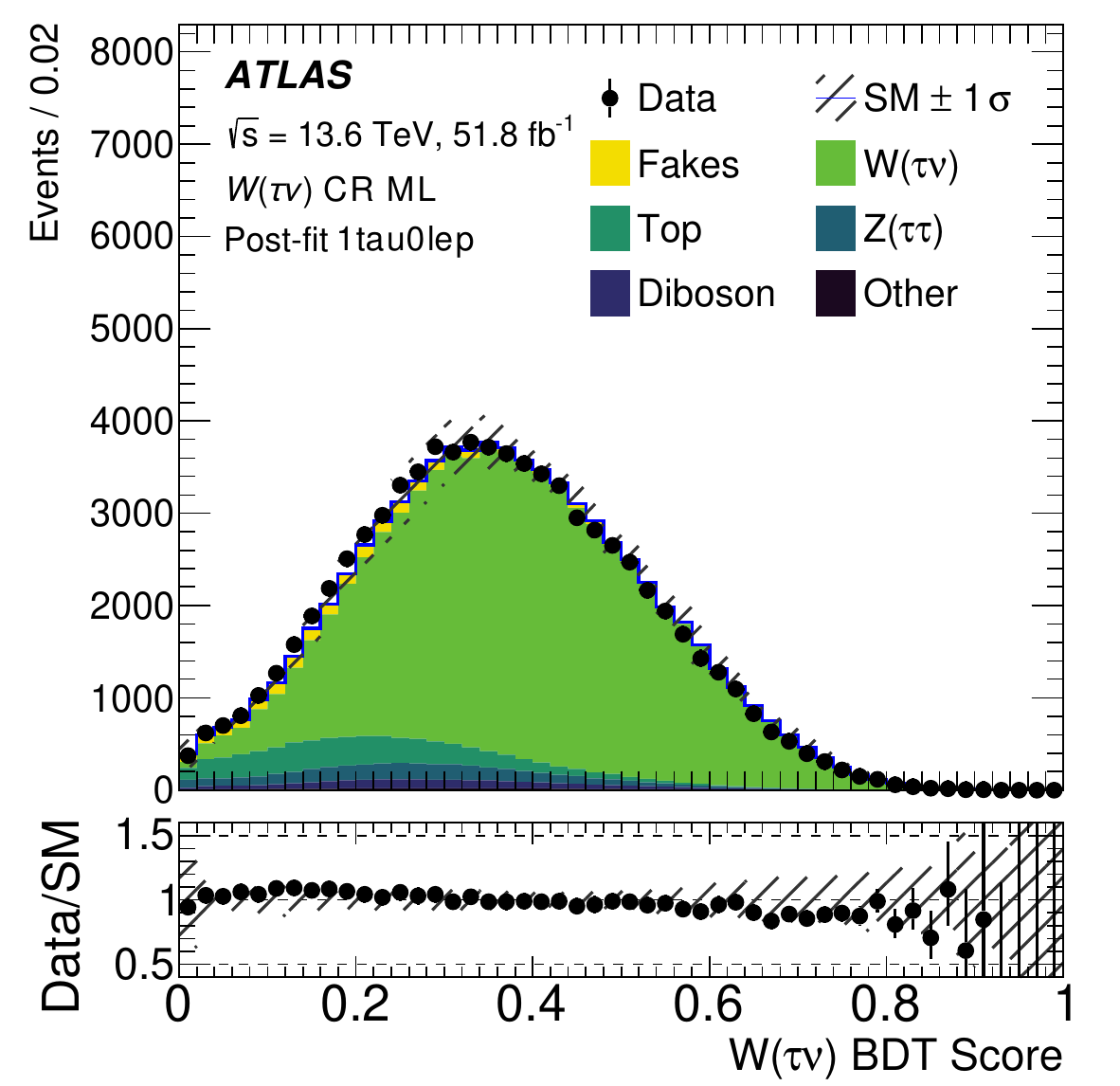}}
\caption{Distributions of important variables used in ML-based CRs in \RunTwo (top row) and \RunThr (bottom row). \BandIsStatSyst
\PostfitML
\RightBinOverflow
\BottomPanelDataSM
}
\label{fig:ml_cr_fullsyst_run2}
\end{figure}

\begin{figure}[htb!]
\centering
\subfloat[\RunTwo, $W(\tau\nu)$ VR \nolep, $p_\mathrm{T}^{\tau_1}$]{\includegraphics[width=0.32\textwidth]{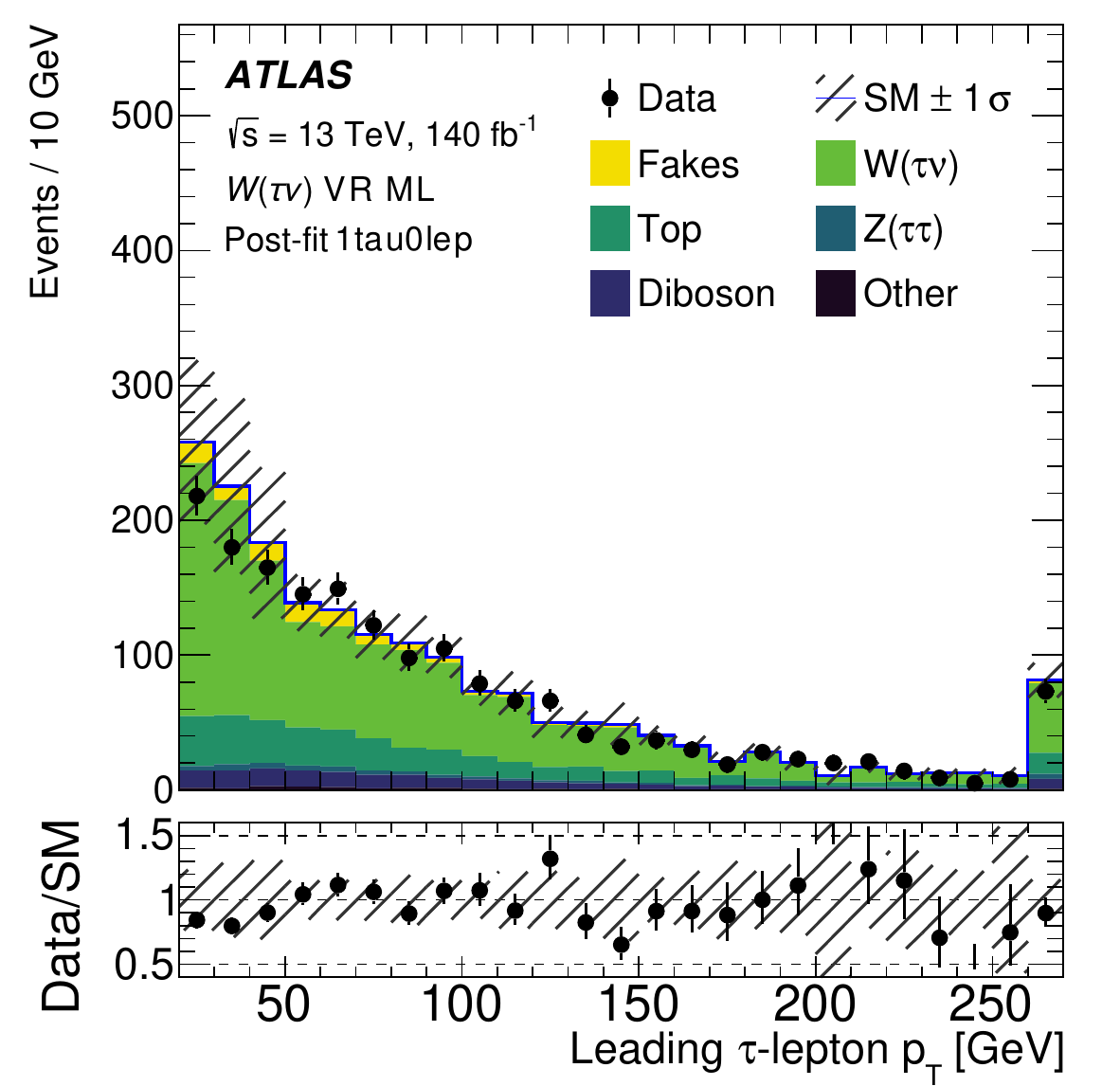}}
\subfloat[\RunTwo, Diboson VR \twotau, $p_\mathrm{T}^{\tau_2}$]{\includegraphics[width=0.32\textwidth]{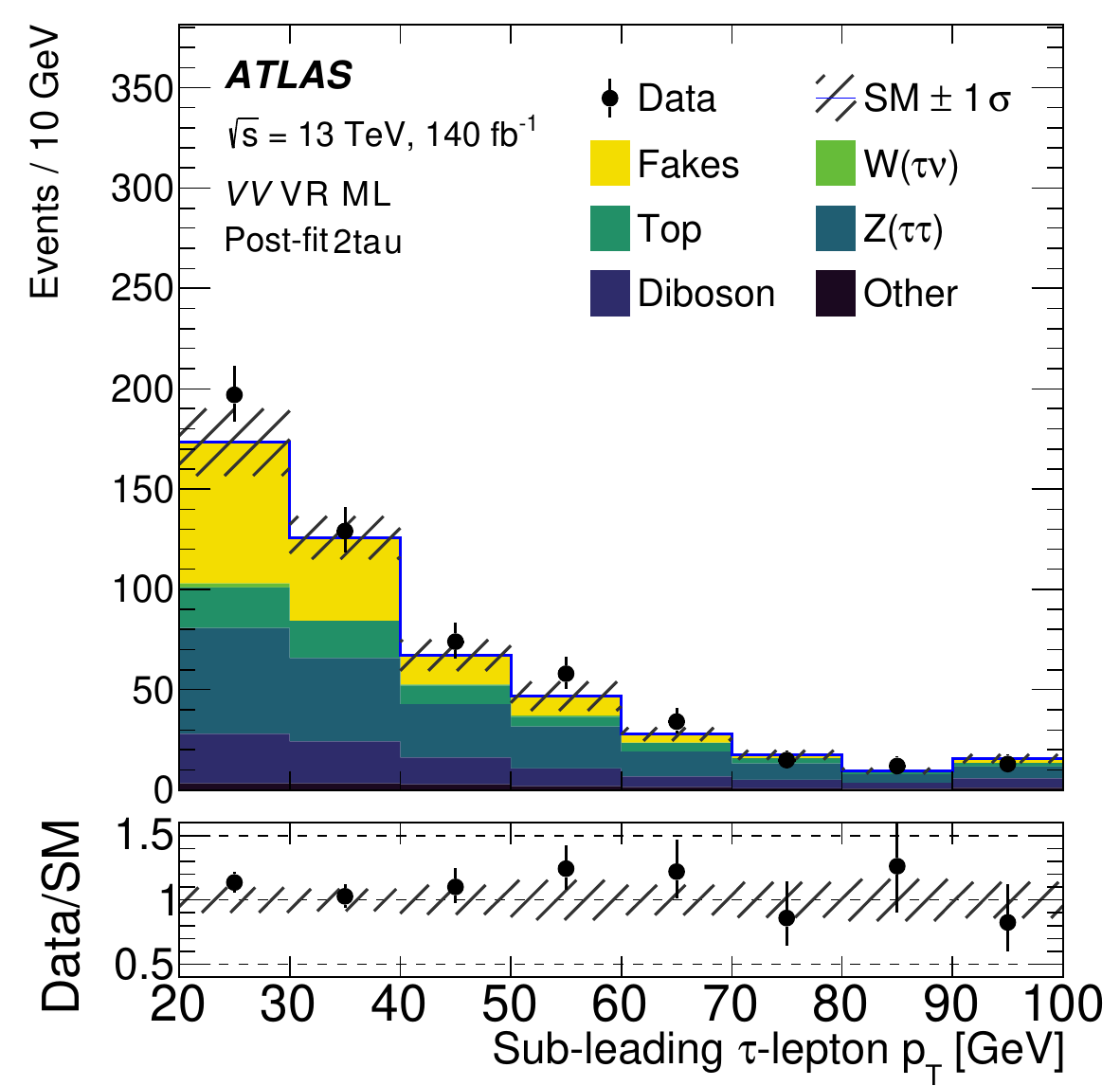}}
\subfloat[\RunTwo, Top VR \nolep, \met]{\includegraphics[width=0.32\textwidth]{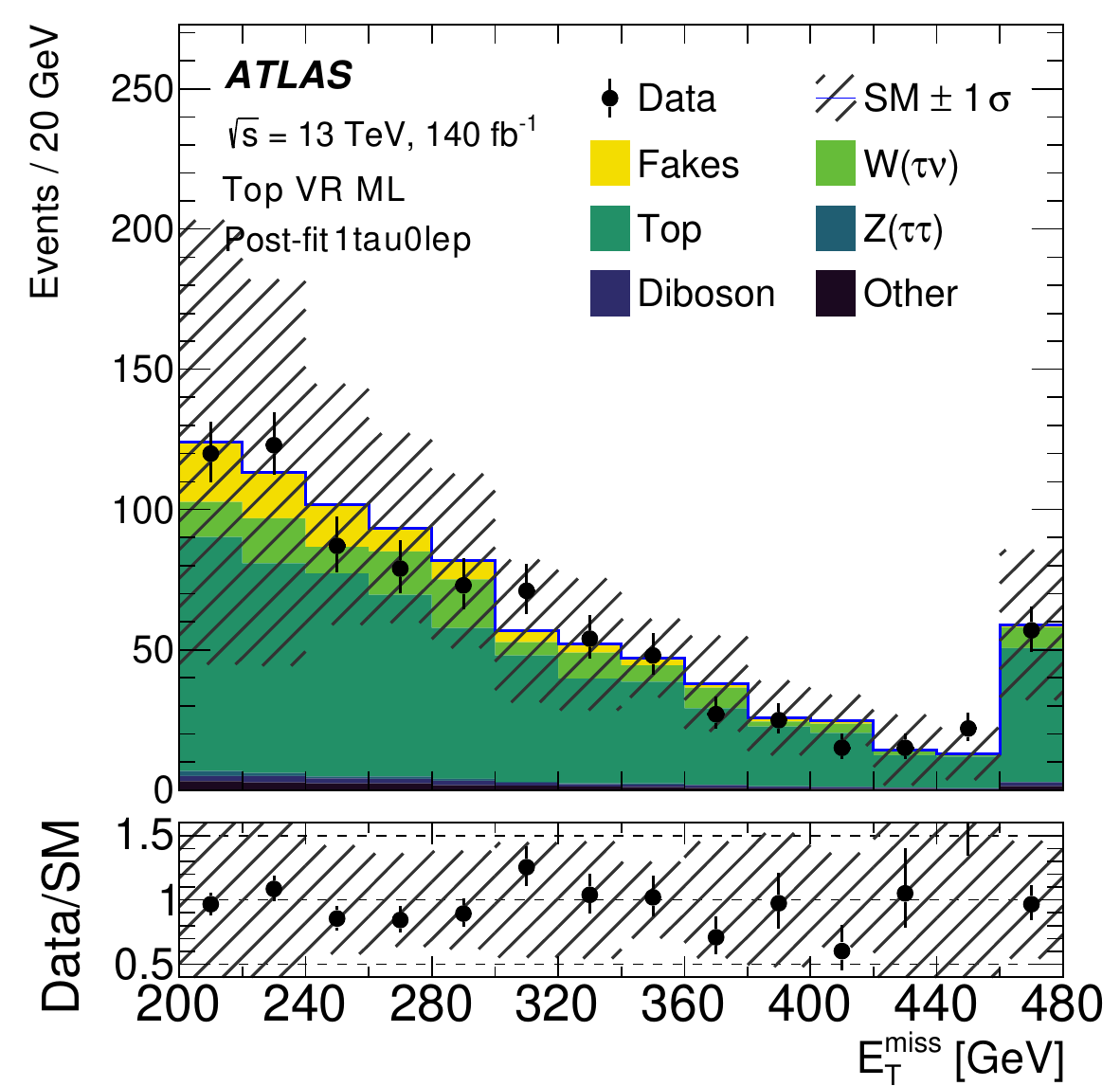}}\\
\subfloat[\RunThr, $W(\tau\nu)$ VR \nolep, $p_\mathrm{T}^{\tau_1}$]{\includegraphics[width=0.32\textwidth]{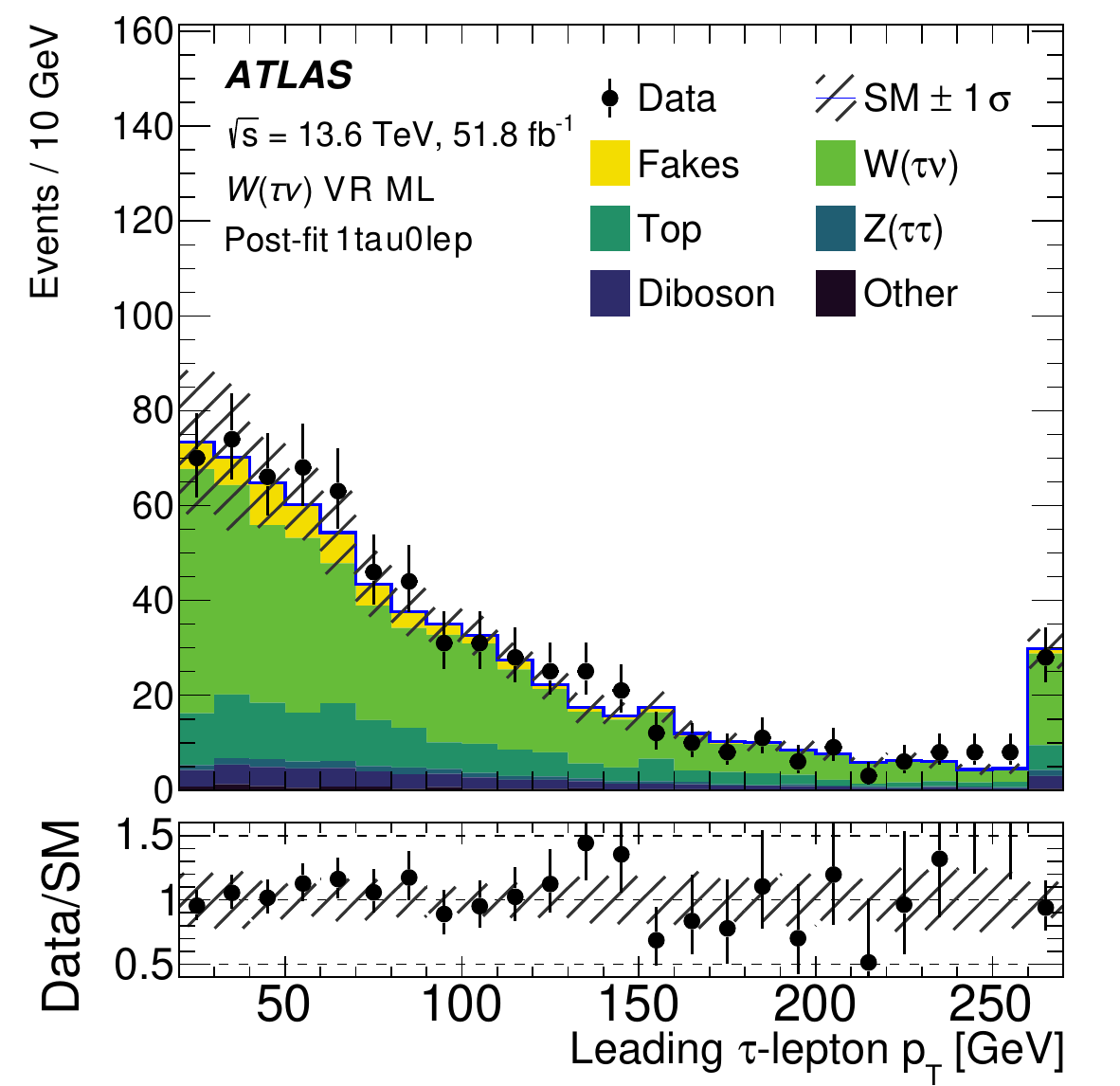}}
\subfloat[\RunThr, Diboson VR \twotau, $p_\mathrm{T}^{\tau_2}$]{\includegraphics[width=0.32\textwidth]{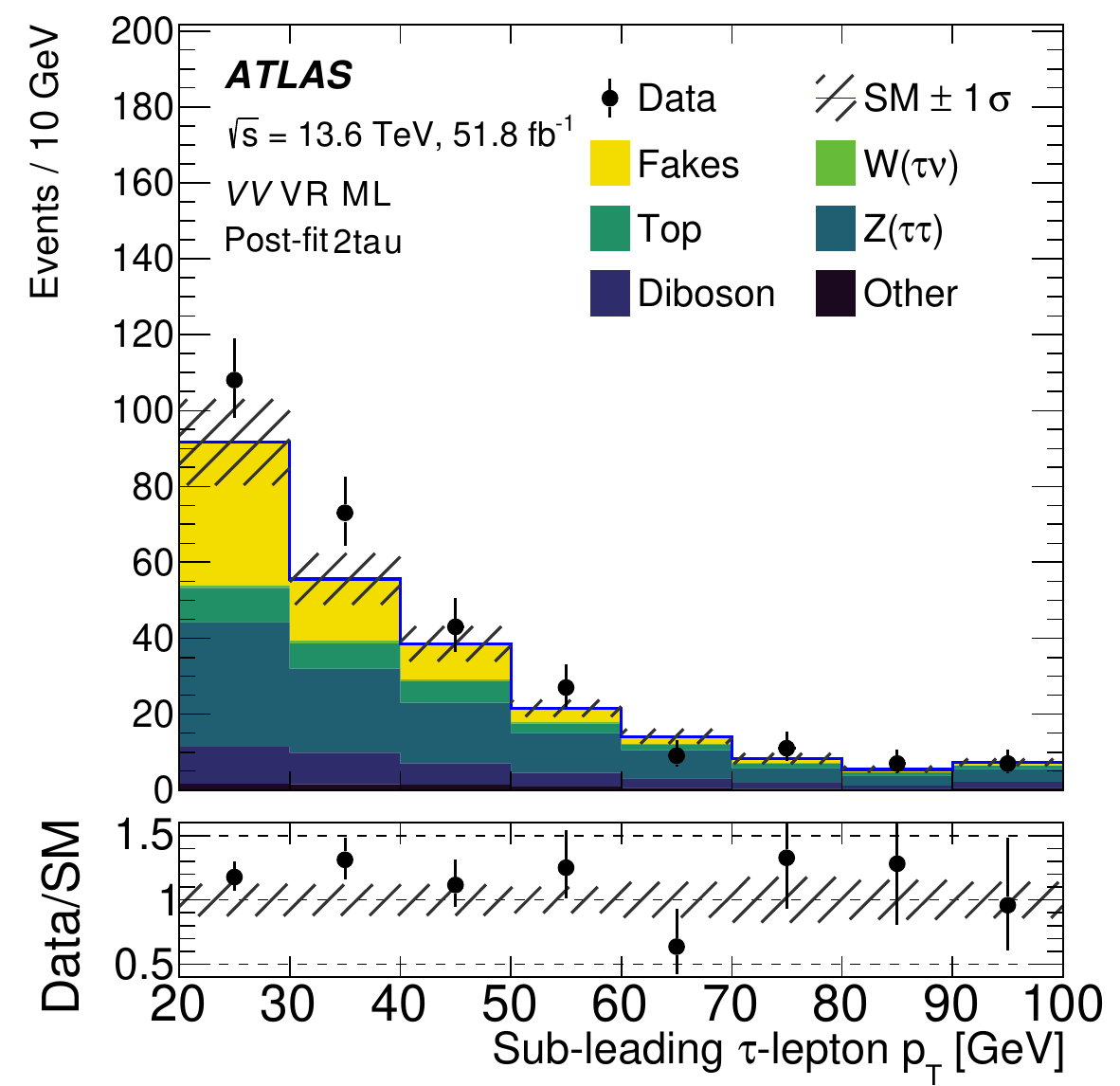}}
\subfloat[\RunThr, Top VR \nolep, \met]{\includegraphics[width=0.32\textwidth]{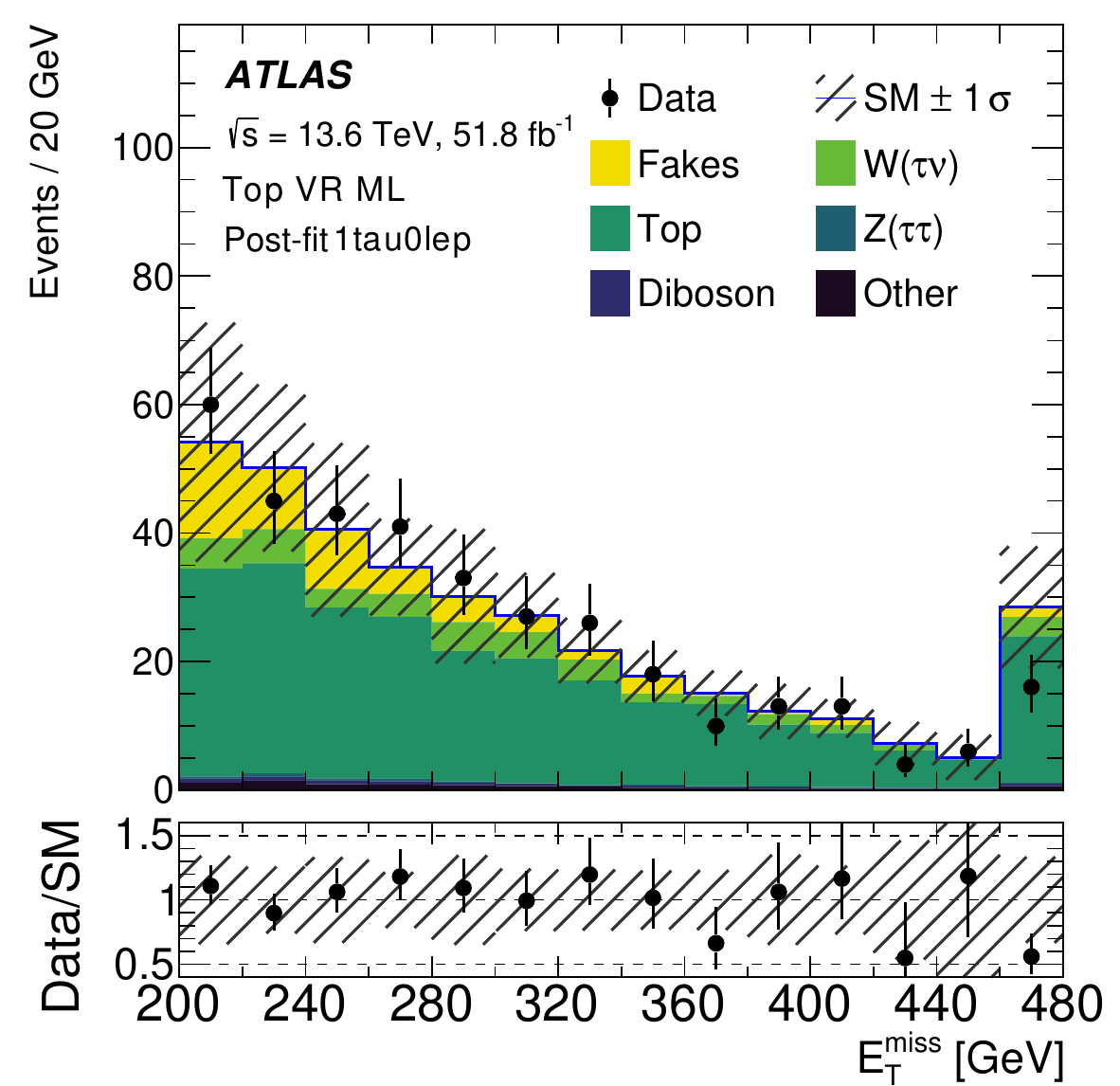}}
\caption{Distributions of important variables used in ML-based VRs in \RunTwo (top row) and \RunThr (bottom row). \BandIsStatSyst
\PostfitML
\RightBinOverflow
\BottomPanelDataSM
}
\label{fig:ml_vr_fullsyst_run2}
\end{figure}

Six CRs are defined for each of the \RunTwo and \RunThr data samples to derive separate sets of NFs. The split allows more freedom to treat different MC generator versions and other experimental differences. All CRs in the \nolep and \onelep channels are required to have signal scores $<0.3$ to suppress potential signal contamination (${<}1\%$ for any signal model not already excluded by previous searches). In the \twotau channel, this requirement is relaxed to $<0.75$, like in the SR definition. The definitions of all the CRs and VRs used in the ML-based approach are summarised in Tables~\ref{tab:ml_tabular_regions_nolep}-\ref{tab:ml_tabular_regions_twotau} and illustrated schematically in Figure~\ref{fig:ml_regions}.

\begin{table}[htb!]
\centering
\caption{Definitions of the ML-based \nolep SRs, CRs and VRs.}
\begin{tabular}{lcccc}
\toprule
& Signal score & $\hat{y} = \{W,Z,VV\}$ & $\hat{y} = \{\text{Top}\}$ & $\hat{y} = \{Fake\}$ \\
\midrule
\nolep SR & $>0.8$ & -- & -- & -- \\
\nolep W CR & $<0.3$ & True & -- & -- \\
\nolep W VR & $\in[0.3,0.8]$ & True & -- & -- \\
\nolep Top CR & $<0.3$ & -- & True & -- \\
\nolep Top VR & $\in[0.3,0.8]$ & -- & True & -- \\
\nolep Fake tau VR & $<0.3$ & -- & -- & True \\
\nolep Fake tight VR & $\in[0.3,0.8]$ & -- & -- & True \\
\bottomrule
\end{tabular}
\label{tab:ml_tabular_regions_nolep}
\end{table}

\begin{table}[htb!]
\centering
\caption{Definitions of the ML-based \onelep SRs, CRs and VRs.}
\begin{tabular}{lcccccc}
\toprule
& Signal score & $\hat{y} = \{Z\}$ & $\hat{y} = \{\text{Top}\}$ & $\hat{y} = \{Fake\}$ & $\hat{y} = \{VV\}$ & Diboson score \\
\midrule
\onelep SR & $>0.8$ & -- & -- & --& -- & -- \\
\onelep Z VR & $<0.8$ & True & -- & -- & -- & -- \\
\onelep Top CR & $<0.3$ & -- & True & -- & -- & --\\
\onelep Top VR & $\in[0.3,0.8]$ & -- & True & -- & -- & --\\
\onelep Fake tau VR & $<0.3$ & -- & -- & True & -- & --\\
\onelep Fake tight VR & $\in[0.3,0.8]$ & -- & -- & True & -- & --\\
\onelep VV CR & $<0.3$ & -- & -- & -- & True & $>0.75$\\
\onelep VV VR \RunTwo & $\in[0.15,0.8]$ & -- & -- & -- & True & $\in[0.5,0.75]$\\
\onelep VV VR \RunThr & $\in[0.05,0.8]$ & -- & -- & -- & True & $\in[0.5,0.75]$\\

\bottomrule
\end{tabular}
\label{tab:ml_tabular_regions_onelep}
\end{table}

\begin{table}[htb!]
\centering
\caption{Definitions of the ML-based \twotau SRs, CRs and VRs.}
\begin{tabular}{lccccc}
\toprule
& Signal score & $\hat{y} = \{Z\}$ & $\hat{y} = \{\text{Top}\}$ & $\hat{y} = \{Fake\}$ & $\hat{y} = \{VV\}$ \\
\midrule
\twotau SR & $>0.75$ & -- & -- & --& --  \\
\twotau Z CR & $<0.75$ & True & -- & -- & -- \\
\twotau Top VR & $<0.75$ & -- & True & -- & --\\
\twotau Fake tau VR & $<0.75$ & -- & -- & True & -- \\
\twotau VV VR & $<0.75$ & -- & -- & -- & True \\

\bottomrule
\end{tabular}
\label{tab:ml_tabular_regions_twotau}
\end{table}

The overall yields of \Wjets in the \nolep channel are so large compared with \Zjets and diboson production that even a relatively small fraction of mislabelled events results in a large contribution to the $\hat{y} = \{Z,VV\}$ regions. Therefore, the \Wjets CR ($W(\tau\nu$) CR \nolep) is defined by the requirement $\hat{y} = \{W,Z,VV\}$. Even with this inclusive selection the purity of the CR is $84\%~(83\%)$ in \RunTwo (\RunThr). The diboson CR is defined in the \onelep channel (as this is the only channel with significant diboson production yields) by requiring a diboson classifier score $>0.75$ (this automatically satisfies $\hat{y} = \{VV\}$ because the sum of scores is normalised to $1$), to suppress Top and fake-tau background contamination. This selection results in a purity of 55\% (49\%) in \RunTwo (\RunThr). The \Zjets CR is then defined in the \twotau channel, where this background makes the largest contribution to the SRs. The requirement for $Z(\tau\tau)$ CR \twotau is $\hat{y} = \{Z\}$. The fraction of \Zjets events in these CRs is $90\%$ in both \RunTwo and \RunThr. The NFs derived for these backgrounds apply to all three channels, even though the CRs are each defined only in one channel. This is motivated by the small $W(\tau\nu)$ and $Z(\tau\tau)$ background yields expected in the SRs other than \nolep and \twotau respectively, and similar kinematic properties of diboson processes observed across the channels.

The combined \ttbar and single-top-quark production yield is large enough to allow a separate CR to be defined for each of the three channels by requiring $\hat{y} = \{\text{Top}\}$. These separate Top CRs allow the fit to accommodate differences between the channels. The purity of these regions is $78\% / 92\% / 71\% $ $(77\% / 90\% / 61\%)$ in \RunTwo (\RunThr) in the \nolep/\onelep/\twotau channels.
The CR distributions of some important kinematic variables in data and MC simulation are shown in Figure~\ref{fig:ml_cr_fullsyst_run2}.

A validation region for \Wjets is defined in the \nolep channel by requiring a tighter selection on the signal score. Validation regions for Top backgrounds are similarly defined in the \nolep and \onelep channels. A diboson validation region is defined in the \onelep channel, by requiring a diboson classifier score $\in [0.5,0.75]$ and a signal score $>0.15~(0.05)$ for \RunTwo (\RunThr).
The \Zjets validation region is defined in the \onelep channel as well, as the yields in the \twotau region are low. This region corresponds to $Z(\tau\tau)$ decays where one of the $\tau$-leptons decays leptonically and the other decays hadronically. Another diboson VR is defined in the \twotau channel by requiring $\hat{y} = {VV}$. This region is used for the validation of the overall modelling in the \twotau channel because it is heavily contaminated by all major backgrounds. The VR distributions of some important kinematic variables in data and MC simulation are shown in Figure~\ref{fig:ml_vr_fullsyst_run2}.

No CRs are defined for fake taus. Instead, five VRs in total are used to validate the performance of the data-driven background modelling. These VRs are defined by requiring $\hat{y} = \{\text{fake tau}\}$. In the \nolep and \onelep channels the events with signal score $\in [0, 0.3]$ are used for fake-tau VRs. Signal scores $\in [0.3,0.8]$ are used for the fake-tau tight VRs, which have significantly fewer events but are more signal-like kinematically. In the \twotau channel, signal scores $\in [0,0.75]$ are used to define the fake-tau VR, as the number of events available is much lower. The fake-tau VR distributions of some important kinematic variables in data and MC simulation are shown later, in Figure~\ref{fig:faketau_plots_fullsyst_run23ml}.
\FloatBarrier

\subsection{Cut-and-count approach}

In the cut-and-count analysis the data are split into three channels -- \nolep, \onelep, and \twotau, similarly to the ML-based approach. The \twotau channel is used inclusively, targeting the higher-mass part of the signal parameter space. In the channels with exactly one $\tau$-lepton, two different types of kinematic signatures are targeted, one corresponding to low \masssplit, and the other to higher \susymass. Unlike the ML approach, the CRs and VRs are used for all three channels inclusively.

The most powerful discriminating variables for the signal models considered in this \paper are found to be related to the missing transverse momentum. For the \nolep channel, \met itself shows the most discrimination power and is used as the multi-bin variable, with selections on \HT and \mttauone providing suppression of \SM backgrounds, while an \met $<400$~\GeV selection is used for CRs and VRs. Two orthogonal SRs are defined -- the Compressed SR (CompSR, targeting the low-\masssplit signal models) and the High Mass SR (HMSR, targeting high-\susymass signal models). Their orthogonality is ensured by a requirement on the $\tau$-lepton \pt. In the CompSR the small mass-splitting between the SUSY decay products results in soft $\tau$-leptons, and the event selection requires $\pt^{\tau_1} < 45$~\GeV. In the HMSR the focus is on sensitivity to low $m_\text{LSP}$ values, corresponding to larger mass-splitting and $\tau$-leptons with higher transverse momentum. As such, all events in the HMSR are required to have $\pt^{\tau_1} > 45$~\GeV, and the requirement on \mttauone is set much higher than in CompSR. Another consequence of the larger mass-splitting in the signal events targeted by the HMSR is a higher jet multiplicity, which allows an $N_\text{jets} \geq 3$ requirement to efficiently reduce the \Wjets background. The exact selections used to define the \nolep SRs are summarised in Table~\ref{tab:signal_regions_1tau}.

In the \onelep channel the kinematic properties of the $\tau$-leptons for the signal models remain approximately the same as in the \nolep channel, and the same splitting into CompSR and HMSR is made by orthogonal selections on $\tau$-lepton \pt. However, use of the $\mttauone + \mtlep$ variable, in addition to \HT and \mttauone, further suppresses the \Zjets and \Wjets backgrounds in the \onelep channel. The main discriminating variable \met is used for binning in both SRs. Due to the low lepton energies in the CompSR, the compressed-topology sensitivity is dominated by the \nolep channel. However, in the HMSR the contribution from the \onelep channel is significant. The exact selections used in the \onelep channel are summarised in Table~\ref{tab:signal_regions_1tau}.

In both the \nolep and \onelep channels, the main backgrounds come from $t\bar{t}$ and \Wjets production. In the \nolep channel, $W(\tau \nu)$+jets is the dominant background, while in the \onelep channel it is $W(\ell \nu)$+jets production with one of the jets misidentified as a $\tau$-lepton. Multijet events with a misidentified $\tau$-lepton are another significant background in the \nolep CompSR.

\begin{table}[htb!]
\centering
\caption{\nolep, \onelep, and \twotau channel SR definitions for the cut-and-count approach.} %
\begin{tabular}{lccccc}
\toprule
& \multicolumn{2}{c}{\textbf{Compressed SR}}                                              & \multicolumn{2}{c}{\textbf{High Mass SR}}    & \textbf{TwoTau SR}   \\
\midrule
Channel          & \nolep    & \onelep    & \nolep & \onelep & \twotau \\
\midrule
$N_e+ N_\mu$   & \multicolumn{1}{c}{$=0$} & \multicolumn{1}{c}{$\geq 1$} & \multicolumn{1}{|c}{$=0$}      &$\geq 1$  & --  \\
$N_\tau$  & \multicolumn{4}{c}{$=1$}  & $2$ \\
\met[GeV] & \multicolumn{4}{c}{$ > 400$ } & $ > 200$  \\
\HT[GeV] & \multicolumn{2}{c}{--}  & \multicolumn{2}{|c}{$ > 1000$}   & $ > 800$  \\
$N_\text{jets}$  & \multicolumn{2}{c}{$ \geq 2$   }  & \multicolumn{2}{|c}{$ \geq 3$}   &  $\geq 3$  \\
$\pt^{\tau_1}$ [GeV] & \multicolumn{2}{c}{$ < 45$}   & \multicolumn{2}{|c}{$ > 45$ }  & --  \\
$\mt^{\tau_1}$ [GeV]  & \multicolumn{1}{c}{$> 80$} & -- & \multicolumn{1}{|c}{$> 250$}     & $> 120$  &  -- \\
$\mt^{\tau_1} + \mt^{\ell_1}$ [GeV]  & --  & \multicolumn{1}{c}{$> 350$ } & \multicolumn{1}{|c}{--} & \multicolumn{1}{c}{$ > 350 $} & --  \\
\mttauone+\mttautwo[GeV]  & --  & \multicolumn{1}{c}{--} & \multicolumn{1}{|c}{--} & \multicolumn{1}{c}{--} & $ > 150$  \\
\bottomrule
\end{tabular}%

\label{tab:signal_regions_1tau}
\end{table}

\begin{table}[htb!]
\centering
\caption{\nolep, \onelep, and \twotau channel SR binning for the cut-and-count approach.} %
\begin{tabular}{lcc}
\toprule
& $\met$ bins [GeV] & $\mttauone+\mttautwo$ bins [GeV] \\
\midrule
\RunTwo \\
\midrule
\nolep Compressed SR & $[400,500,600,700,\infty]$ & -- \\
\onelep Compressed SR & $[400,550,700,\infty]$  & -- \\
\nolep High Mass SR & $[400,500,600,700,\infty]$  & -- \\
\onelep High Mass SR & $[400,550,700,\infty]$  & -- \\
\twotau TwoTau SR & -- & $[150,200,250,300,400,500,600,\infty]$\\
\midrule
\RunThr \\
\midrule
\nolep Compressed SR & $[400,500,600,700,\infty]$ & -- \\
\onelep Compressed SR & $[400,550,\infty]$  & -- \\
\nolep High Mass SR & $[400,500,600,\infty]$  & -- \\
\onelep High Mass SR & $[400,550,\infty]$  & -- \\
\twotau TwoTau SR & -- & $[150,200,250,300,400,500,\infty]$\\
\bottomrule
\end{tabular}

\label{tab:signal_regions_bins}
\end{table}

Finally, for the \twotau channel, the $\mttauone + \mttautwo$ variable provides the most discriminating power and is used for binning in the SR\@. This variable is highly correlated with \met, and no additional \met selection beyond the preselection improves the sensitivity. Together with the \HT and $N_\text{jets}$ requirements, it suppresses the \SM backgrounds (mainly \Zjets). The selection of the two $\tau$-leptons reduces the signal yields in the compressed region, so no splitting of the SRs is considered and the optimisation of the SR focuses on the high-\susymass scenarios only. The selections used to define the TwoTauSR are summarised in Table~\ref{tab:signal_regions_1tau}. All SRs are further binned in the most sensitive variables, $\met$ for \nolep and \onelep channels, $\mttauone+\mttautwo$ for the \twotau channel, with the exact values shown in Table~\ref{tab:signal_regions_bins}.

\FloatBarrier

\begin{table}[htb!]
\centering
\caption{Requirements for the cut-and-count CRs and VRs.} %
\begin{tabular}{lcccccc}
\toprule
& \textbf{Z CR} & \textbf{W CR}  & \textbf{Top CR} & \textbf{Z VR} & \textbf{W VR}  & \textbf{Top VR}  \\
\midrule
\HT[GeV] & $\leq 800$ & $\leq 800$   & $\leq 800$   & $\leq 800$   & \multicolumn{2}{c}{$[600,1000]$} \\
\met[GeV] & -- &  $\leq 300$  & $\leq 300$  & $\leq 300$  &  \multicolumn{2}{c}{$[300,400]$ } \\
\Ntau &  $=2$ & $=1$ & $=1$ & $=1$ & $=1$ & $=1$ \\
Charge  &  Opposite $\tau$-lepton charge  & --& --& --& --& -- \\
\Nmu + $N_e$ & -- &  $=0$ & $=0$  & $\geq 1$ &  $=0$ & $=0$\\
\Nbjets & $=0$ & $=0$ &  $\geq 1$ & $=0$ & $=0$ &  $\geq 1$ \\
\Njets & $\geq 2$  & $\geq 3$ & $\geq 3$ & $\geq 2$  & $\geq 3$ & $\geq 3$  \\
\mttauone[GeV]& -- & $\leq 80$ & $\leq 80$  & -- & $\leq 80$ & $\leq 80$\\
\mttauone + \mttautwo[GeV]& $\leq 100$ &  --& --& --& --& -- \\
\mttauone + $\mt^{\ell_1}$ [GeV]& -- & -- & -- & $\leq 100$ & -- & --  \\
\mttwo[GeV]& $\leq 70$  &  --& --& --& --& -- \\
$m(\tau_1,\ell_1)$ [GeV]& -- & -- & -- &$<60$ & -- & -- \\
\bottomrule
\end{tabular}
\label{tab:cutbased_cr}
\end{table}

Control regions enriched in \Zjets (Z CR) are selected by requiring the presence of exactly two $\tau$-leptons with opposite charges. A veto on $b$-jets is applied to suppress Top backgrounds. To increase the purity of \Zjets\ and to reduce signal contamination, additional requirements on $\HT$, $\mttauone+\mttautwo$, and $\mttwo$ variables are applied. The Z CRs have $82\%~(83\%)$ purity in \RunTwo (\RunThr) and include most of the \Zjets available with the analysis preselection. No additional selection is imposed on the number of leptons -- the contributions of such events were studied and found to be less than $1\%$ of the total yields.

\begin{table}[htb!]
\centering
\caption{Requirements for the fake-tau validation regions in the cut-and-count channel.} %
\begin{tabular}{lccc}
\toprule
& \textbf{Z($vv$) VR} & \textbf{W fake VR} & \textbf{Top fake VR} \\
\midrule
\HT[GeV]& \multicolumn{3}{c}{$\leq 800$ } \\
\met[GeV]&  \multicolumn{3}{c}{$\leq 300$ }\\
\Ntau &  $=1$ & $=1$ & $=1$\\
\Nmu + $N_e$ & $=0$ & $=1$ & $=1$  \\
\Nbjets & $=0$ & $=0$ & $=1$  \\
\mttauone[GeV]& $\in [100,200]$  & -- & -- \\
$m_\mathrm{T}^{\ell_1}$ [GeV]& -- & $\leq 100$ & $\leq 100$  \\
$m(\tau_1,\ell_1)$ [GeV]& -- &$>60$ & -- \\
\midrule
$\met /(\met+\HT)$ & $>0.3$ & -- & -- \\
$\Delta \phi(\text{jet}_{1}, \pt^\mathrm{miss})$ & $>2.0$ & -- & -\\
$\Delta \phi(\tau_{1}, \pt^\mathrm{miss})$ & $>1.0$ & -- & -- \\
\bottomrule
\end{tabular}
\label{tab:cutbased_fake}
\end{table}

Due to the potential signal contamination, no reasonably populated validation region can be defined for \Zjets in the \twotau channel. Instead, the \onelep channel is used. The events are required to have exactly one lepton and one $\tau$-lepton, and a $b$-jet veto is applied. Additional requirements on $\met$, $\HT$, and the sum of transverse masses of the $\tau$-lepton and $\ell$ are used to suppress signal and \ttbar contamination. Finally, a selection on the visible invariant mass, $m(\tau,\ell) < 60$~\GeV, suppresses the non-resonant contributions and keeps the region orthogonal to the fake-tau validation regions (described below).

Control regions enriched in \Wjets background (W CR) are defined by requiring the presence of at least three jets and exactly one $\tau$-lepton. A veto on $b$-jets and leptons is applied. Signal and fake tau contamination is suppressed by requirement on $\HT$, $\met$, and $\mttauone$ variables. This results in highly populated regions with a purity of $82\%~(79\%)$ in \RunTwo (\RunThr). The regions used to validate the \Wjets corrections (W VR) are defined by tightening the $\met$ and $\HT$ requirements of the W CR, keeping the selection orthogonal to the CR while ensuring that the kinematic selection is sufficiently similar to the SRs.

\begin{figure}[htb!]
\centering
\subfloat[\RunTwo, W VR, \met]{\includegraphics[width=0.32\textwidth]{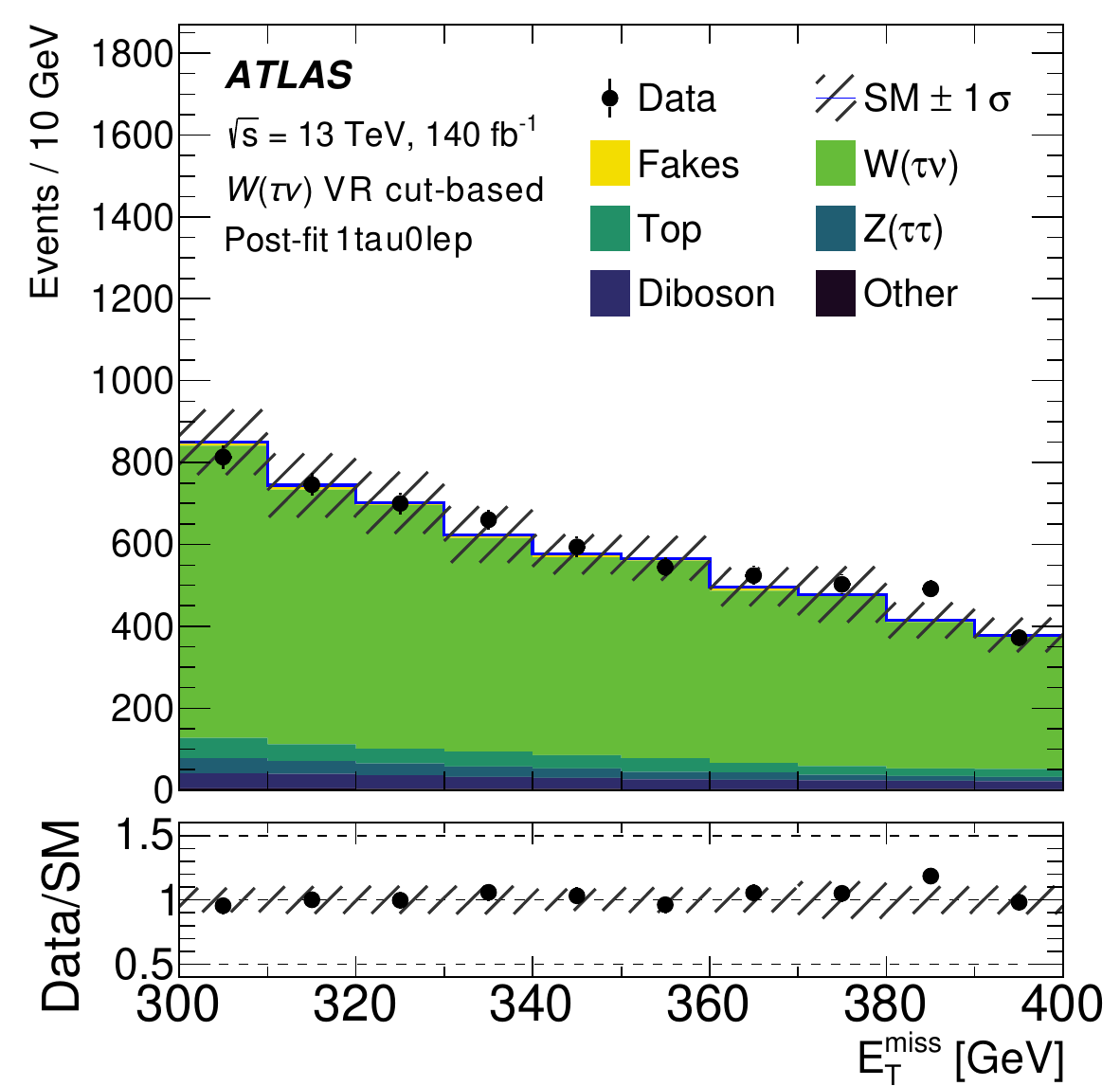}}
\subfloat[\RunTwo, Top VR, $\mttauone$]{\includegraphics[width=0.32\textwidth]{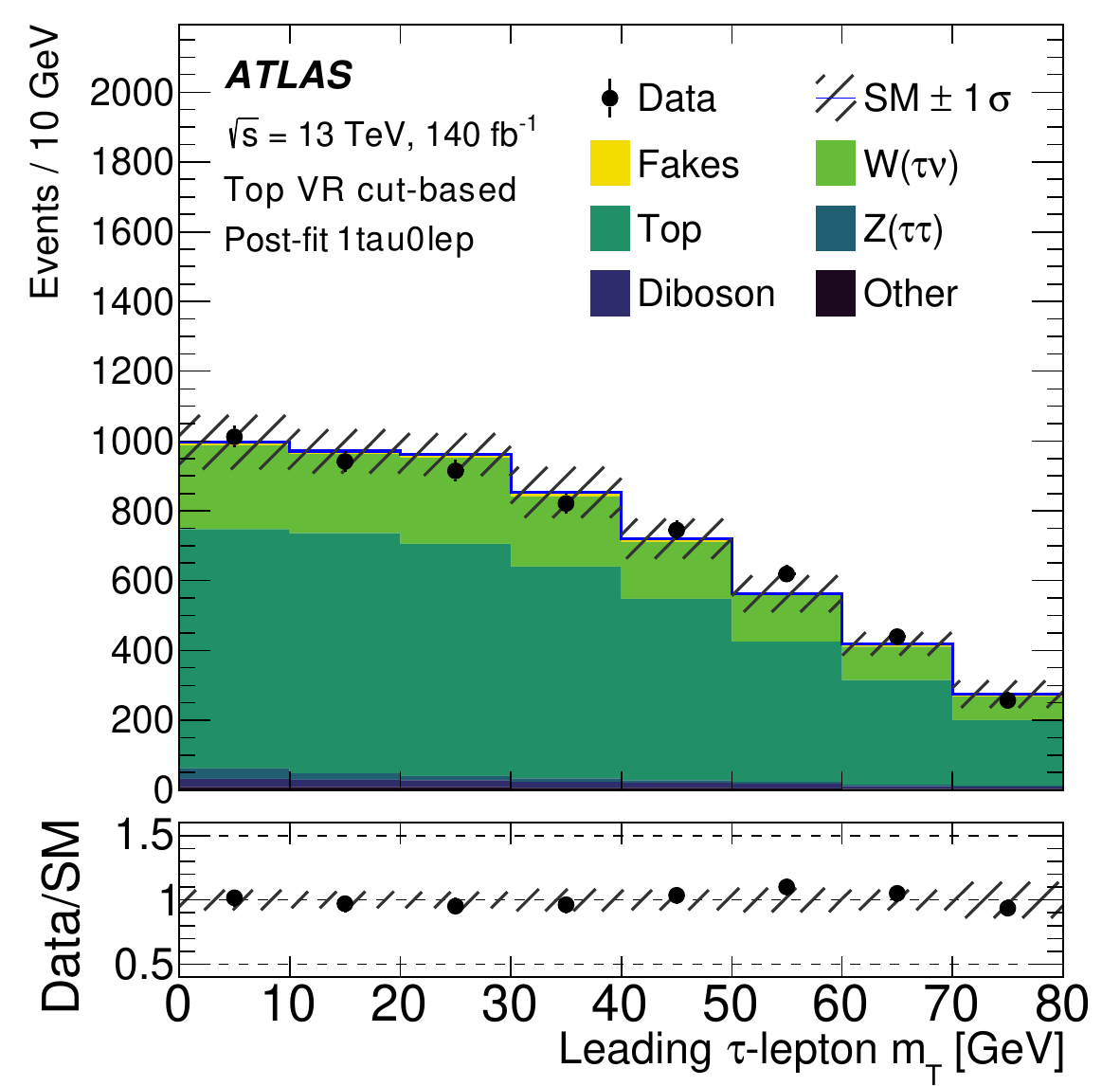}}
\subfloat[\RunTwo, Z VR, $\pt^{\tau_1}$]{\includegraphics[width=0.32\textwidth]{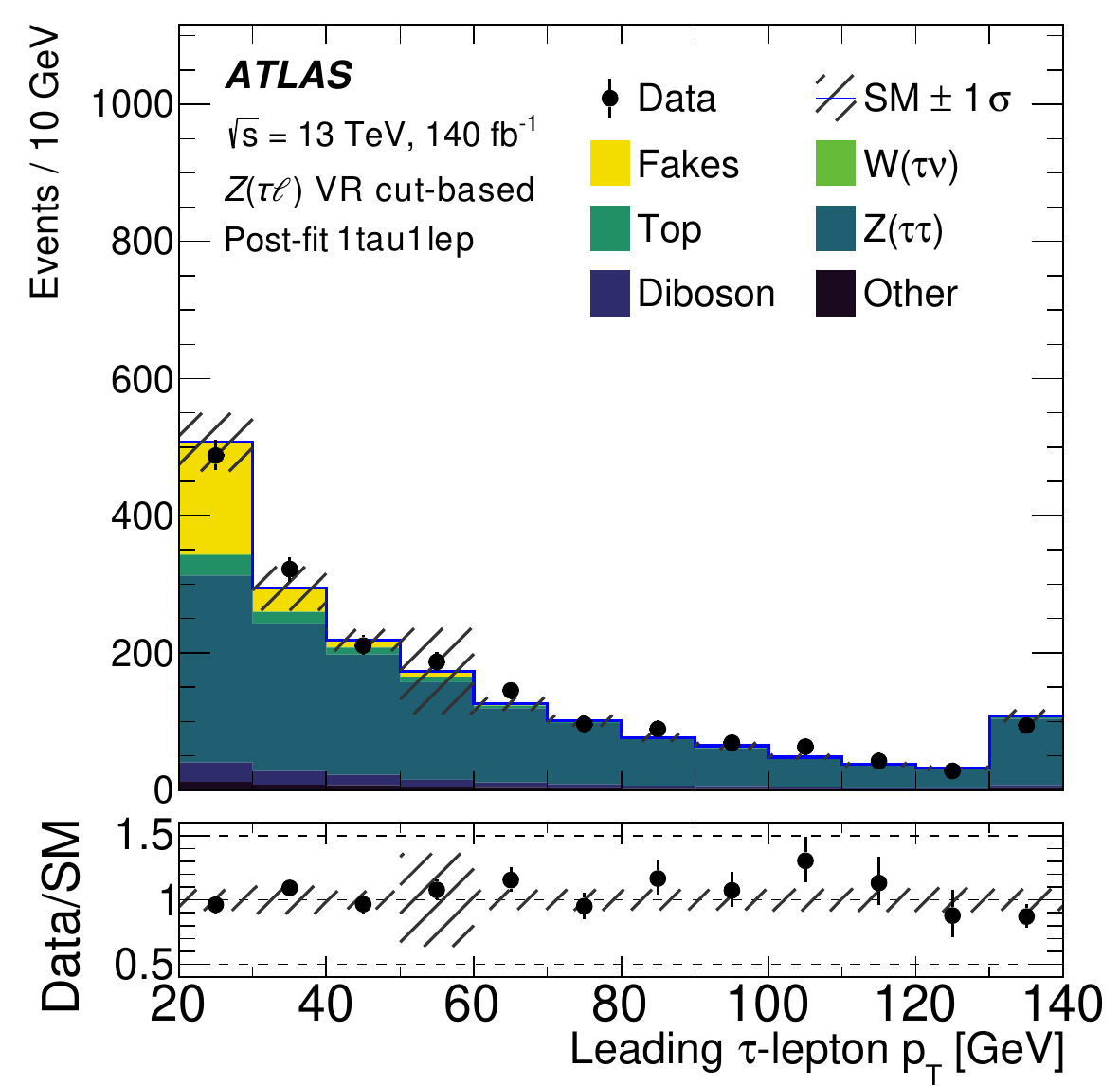}}\\
\subfloat[\RunThr, W VR, \met]{\includegraphics[width=0.32\textwidth]{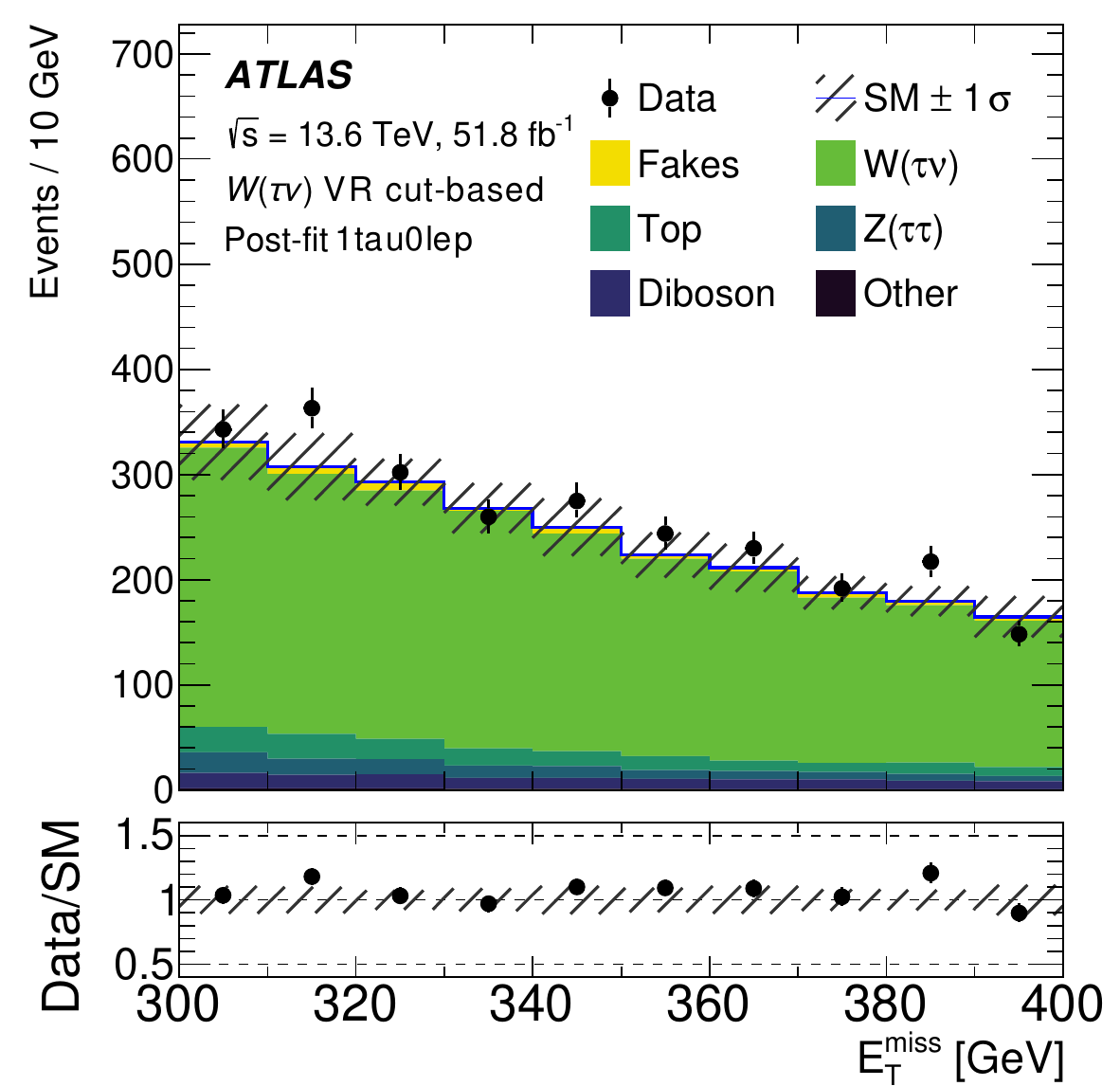}}
\subfloat[\RunThr, Top VR, $\mttauone$]{\includegraphics[width=0.32\textwidth]{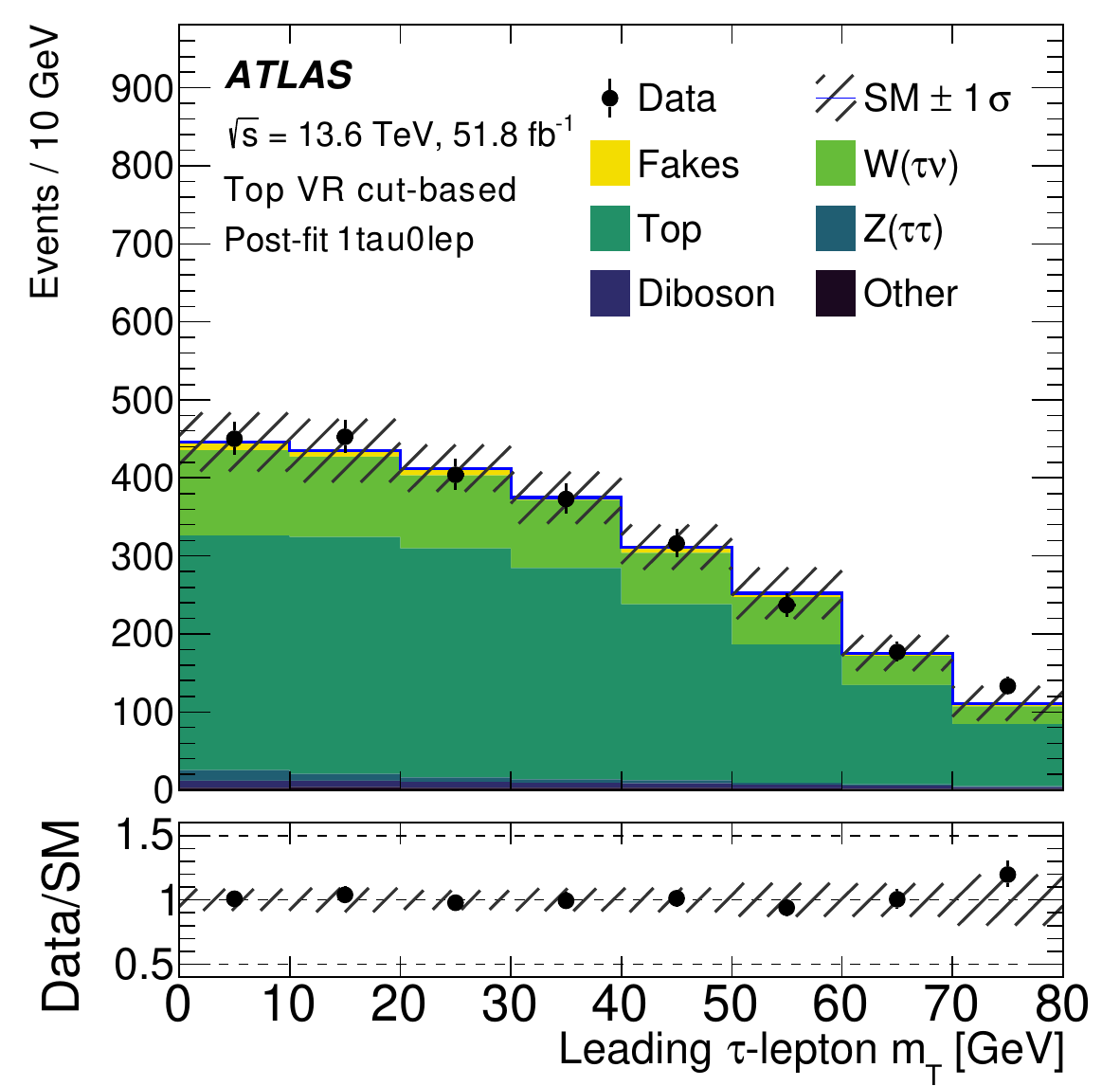}}
\subfloat[\RunThr, Z VR, $\pt^{\tau_1}$]{\includegraphics[width=0.32\textwidth]{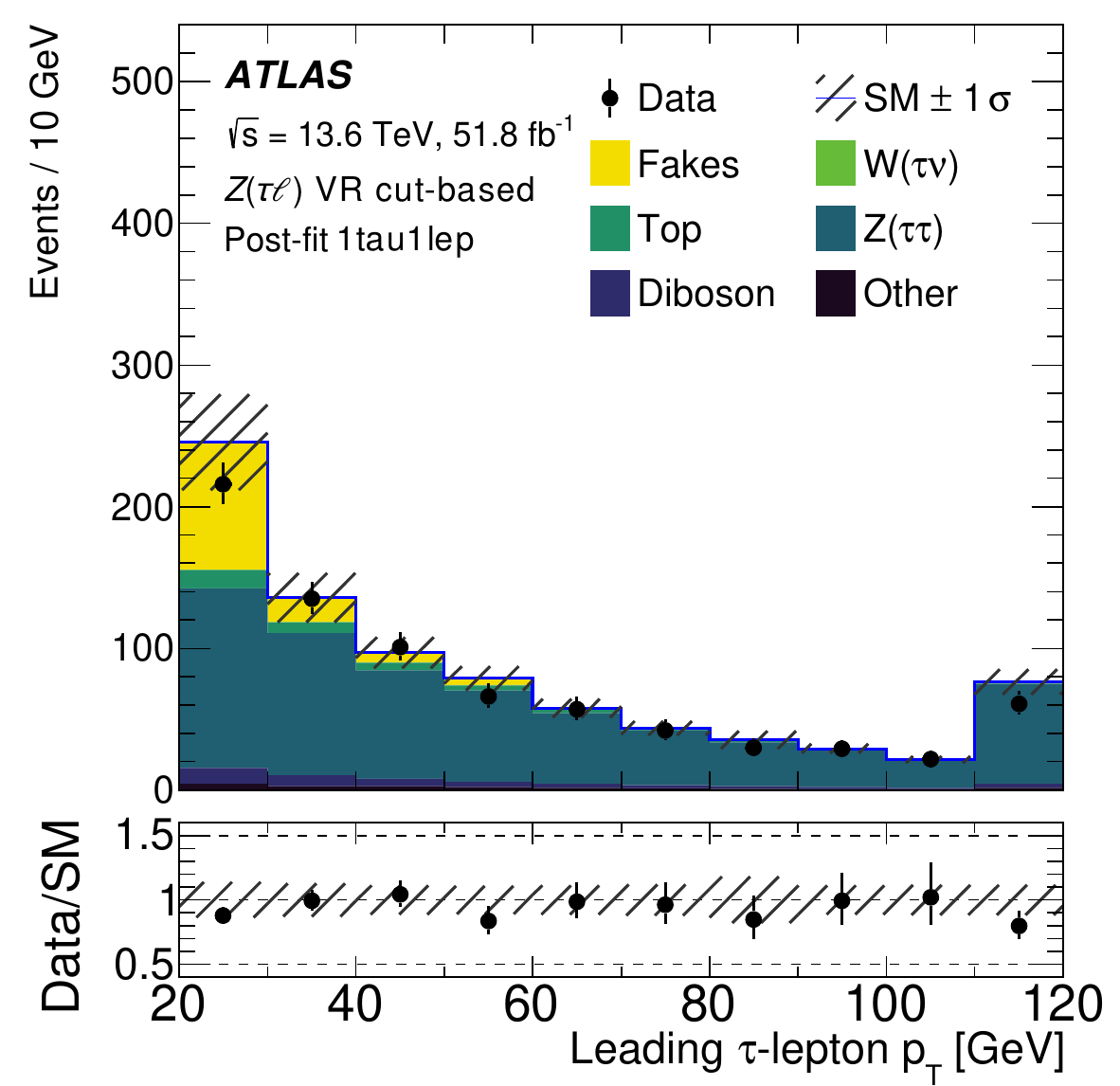}}
\caption{Distributions of important variables used in cut-and-count VRs in \RunTwo (top row) and \RunThr (bottom row).   \BandIsStatSyst
\Postfit
\RightBinOverflow
\BottomPanelDataSM
}
\label{fig:CRVR_plots_fullsyst_run2}
\end{figure}

Another pair of CRs are used to correct the inclusive modelling of \ttbar\ and single-top-quark backgrounds (Top CR). The dominant contribution comes from \ttbar production, with single-top-quark production contributing $10\%{-}20\%$ in all the analysis regions. A common normalisation factor is derived for the two backgrounds. The Top CR is defined similarly to the W CR, but with a requirement of at least one $b$-jet added. The purity of the Top CR is $72\%~(72\%)$ in \RunTwo (\RunThr).
The validation regions for Top backgrounds (Top VR) are defined similarly, tightening the Top CR $\met$ and $\HT$ requirements.
The CR and VR definitions are the same in \RunTwo and \RunThr. The definitions of the CRs and VRs are summarised in Table~\ref{tab:cutbased_cr}, and the modelling of important kinematic variables is shown in Figure~\ref{fig:CRVR_plots_fullsyst_run2}. The potential signal contamination in all CRs is below $1\%$ for all the signal models not excluded by the previous searches.

A set of additional validation regions is defined in order to monitor the data-driven modelling of the fake-tau backgrounds.
W fake-tau (Top fake-tau) VRs are defined to be kinematically similar to the W(Top) CRs. The events are required to have exactly one $\ell$ and one $\tau$-lepton, with additional kinematic selections on $\met$, $\HT$, and the lepton transverse mass $\mt^{\ell_1}$ to suppress signal and true tau contamination. The W(Top) fake-tau VRs are also required to have exactly $0(\geq 1)$ $b$-tagged jets. Finally, the dilepton visible invariant mass has to exceed $60$~\GeV in the W fake-tau VR, suppressing $Z(\tau_\text{had} \tau_\text{lep})$ contributions.
The $Z(\nu \nu)$ fake-tau VR is defined by selecting events with large $\mttauone$, $\met / (\met + \HT)$, and angular separation between the $\tau$-lepton/leading jet and \met in the transverse plane, ensuring that the main $\met$ contribution comes from the invisible decay of $Z$ boson. Events containing leptons or $b$-jets are vetoed while $\met$ and $\HT$ requirements suppress signal contamination. The selection criteria for the fake-tau VRs are summarised in Table~\ref{tab:cutbased_fake}. The modelling of some important kinematic variables in data and simulation is shown in Figure~\ref{fig:faketau_plots_fullsyst_run23}.


%
\FloatBarrier
\section{Modelling of jets misidentified as $\tau$-leptons}
\label{sec:faketau}

%

All of the events used by the analysis are required to have at least one object satisfying the selection criteria for a $\tau$-lepton. These objects are often actual, true $\tau$-leptons, but some light leptons and jets can also satisfy the selection (fake taus). Events with $\tau$-lepton candidates coming from all three lepton generations are modelled with MC simulation, as the multiplicity of such objects is well predicted. The multiplicity of jets in an event, on the other hand, is harder to model. Therefore, a data-driven approach is used to model such events inclusively, regardless of the \SM process producing them. The method is based on the Universal Fake Factor (UFF) approach~\cite{TAUP-2023-01}. %
A notable advantage of the data-driven approach is that the backgrounds estimated in this way include contributions from multijet production, beam noise, and other hadronic backgrounds that are traditionally hard to model with MC simulation.

The UFF approach uses dedicated fake-tau-enriched regions that are independent of the analysis phase-space to measure the fake factors (FFs), defined as
$$\text{FF} = \frac{\text{Number of fake taus passing medium tau ID criteria}}{\text{Number of fake taus not passing medium tau ID criteria}}$$
as a function of several $\tau$-lepton kinematic variables -- \pt, $|\eta|$, and decay mode.
The main assumption is that the probability of an object to satisfy the $\tauhad$ ID criteria depends only on the properties of the object itself, so it can be extrapolated from any measurement region.
This allows the use of the same technique and FF measurements for the two different search strategies, cut-and-count and ML-based.

Three FF measurement regions are used. The first one is a back-to-back dijet selection (Multijet) with an additional low-\pt (relative to the leading jet) $\tau$-lepton candidate. The second region ($Z(\ell \ell)$) requires two same-flavour opposite-charge leptons consistent with a $Z$-boson decay with an additional $\tau$-lepton candidate present. Finally, the third region is based on a same-charge muon and $\tau$-lepton selection ($W(\mu \nu)$). The $W(e\nu)$ process was also considered, but ultimately rejected due to the large $Z(ee)$ contamination. An additional selection is imposed on the NNJVT score of the jet geometrically matched to the $\tau$-lepton candidate before the OR procedure is applied. This is done to further suppress/enhance contributions from the \pileup jets. The details of the FF measurement regions are gathered in Table~\ref{tab:ff_regs}. In particular, the $\met < 50$~\GeV\ requirement guarantees orthogonality to the analysis phase-space.

For each analysis CR, VR, and SR, an anti-ID region is defined by replacing the requirement on the number of $\tau$-leptons with one on anti-ID taus. The contributions from $\tau$-leptons modelled with the MC simulation are subtracted. The resulting data-minus-simulation mixture is scaled by the FFs as a function of anti-ID tau kinematic properties, giving the fake-tau background estimate in the nominal region. In the case of events with more than one $\tau$-lepton,
the data-driven approach is used if at least one of the two leading $\tau$-leptons is a fake tau.
The method described in Ref.~\cite{TAUP-2023-01} is used to avoid double-counting in events with two fake taus.

\begin{table}[htb!]
\centering
\caption{Fake-factor measurement regions. The $\pt^{\text{jet}_1} / \pt^{\tau_1}$ requirement is relaxed to $>3$ for $\pt^{\tau_1} > 150$~\GeV.}\label{tab:ff_regs}
\begin{tabular}{lccc}
\toprule
Requirement & Multijet & $Z(\ell \ell)$ & $W(\mu \nu)$ \\
\midrule
Single-jet trigger & Pass & -- & -- \\
Single-lepton trigger & -- & Pass & Pass \\
\midrule
$N_\tau$ & $=1$ & $\geq 1$ & $=1$ \\
$N_\text{jet}$ & $\geq 2$ & -- & -- \\
\Nbjets & $=0$ & $=0$ & $=0$ \\
$\Delta R(\text{jet}_1,\text{jet}_2)$ & [$2.7,3.5$] & -- & -- \\
$\pt^{\text{jet}_1} / \pt^{\tau_1}$ & $> 4$ & -- & -- \\
$N_\ell$ & -- & $2e$ or $2\mu$ & $0e$ and $1\mu$ \\
$m(\ell_1,\ell_2)$ [GeV]& -- & [$70,110$] & -- \\
Charge($\tau_1, \mu_1$) & -- & -- & same-sign \\
\midrule
NNJVT score  & $> 0.8$ & $>0.8$ &$< 0.8$ \\
\midrule
\met[GeV]& $< 50$ & $< 50$ & $< 50$ \\
\met significance & $<2$ & $<2$ & $<2$ \\
$\Delta \phi(\tau_1, \pt^\mathrm{miss})$ & $>1.0$ & $>1.0$ & $>1.0$ \\
\bottomrule
\end{tabular}
\end{table}

In the UFF approach, the most important variable to control is the jet origin of the fake tau. The jets can originate from gluons, light- or heavy-flavour quarks, and \pileup, each having distinct properties. Several regions are used to measure FFs such that the mixture of fake tau origins is different in each of them. Templates of the track-based $\tau$-lepton width variable
\[\frac{\Sigma \pt^{\text{track}_i} \times \Delta R(\tau, \text{track}_i)}{\Sigma \pt^{\text{track}_i}},\]
which is sensitive to the jet origin mixture are also measured with the same binning as the FFs.
Only the tracks matched to the $\tauhadvis$ vertex or jets arising from the hard scattering interactions by a dedicated RNN-based track classifier are considered.
The combined FF is determined in template fits using the track-based $\tau$-lepton width.

\FloatBarrier

\begin{figure}[htbp!]
\centering
\subfloat[\RunTwo, fake-tau VR \nolep, \HT]{\includegraphics[width=0.32\textwidth]{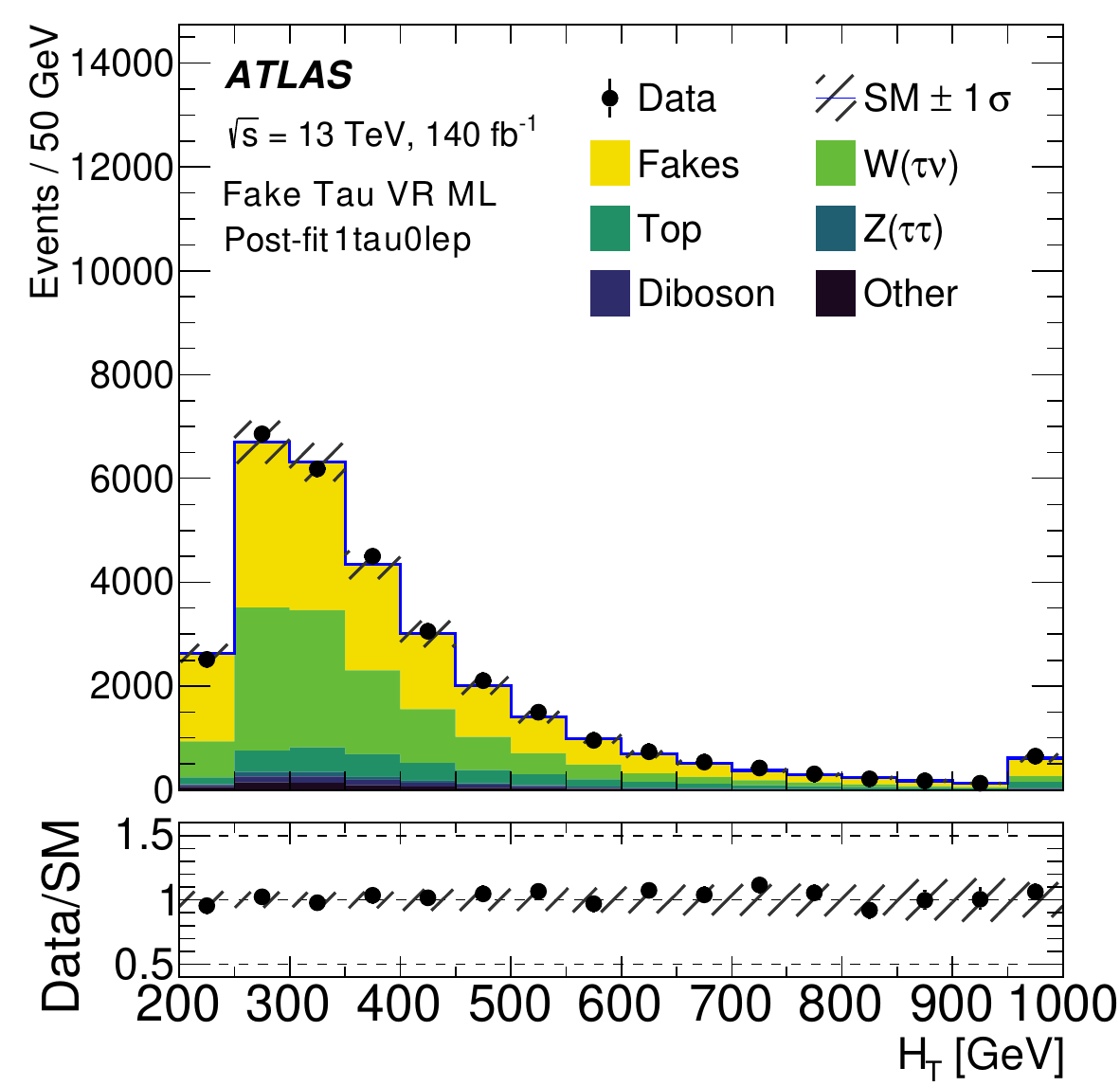}}
\subfloat[\RunTwo, fake-tau VR \onelep, $m_\mathrm{T}^{\mu_1}$]{\includegraphics[width=0.32\textwidth]{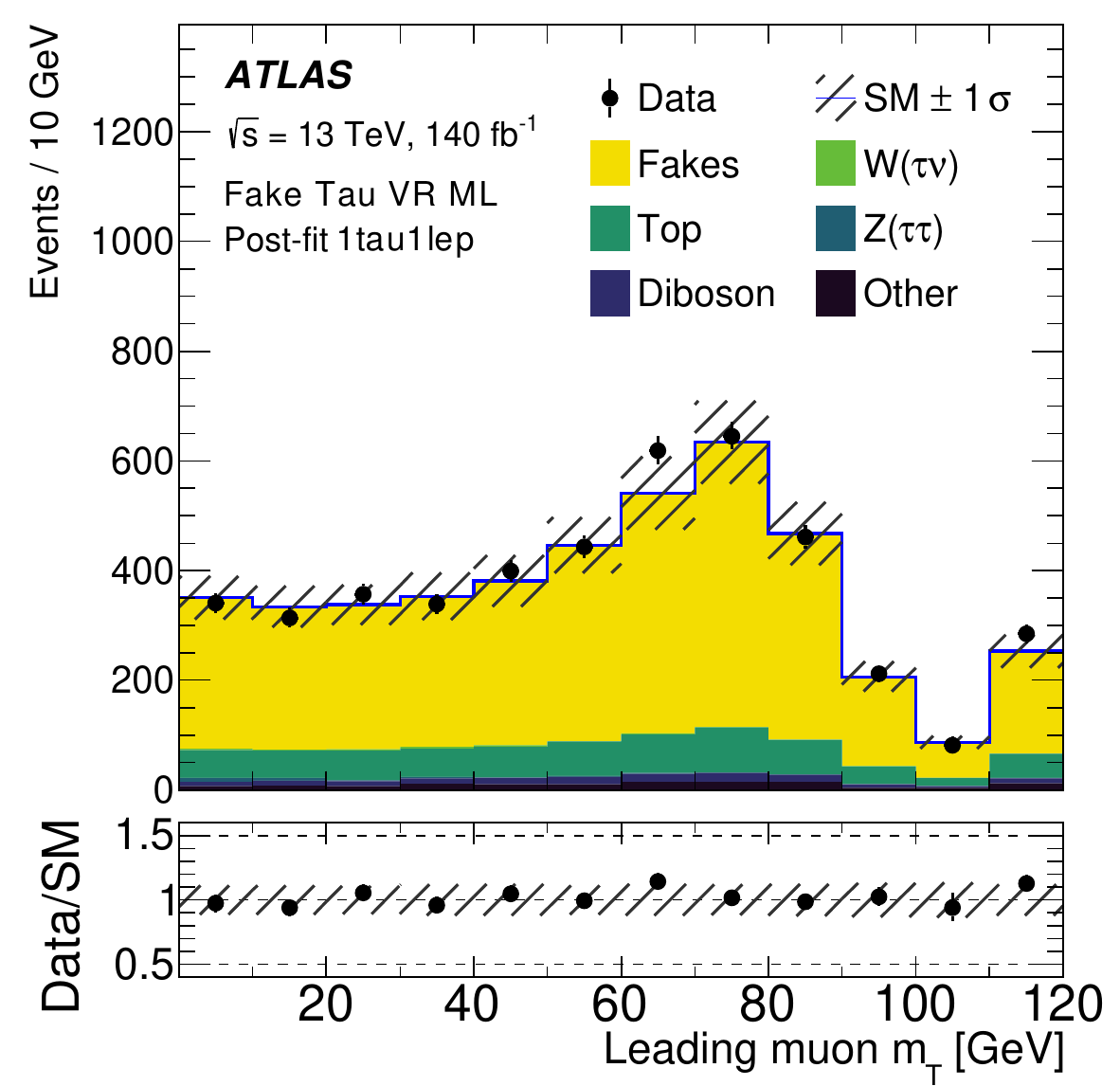}}
\subfloat[\RunTwo, fake-tau VR \twotau, $p_\mathrm{T}^{\tau_1}$]{\includegraphics[width=0.32\textwidth]{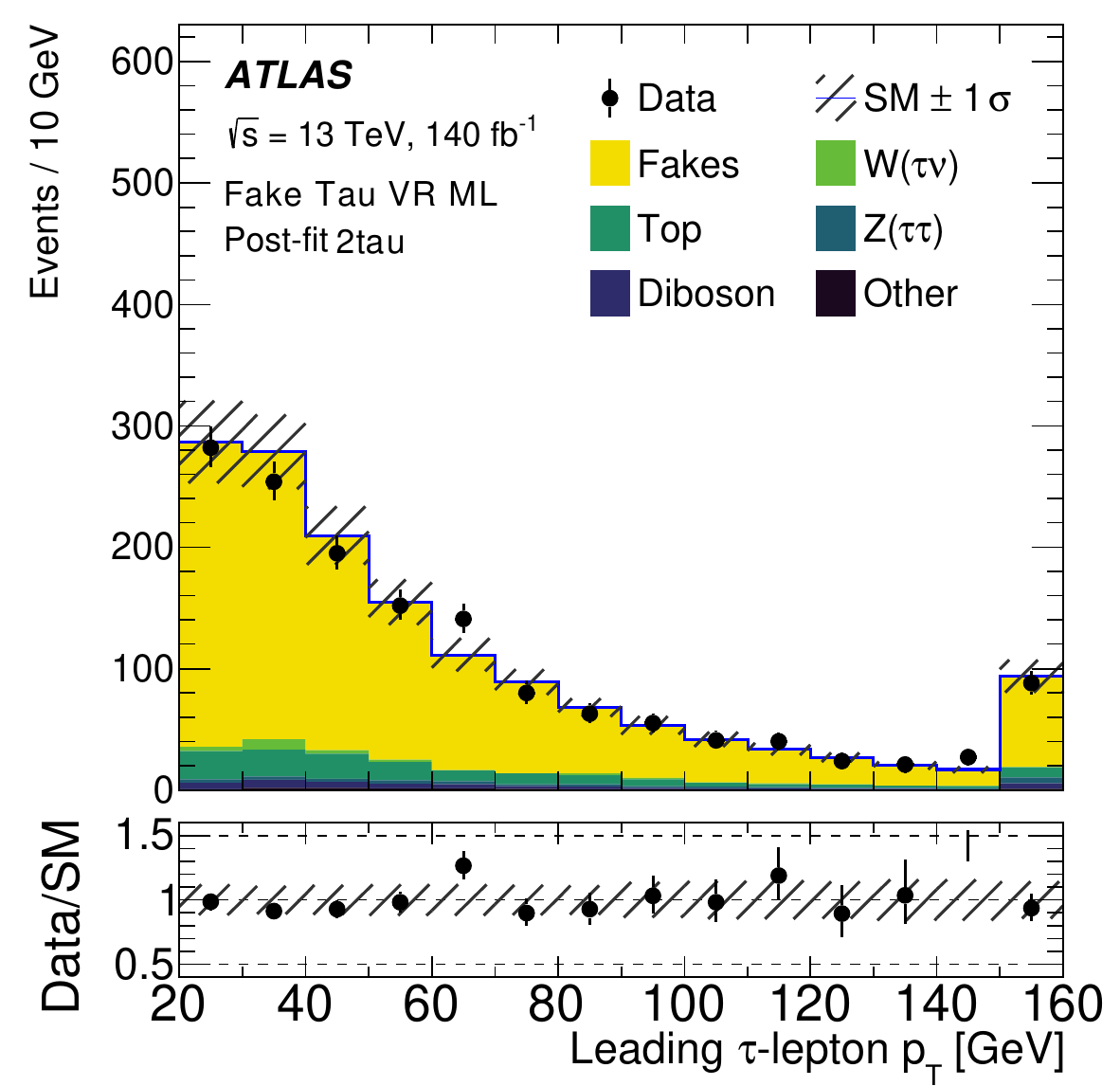}}\\
\subfloat[\RunThr, fake-tau VR \nolep, \HT]{\includegraphics[width=0.32\textwidth]{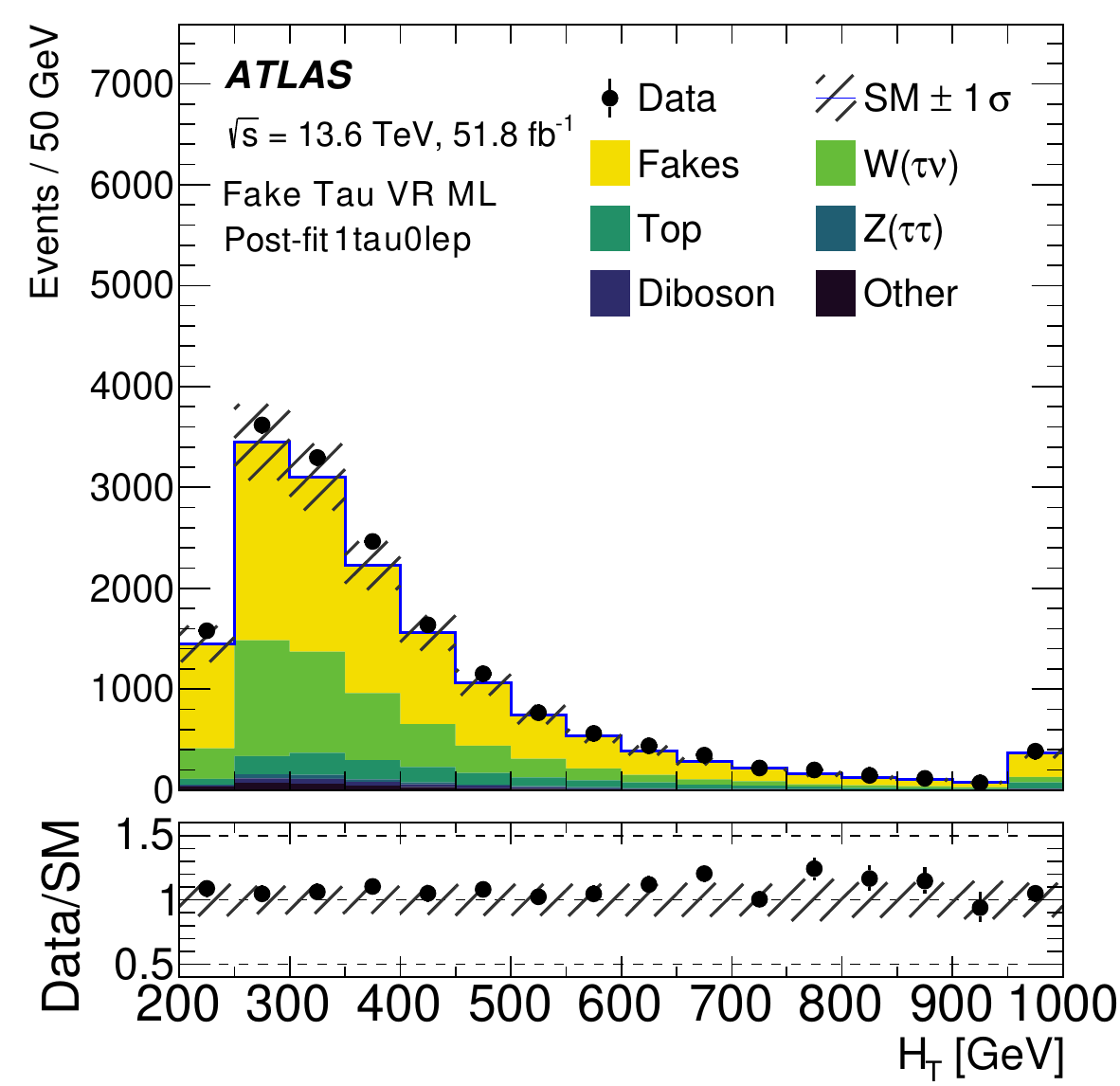}}
\subfloat[\RunThr, fake-tau VR \onelep, $m_\mathrm{T}^{\mu_1}$]{\includegraphics[width=0.32\textwidth]{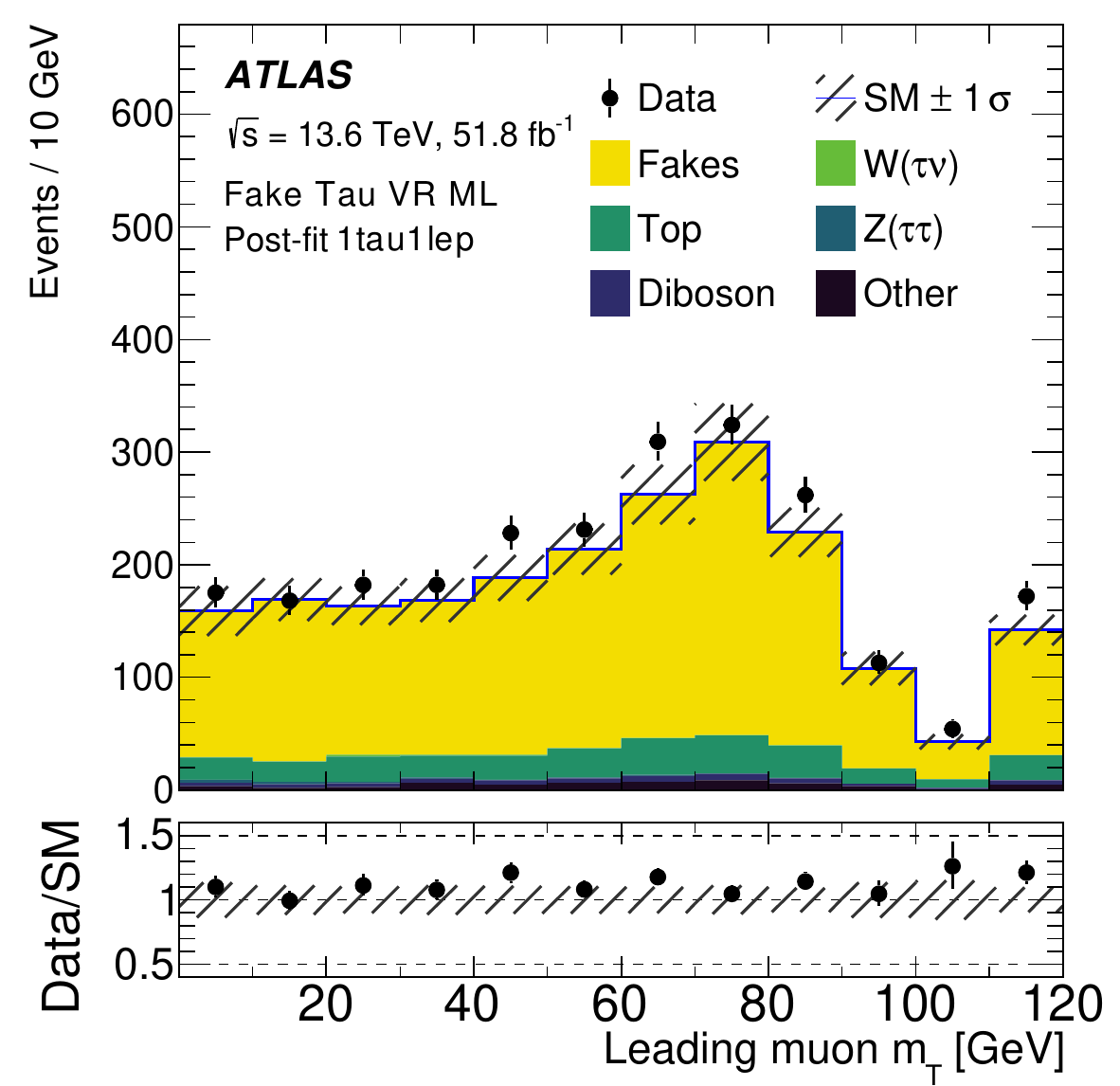}}
\subfloat[\RunThr, fake-tau VR \twotau, $p_\mathrm{T}^{\tau_1}$]{\includegraphics[width=0.32\textwidth]{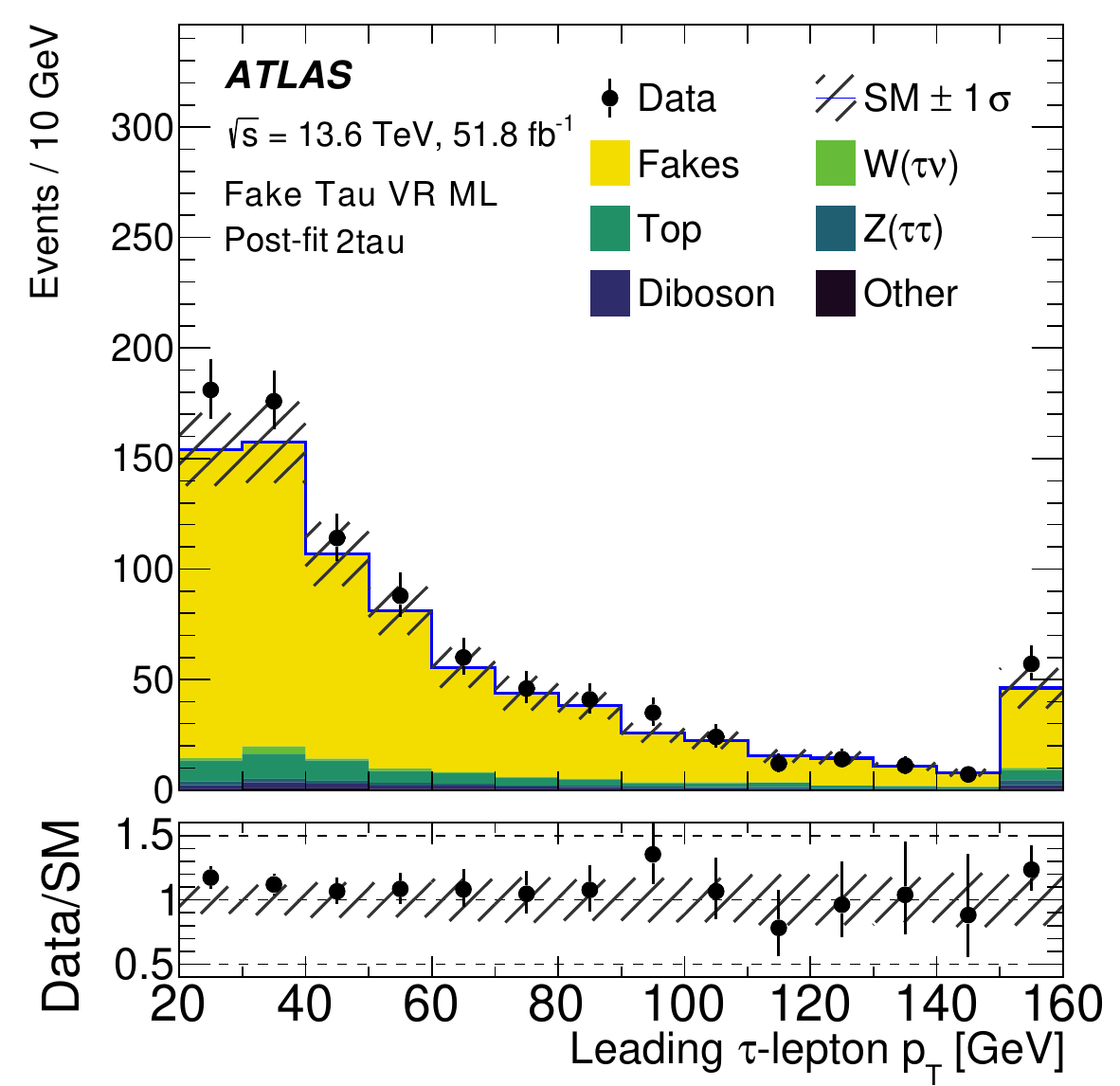}}
\caption{Distributions of important variables used in the ML-based fake-tau VRs in \RunTwo (top row) and \RunThr (bottom row).   \BandIsStatSyst
\PostfitML
\RightBinOverflow
\BottomPanelDataSM
}
\label{fig:faketau_plots_fullsyst_run23ml}
\end{figure}

\begin{figure}[htbp!]
\centering
\subfloat[\RunTwo, W fake-tau VR, \met]{\includegraphics[width=0.32\textwidth]{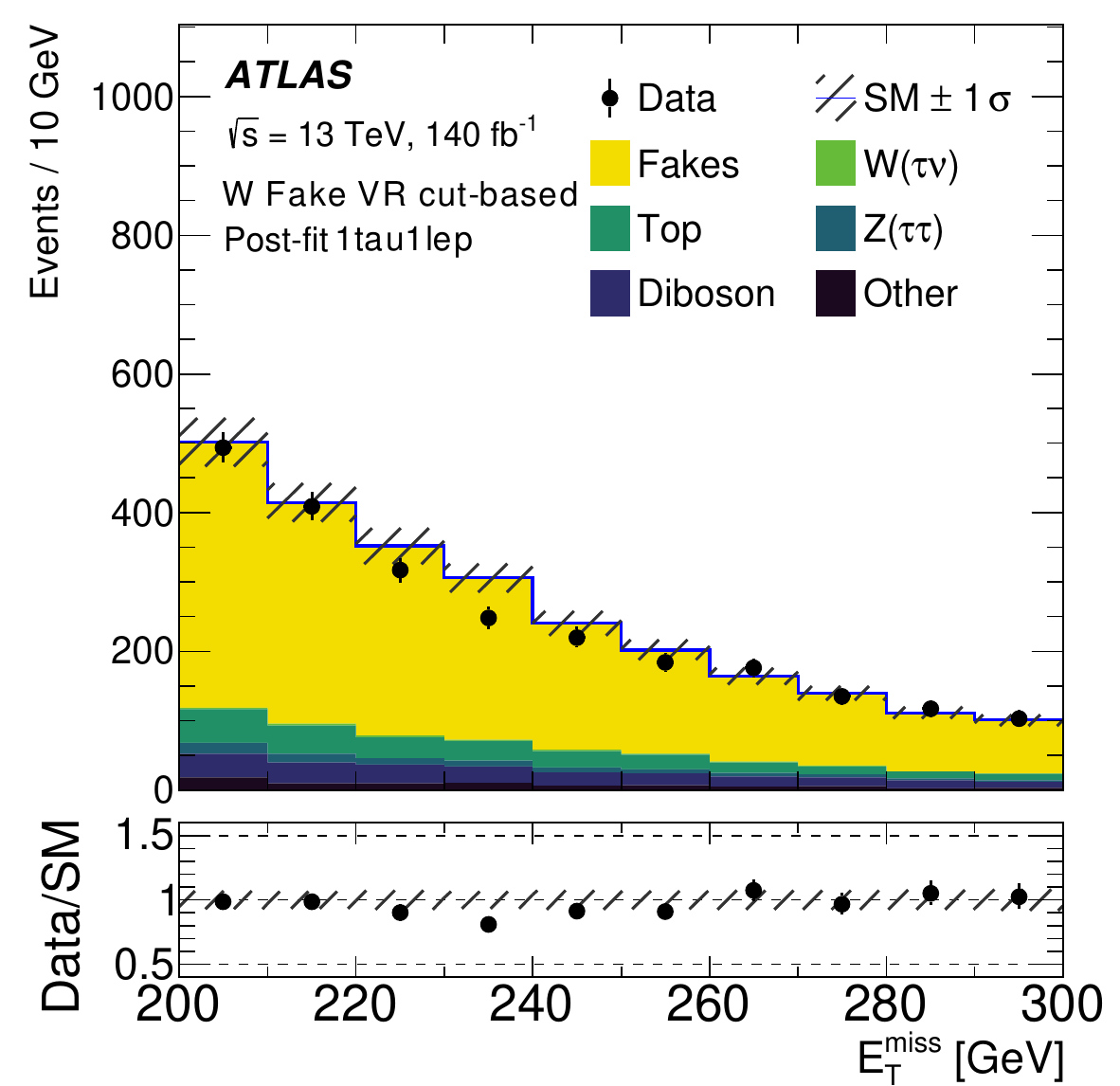}}
\subfloat[\RunTwo, Top fake-tau VR, $m_\mathrm{T}^{\tau_1}$]{\includegraphics[width=0.32\textwidth]{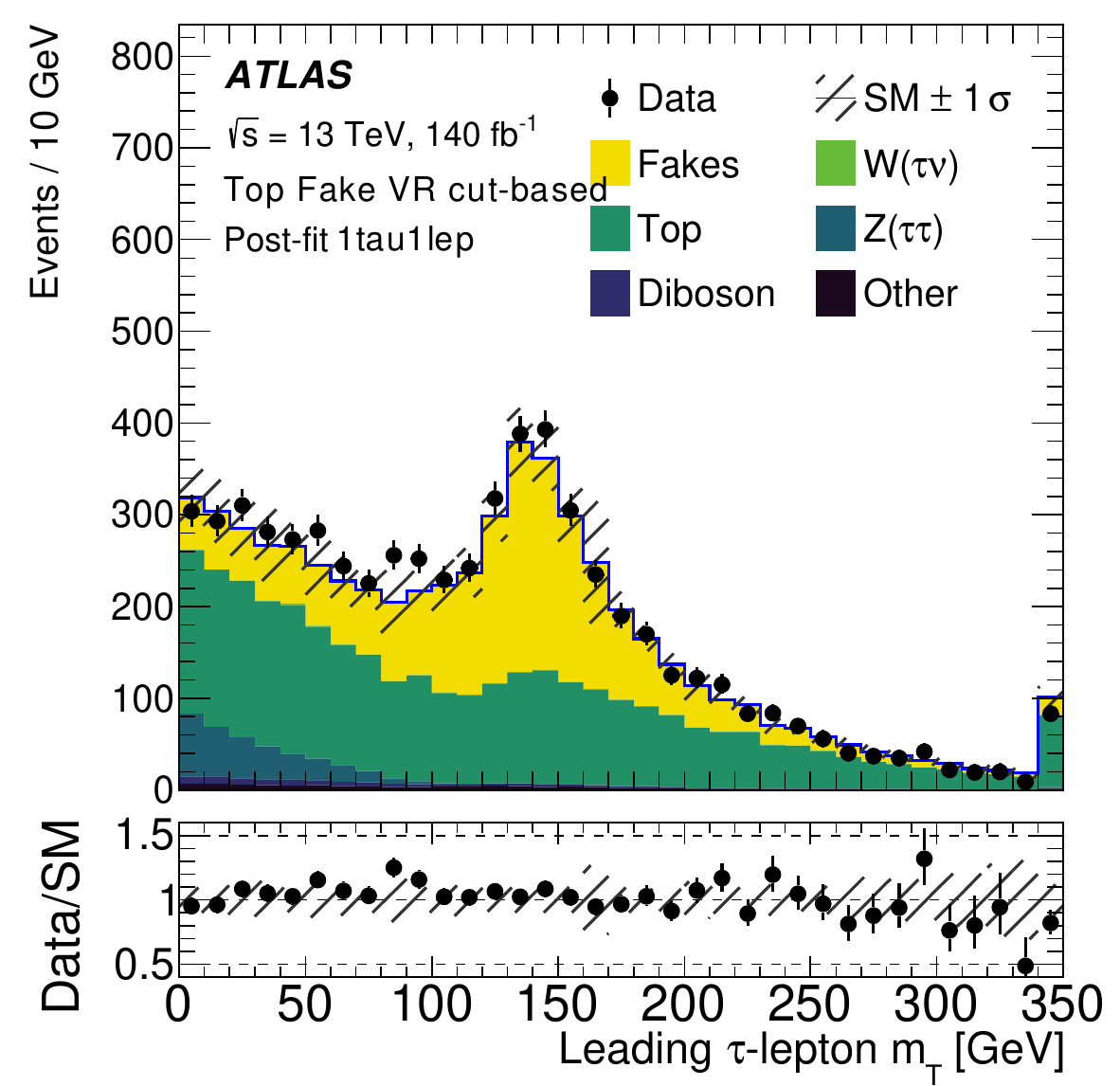}}
\subfloat[\RunTwo, Z($\nu\nu$) VR, $\Delta \phi(\tau_1, \pt^\mathrm{miss})$]{\includegraphics[width=0.32\textwidth]{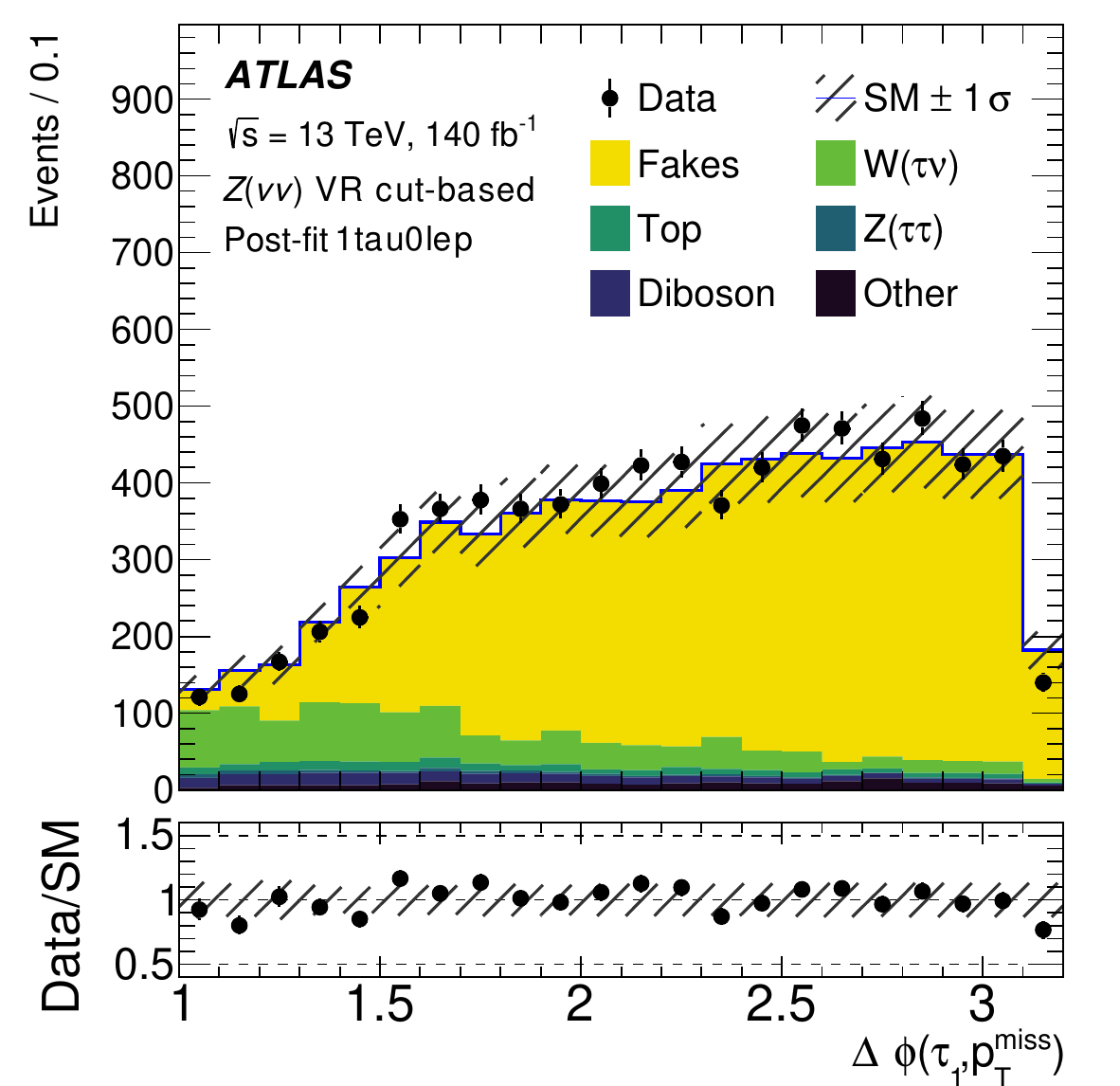}} \\
\subfloat[\RunThr, W fake-tau VR, \met]{\includegraphics[width=0.32\textwidth]{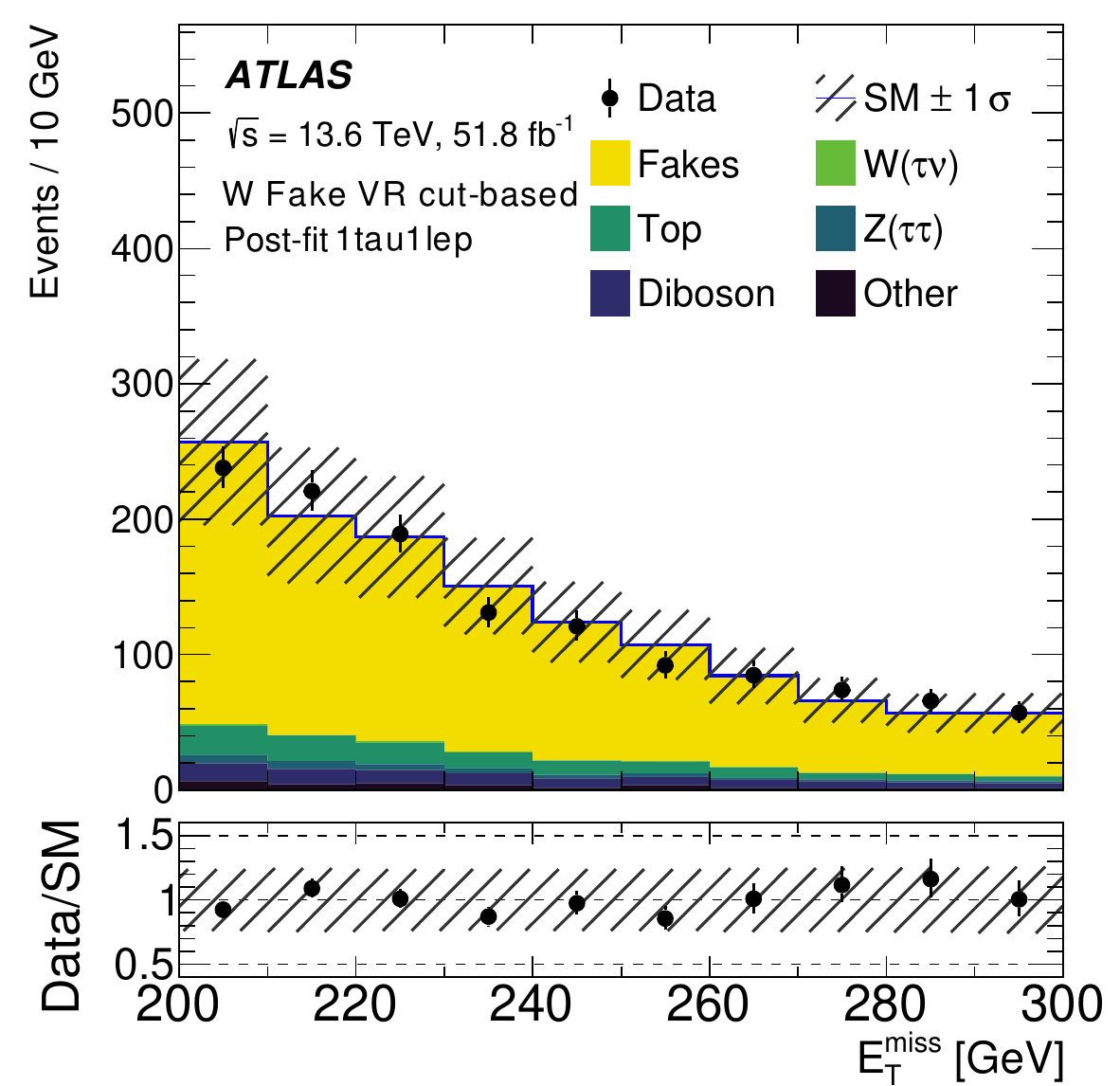}}
\subfloat[\RunThr, Top fake-tau VR, $m_\mathrm{T}^{\tau_1}$]{\includegraphics[width=0.32\textwidth]{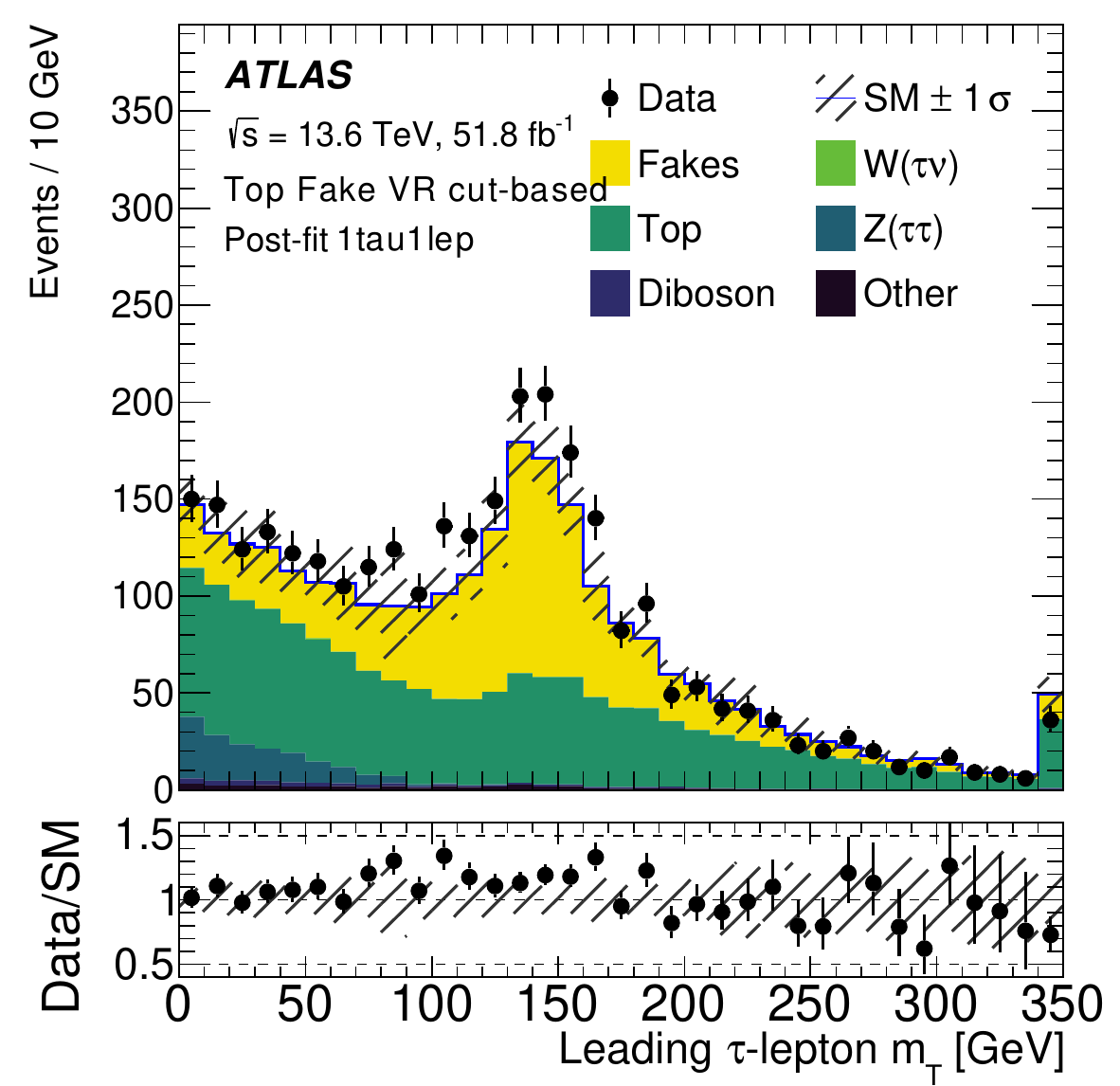}}
\subfloat[\RunThr, Z($\nu\nu$) VR, $\Delta \phi(\tau_1, \pt^\mathrm{miss})$]{\includegraphics[width=0.32\textwidth]{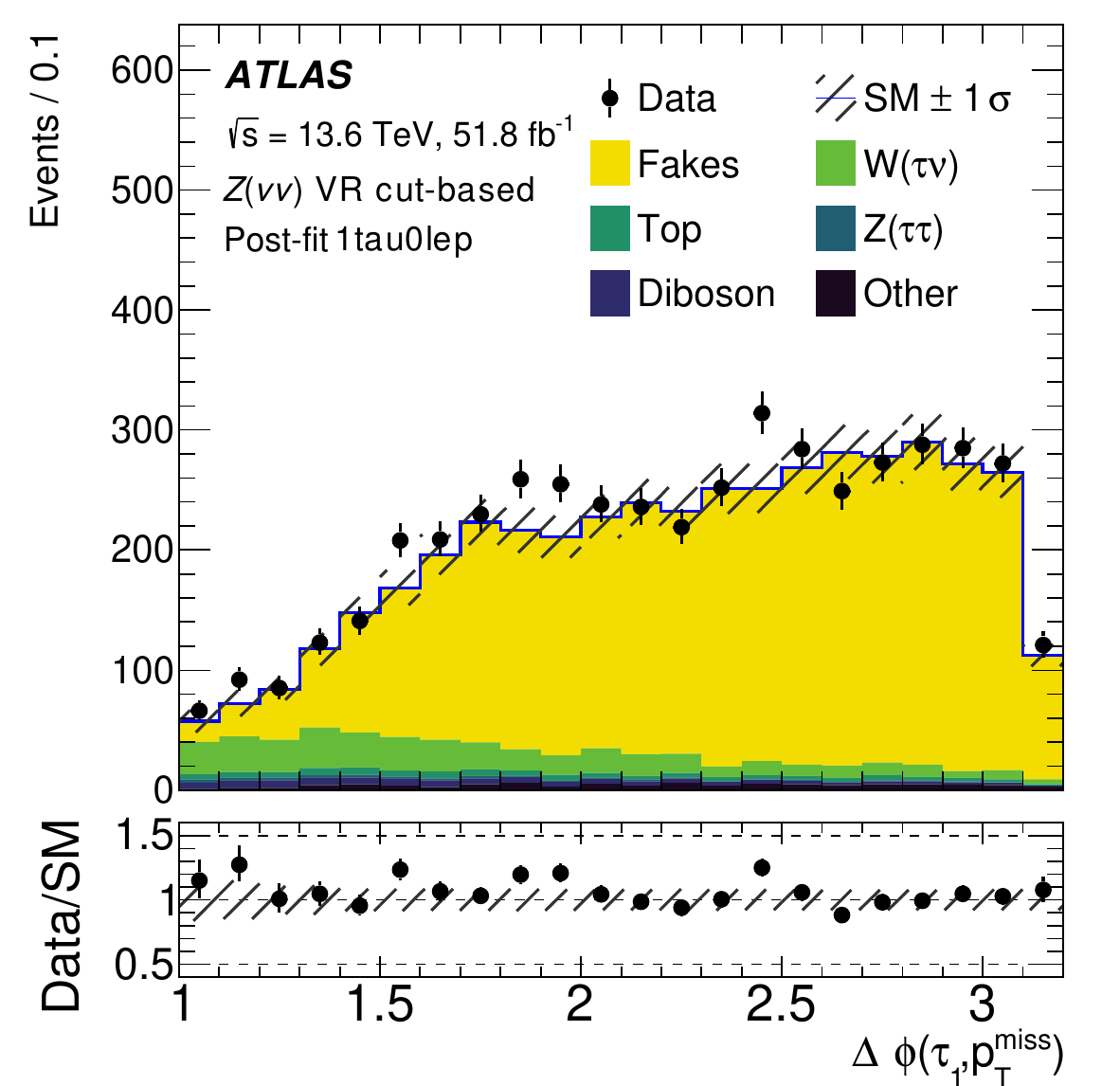}} \\
\caption{Distributions of important variables used in cut-and-count fake-tau VRs in \RunTwo (top row) and \RunThr (bottom row).   \BandIsStatSyst
\Postfit
\RightBinOverflow
\BottomPanelDataSM
}
\label{fig:faketau_plots_fullsyst_run23}
\end{figure}

Several validation regions are defined to access the performance of the fake-tau modelling, as described in Section~\ref{sec:analysis_strategy}. The modelling of a few important kinematic variables is shown in Figures~\ref{fig:faketau_plots_fullsyst_run23ml} and~\ref{fig:faketau_plots_fullsyst_run23} for the ML-based and cut-and-count approaches, respectively. The normalisation of the fake taus is not fitted to the data, but some of the VRs have significant contributions from \ttbar and \Wjets processes, which are corrected in the fits. The systematic uncertainties of this approach are described in Section~\ref{sec:systematics}.

%
%
%

%
%
%
%

%


\FloatBarrier
\section{Statistical analysis and systematic uncertainties}
\label{sec:systematics}

%
%
%

For the statistical analysis of the data the \histfitter workflow~\cite{Baak:2014wma} is followed by using \textsc{pyhf}~\cite{pyhf_joss}, the Python implementation of \textsc{HistFactory}. Diagnostic utilities from the \textsc{cabinetry} package~\cite{Cranmer:2021oxr} are also used. Two types of minimisation routines are used by \textsc{pyhf}: \textsc{Minuit} (Migrad)~\cite{James:1994vla} with \textsc{numpy} backend, and minimize from the \textsc{scipy} library~\cite{2020SciPy-NMeth} with \textsc{jax} backend. As described in Section~\ref{sec:analysis_strategy}, dedicated CRs and SRs are defined and included in simultaneous likelihood fits. The binned-fit procedure is based on the \textsc{HistFactory}~\cite{Cranmer:2012sba} algorithm. It is used to build a parametric model describing the observed data, the nominal predictions, and the systematic and statistical uncertainties of these predictions, in all regions of interest. A general likelihood function $L$ is constructed as a product of Poisson distributions, corresponding to the event counts in the binned regions, and constraint terms (Gaussian or Poisson) describing the systematic uncertainties.
Signal rate(s) and normalisation factors for background processes are free within bounds in the fit, while the systematic uncertainties are subject to the constraints. The fit then minimises the value of $-\log(L)$.
For the current analysis the asymptotic approximation~\cite{Cowan:2010js} is assumed to be valid (and is validated by direct comparison with MC experiments). Three main fit types are used by the search, differing by the presence of different regions in the likelihood and the assumptions made about the signal model and rate.

In background-only fits, the signal contribution is assumed to be zero everywhere and the signal rate is thus decoupled from the fit. Only CRs are included in the fit, and the results are extrapolated to the VRs and SRs. The purpose of these fits is to evaluate the quality of background modelling and to allow model-independent statements to be made. All direct comparisons of data and MC modelling shown in this \paper are either done before any fit or come from background-only fits.

Model-dependent (or exclusion) fits include both the CRs and SRs, which allows systematic uncertainties to be constrained more strongly. A particular signal model is assumed and the signal cross-section is included as the parameter of interest in the fit. For each of the models, the background-only and background-plus-signal hypotheses are tested and the \CLs prescription~\cite{Read:2002hq} is used to obtain results corresponding to an exclusion region defined by $\CLs < 0.05$.
These fits are used to construct the exclusion contours that are presented in Section~\ref{sec:result}.

Finally, model-independent fits are used to make qualitative (and model-independent) statements about whether the measurements in single-bin regions are consistent with the \SM predictions or not.
All the CRs and one chosen SR are included in each of these fits. A signal contribution of zero is assumed in the CRs, and a signal contribution equal to the fitted signal strength is assumed in the SR\@. The signal rate derived from the fit is then interpreted as the allowed/excluded cross-section of the signal. The selection of the regions tested and the results are further discussed in Section~\ref{sec:result}.

For all of the fits, \RunTwo and \RunThr data samples are first treated separately in individual fits. Combined \RunTwo and \RunThr exclusion fits are then performed to obtain the most stringent limits.

\FloatBarrier

The analysis considers three main types of systematic uncertainties. The experimental uncertainties include systematic uncertainties related to the reconstruction, identification, calibration, and corrections applied to the physical objects used in the analysis. Theoretical uncertainties comprise uncertainties of the MC generators and PDF sets. The third type of uncertainties considered are statistical uncertainties, related to the limited number of MC simulated events or limited number of data events used in the data-driven fake-tau estimations. The following correlation scheme is used:
\begin{itemize}
\item detector uncertainties are correlated across regions and samples;
\item theory uncertainties are correlated across regions, but not samples;
\item fake-tau uncertainties are uncorrelated across regions and samples, as they are largely statistical;
\item no uncertainties are correlated between \RunTwo and \RunThr.
\end{itemize}
In the cases where correlation between \RunTwo and \RunThr is expected (primarily jet and theory uncertainties), both correlation schemes were tested and no significant difference was observed. Thus, it is more practical to treat them as uncorrelated, simplifying the fitting procedures.

The experimental uncertainties are grouped according to the object they are applied to. Jet uncertainties include uncertainties from the energy scale calibration~\cite{PERF-2016-04} and resolution~\cite{JETM-2018-05}, and the NNJVT jet-vertex-tagging uncertainties~\cite{ATLAS-CONF-2014-018}. The jet flavour-tagging uncertainties~\cite{FTAG-2018-01,ATLAS-CONF-2018-001,ATLAS-CONF-2018-006} are also considered. Experimental uncertainties for $\tau$-leptons consist of uncertainties from reconstruction and identification algorithms, energy scale calibrations, and the electron veto~\cite{ATL-PHYS-PUB-2015-045,ATLAS-CONF-2017-029}. The experimental uncertainties for electrons and muons include uncertainties from reconstruction, isolation, and identification algorithms~\cite{EGAM-2018-01,EGAM-2021-02,MUON-2018-03,MUON-2022-01}.
The resolution and energy scale uncertainties for all the objects used in the analysis are propagated into the \met calculations, and uncertainties in the modelling of the underlying event and tracks not matched to any of the reconstructed objects are used to compute dedicated soft-term uncertainties~\cite{JETM-2020-03}.

The uncertainty in the combined \RunTwo(\RunThr) integrated luminosity is $0.83\%~(2.0\%)$~\cite{LUCID2,DAPR-2021-01,ATL-DAPR-PUB-2024-001}, obtained from luminosity measurements with the LUCID-2 detector. \Pileup reweighting is applied to all simulated events to match the observed conditions in data and is also assigned systematic uncertainties.

Dedicated uncertainties are applied to the fake-factor method for fake taus. The statistical uncertainty of the FFs due to the limited number of events in the measurement regions is one such uncertainty. The statistical uncertainty due to the limited number of data events in the anti-ID regions used for the measurements is also considered. The MC-simulated number of true $\tau$-leptons in the anti-ID regions is varied by $\pm 30\%$ (as a conservative estimate of expected detector uncertainty), and the effect of this variation is included as a separate uncertainty. The shape uncertainty of the templates used to determine the flavour composition of the fake-tau seed jets and the uncertainty of the flavour composition itself are also considered.
Given that most of the fake-tau uncertainties are statistical, no correlation between regions is assumed. Alternative correlation schemes were tested, and no significant difference was observed, so the no-correlation scheme is selected as the most conservative and straightforward approach.

Variations of the factorisation and renormalisation scales (\muF and \muR) are used to estimate uncertainties due to the lack of higher-order corrections in the simulations. The standard $7$-point scale-variation scheme is used, where $\{\muR, \muF\}$ are varied from their nominal values by factors of $\{0.5,0.5\}$, $\{0.5,1\}$, $\{1,0.5\}$, $\{2,1\}$, $\{1,2\}$, and $\{2,2\}$. These variations are implemented as alternative weights. The envelope of the effects of the six variations is used as the final uncertainty.
For the \NNPDF sets the standard prescription is followed~\cite{Butterworth:2015oua}, using the \textsc{LHAPDF} library to evaluate the effect of $100$ alternative PDF variations. The effect of the PDF uncertainties is then combined in quadrature with the effect of varying the strong coupling constant \alphas. Theoretical systematic uncertainties related to the PDF and scale variations are applied to all samples except the signal samples, for which dedicated higher-order cross-section uncertainties are available.

\begin{figure}[htbp!]
\centering
\subfloat[\RunTwo ML-based SRs]{\includegraphics[width=1.\textwidth]{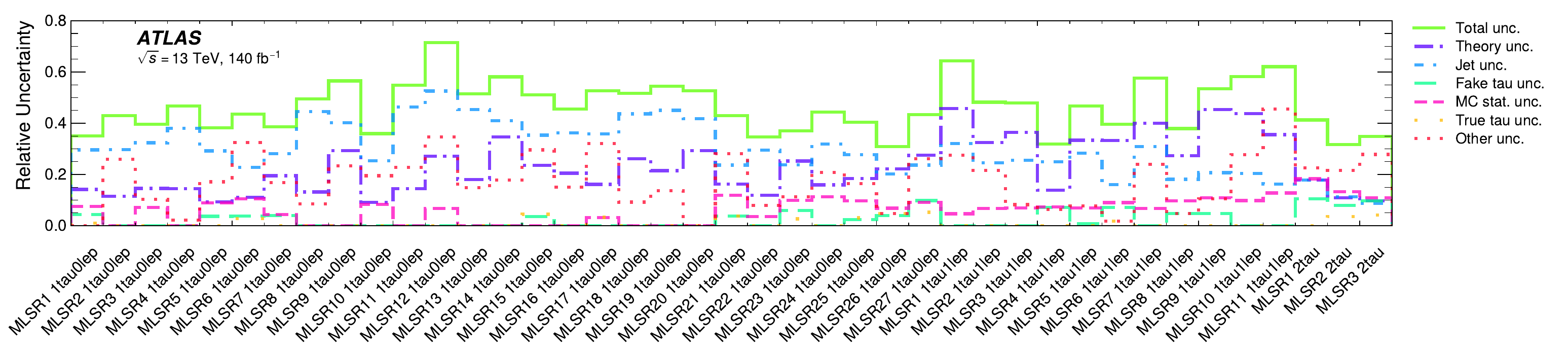}}\\
\subfloat[\RunThr ML-based SRs]{\includegraphics[width=1.0\textwidth]{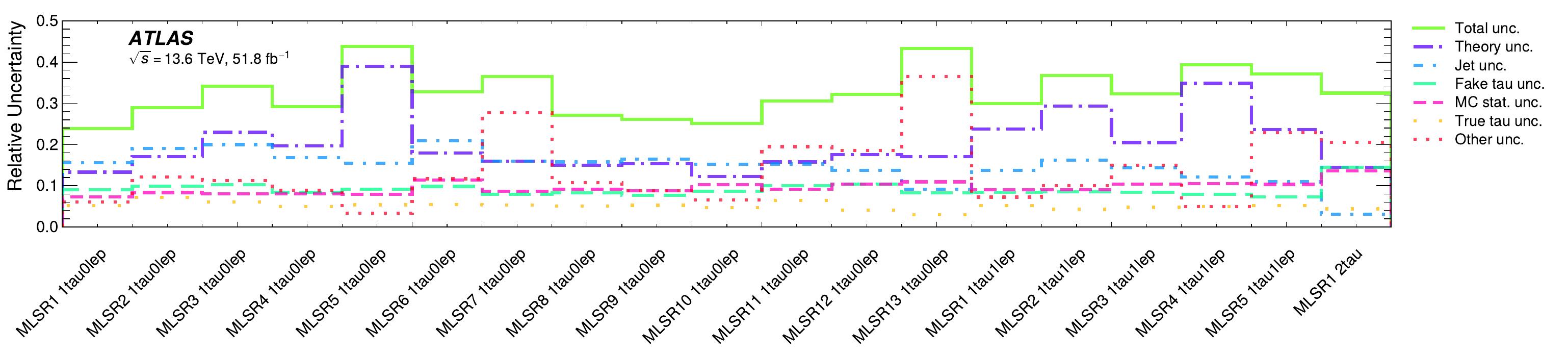}}\\
\subfloat[\RunTwo ML-based VRs]{\includegraphics[width=0.475\textwidth]{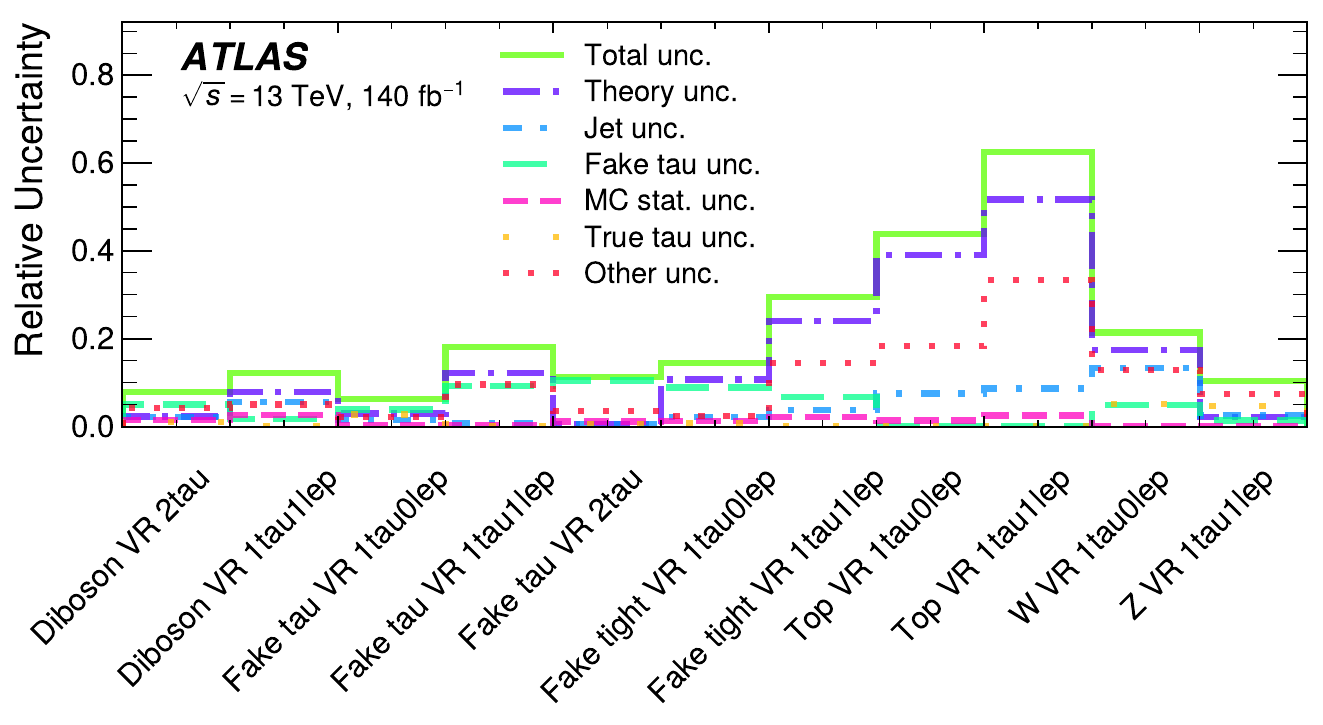}}
\subfloat[\RunThr ML-based VRs]{\includegraphics[width=0.475\textwidth]{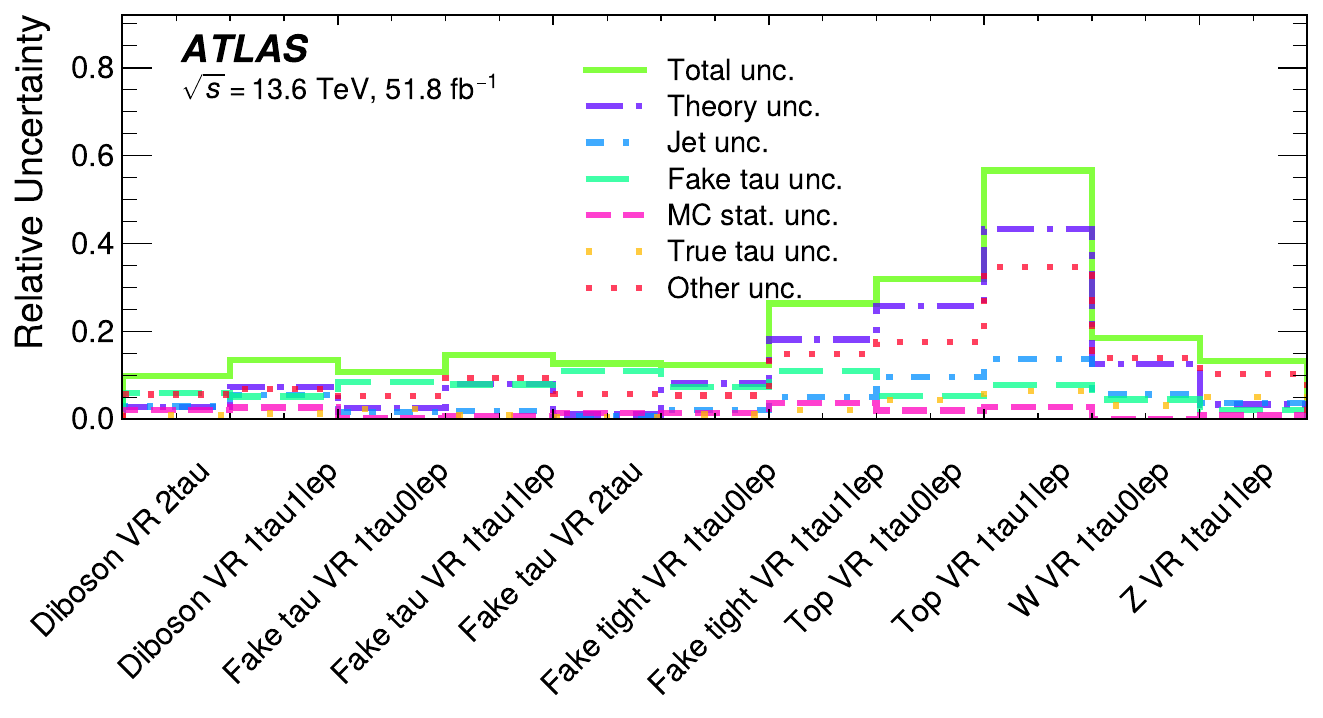}}
\caption{Comparison of the dominant groups of systematic uncertainties in the total background prediction after the background-only fit for the ML-based approach, relative to the total yields. The total uncertainty is calculated using all the available systematic and statistical uncertainties. \enquote{Other} refers to \met, muon, electron, and \pileup uncertainties. \enquote{True tau} refers to the experimental uncertainties of $\tau$-leptons. The individual uncertainties can be correlated and do not necessarily add in quadrature to the total uncertainty.
}
\label{fig:syst_ml}
\end{figure}

\begin{figure}[htbp!]
\centering
\subfloat[\RunTwo cut-and-count SRs and VRs]{\includegraphics[width=1.\textwidth]{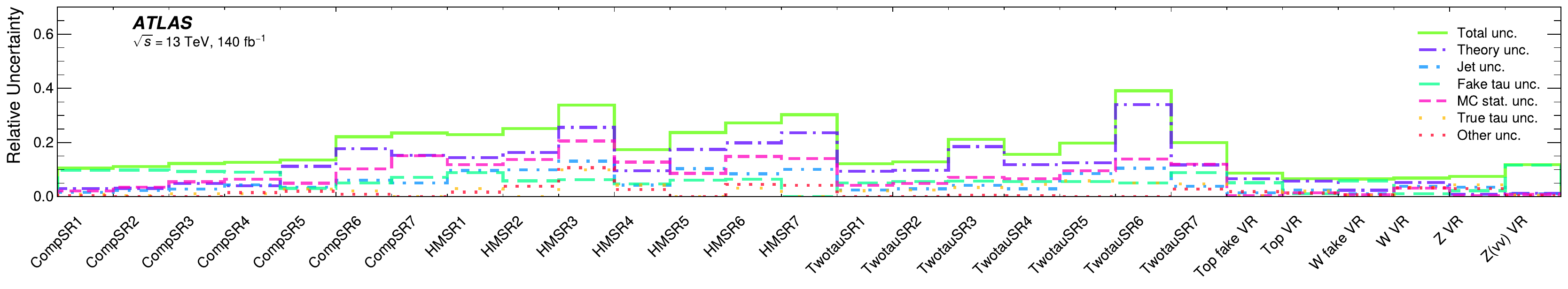}}\\
\subfloat[\RunThr cut-and-count SRs and VRs]{\includegraphics[width=1.\textwidth]{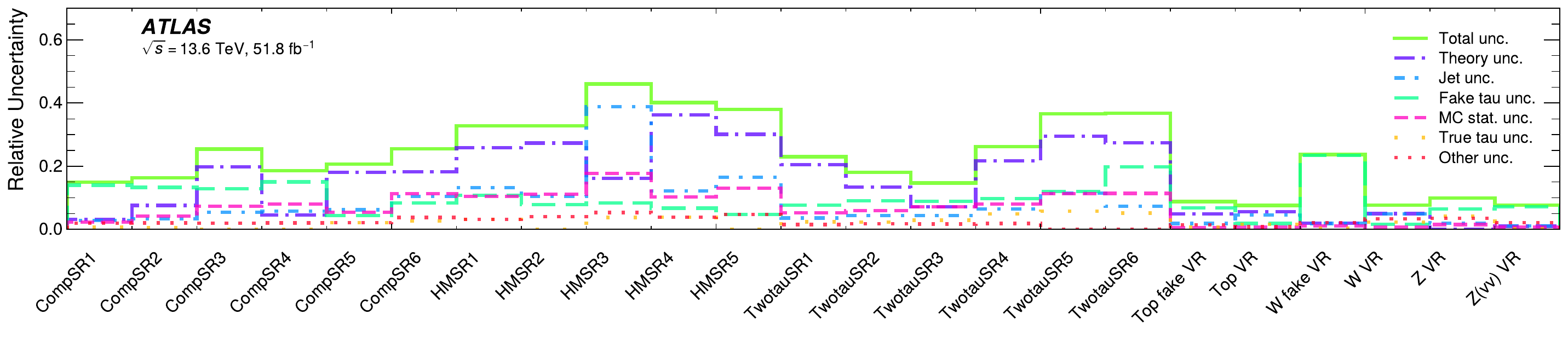}}
\caption{Comparison of the dominant groups of systematic uncertainties in the total background prediction after the background-only fit for the cut-and-count approach, relative to the total yields. The total uncertainty is calculated using all the available systematic and statistical uncertainties. \enquote{Other} refers to \met, muon, electron, and \pileup uncertainties. \enquote{True tau} refers to the experimental uncertainties of $\tau$-leptons. The individual uncertainties can be correlated and do not necessarily add in quadrature to the total uncertainty.
}
\label{fig:syst_cb}
\end{figure}

Additional theory uncertainties related to final-state radiation variations, variations of the A14 tune due to varying \alphas for initial-state radiation, and variations of the merging parameter and resummation scale in \SHERPA-generated samples~\cite{Catani:2001cc,Hoeche:2009rj} are considered where appropriate. For \ttbar samples, a direct comparison is made with alternative samples generated such that the $\pt$ of the \POWHEG emission is tested against all other incoming and outgoing partons, and the minimum value is chosen as the \textit{pThard} scale (i.e.~the \textit{pThard} parameter is set to option $1$ in \PYTHIA[8]).
Another comparison uses \ttbar samples generated with the $h_\mathrm{damp}$ (first-emission damping parameter) value set to $3 \times m_\text{top}$ instead of the nominal $1.5 \times m_\text{top}$. The uncertainty from the parton-shower approach is estimated by comparing the nominal \ttbar samples (\POWPY) with \POWHER samples~\cite{Bahr:2008pv,Bellm:2015jjp,Bellm:2015jjp}. For single-top-quark production a comparison is made between samples produced using the diagram removal (nominal) and diagram subtraction schemes, which are used to remove the overlap with the \ttbar production in the next-to-leading-order calculations.

For signal MC production, general simulation variations not affected by the cross-section uncertainties are also considered. These uncertainties include the radiation uncertainty and the effect of varying the factorisation and renormalisation scales, merging scales, and parton-shower tuning. The combined uncertainty from the signal theory uncertainties is estimated by using envelopes around the nominal result. From the comparisons the $68\%~(95\%)$ confidence intervals generally lie within the $1\sigma$~$(2\sigma)$ Poisson uncertainty of the nominal sample, indicating that no statistically significant effect is expected. A conservative upper limit of $10\%$ is used for all signal models.

The largest individual contributions to the total uncertainty in the ML-based SRs come from the factorisation and renormalisation scale uncertainties for \Wjets and \Zjets backgrounds, and comparisons with alternative generators for \ttbar and single-top-quark MC samples. In the \nolep channel the cumulative effect of experimental jet uncertainties is sizeable. Statistical uncertainties and modelling of the $\tau$-leptons (true and fake) also contribute significantly in most SRs. In most of the cut-and-count SRs, Top theory uncertainties (mainly from comparisons with alternative generators) are dominant. In the HMSR \nolep channel, factorisation and renormalisation scale uncertainties for \Wjets backgrounds are important as well. The CompSR \nolep channel is dominated by fake taus, leading to a large contribution from the fake-tau uncertainties.

The breakdown of the effects of different sources of systematic uncertainty in the VRs and SRs is shown in Figures~\ref{fig:syst_ml} and~\ref{fig:syst_cb}. To find the effect of a group of uncertainties, its square is evaluated as the difference between the squares of the total uncertainty and the uncertainty computed without this group. In the ML-based analysis the \RunTwo SRs tend to have larger total uncertainties than the \RunThr ones, due to the effects of several experimental jet and \met uncertainties in the \nolep channel. The contribution of theory uncertainties is on average larger in \RunTwo than in \RunThr in all channels. This can also be seen in the uncertainties of the Top VRs of \RunTwo and \RunThr. In the cut-and-count SRs, the Top theory uncertainties and jet non-closure uncertainties are more dominant in the \RunThr SRs, leading to larger total uncertainties overall. The fake-tau uncertainties are also larger in \RunThr because fewer events are available for measurements of the fake factors; this is the leading contribution in the CompSR \nolep channel.


\FloatBarrier
\section{Results}
\label{sec:result}

\begin{figure}[t]
\centering
\subfloat[\RunTwo \nolep]{\includegraphics[width=0.95\textwidth]{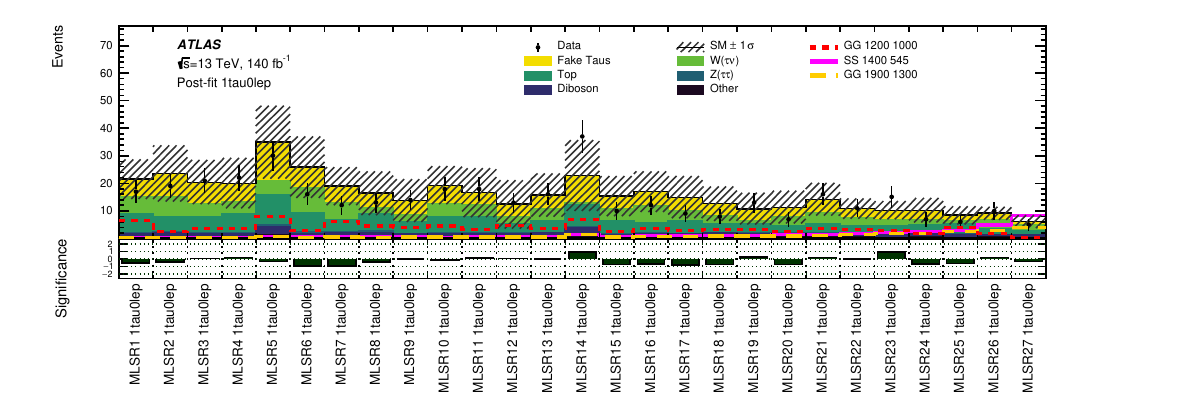}\label{fig:sr_ml_run2}}\\
\subfloat[\RunTwo \onelep and \twotau]{\includegraphics[width=0.405\textwidth]{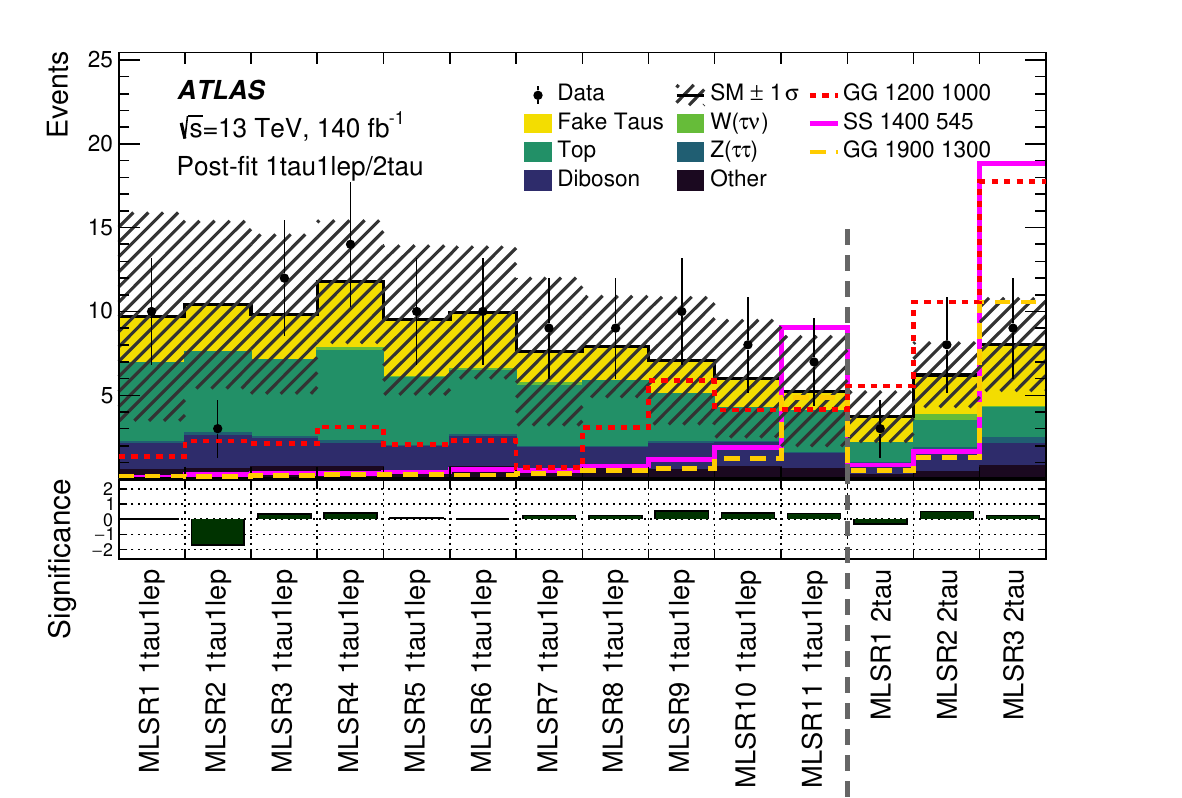}\label{fig:sr2_ml_run2}}
\subfloat[\RunThr]{\includegraphics[width=0.57\textwidth]{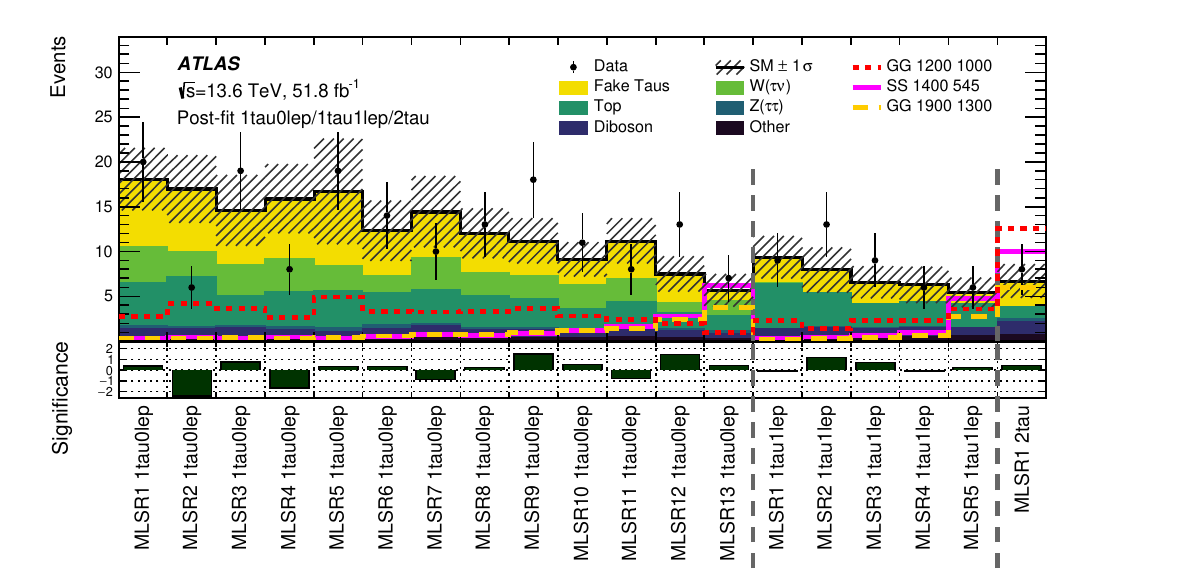}\label{fig:sr_ml_run3}}

\caption{A comparison of the observed and expected yields in the bins of the
\protect\subref{fig:sr_ml_run2} \RunTwo \nolep,
\protect\subref{fig:sr2_ml_run2} \RunTwo \onelep and \twotau, and
\protect\subref{fig:sr_ml_run3} \RunThr ML SRs used in the analysis is shown in the upper panel.
\BottomPanelSig
\Postfit
\BandIsStatSyst
Benchmark signal models are overlayed.
}
\label{fig:sr_ml}
\end{figure}

\begin{figure}[t]
\centering
\subfloat[\RunTwo]{\includegraphics[width=0.8\textwidth]{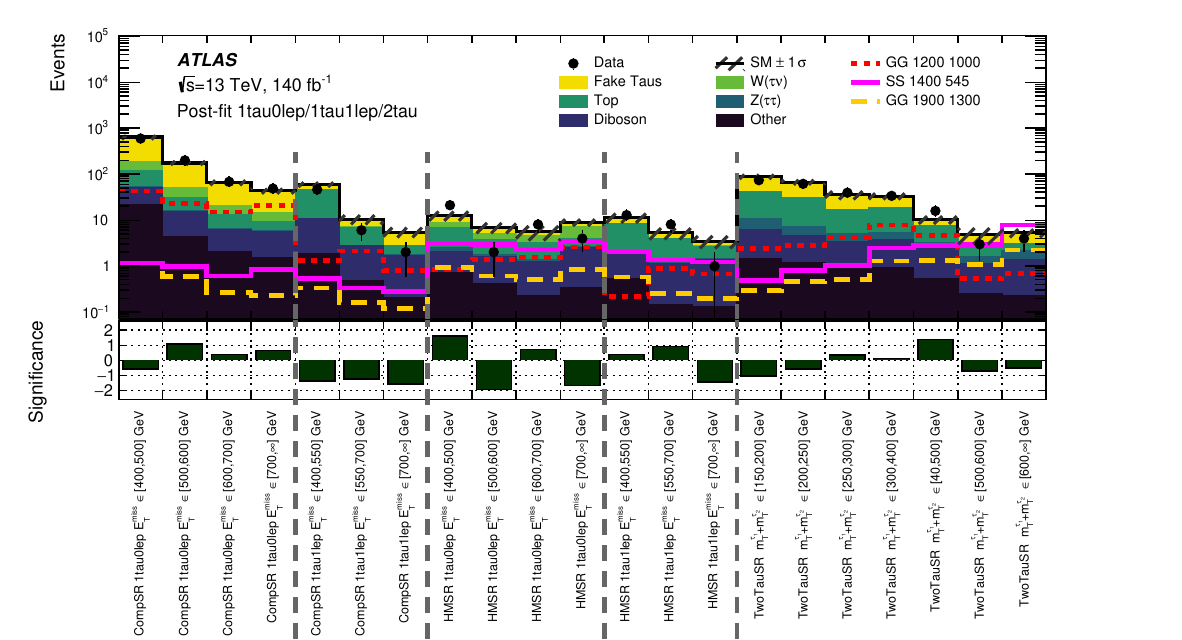}\label{fig:sr_cb_run2}} \\
\subfloat[\RunThr]{\includegraphics[width=0.8\textwidth]{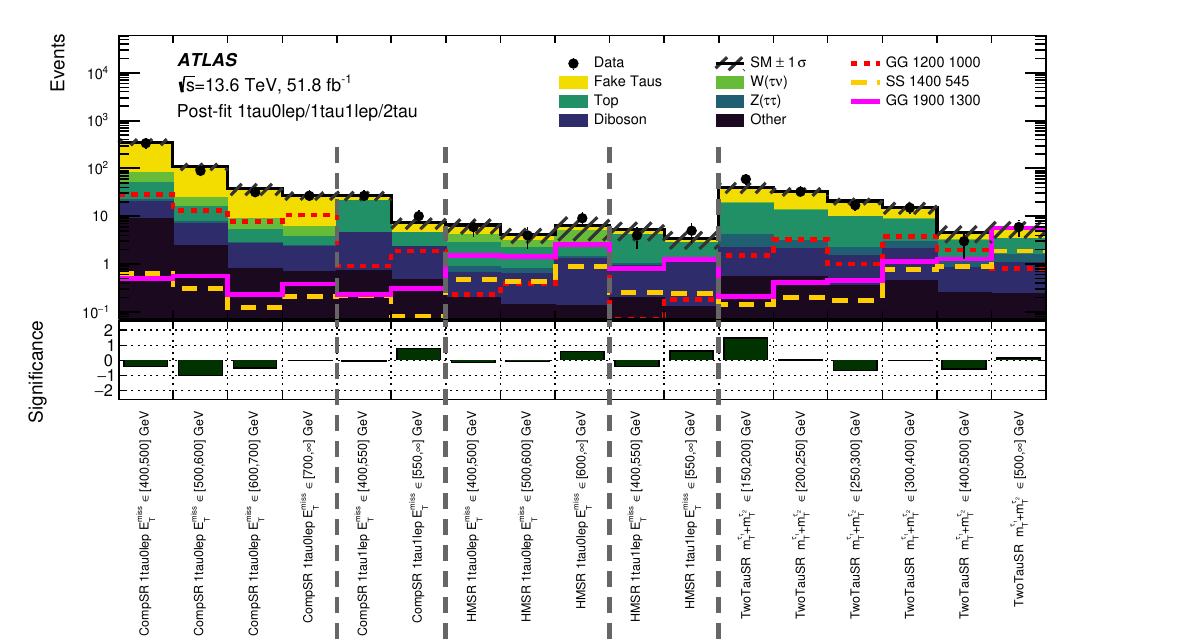}\label{fig:sr_cb_run3}}
\caption{A comparison of the observed and expected yields in the bins of the
\protect\subref{fig:sr_cb_run2} \RunTwo and
\protect\subref{fig:sr_cb_run3} \RunThr cut-and-count SRs used in the analysis is shown in the upper panel.
\BottomPanelSig
\Postfit
\BandIsStatSyst
Benchmark signal models are overlayed.
}
\label{fig:sr_cb}
\end{figure}

The observed and expected yields in all the SRs used by the analysis are shown in Figures~\ref{fig:sr_ml} and~\ref{fig:sr_cb}. The NFs are determined with the background-only fit procedure described in Section~\ref{sec:systematics}. No significant deviations from the \SM are observed. The largest excess of events has less than $2 \sigma$ significance, in the \RunTwo HMSR \nolep $\met \in [400,500]$ \GeV bin. The largest deficit of events has just over $2 \sigma$ significance, in the second bin of the \RunThr \nolep ML SR\@. Benchmark signal models are overlayed, showing that the minor excesses seen are not compatible with the expected SUSY signal yields. A slight tendency to overestimate the \SM background is seen in the \RunTwo CompSR \onelep.

\begin{figure}[t]
\centering
\subfloat[Gluino models]{\includegraphics[width=0.40\textwidth]{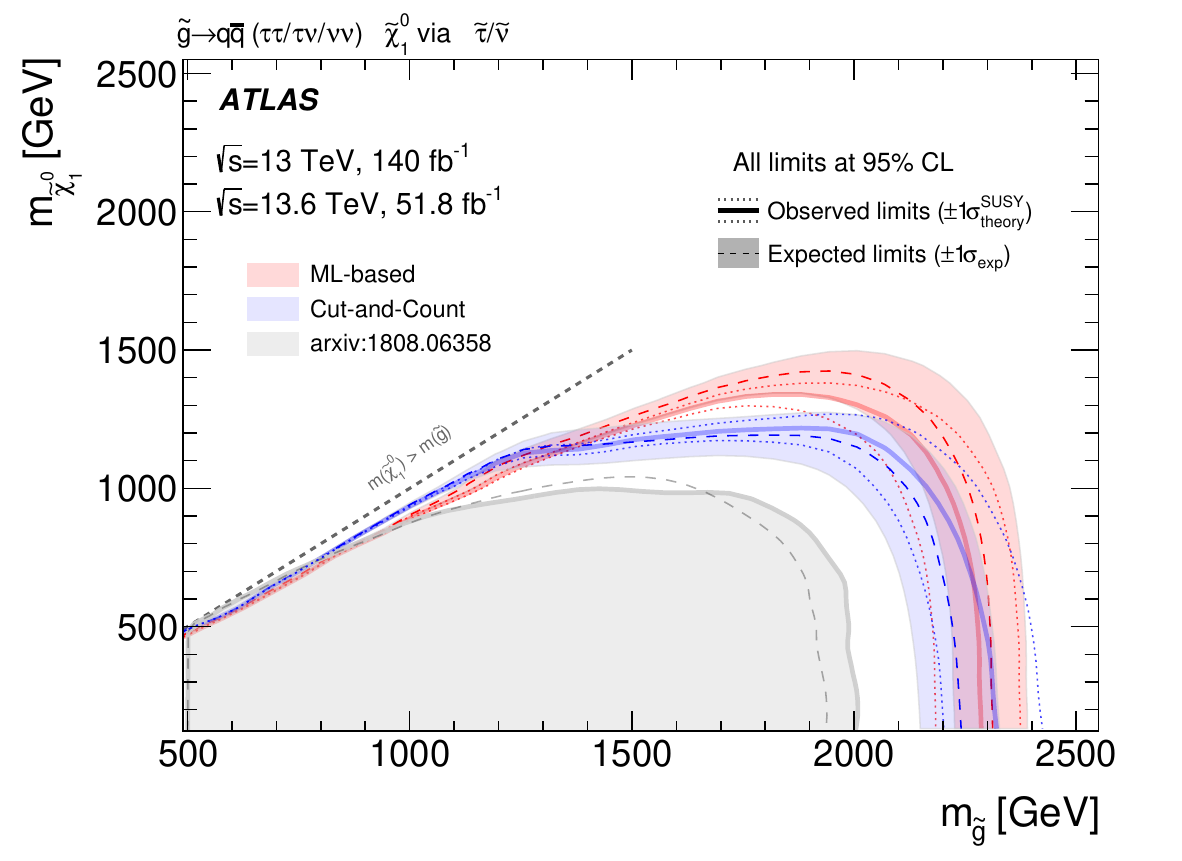}\label{fig:gg_cls}}
\subfloat[Squark models]{\includegraphics[width=0.40\textwidth]{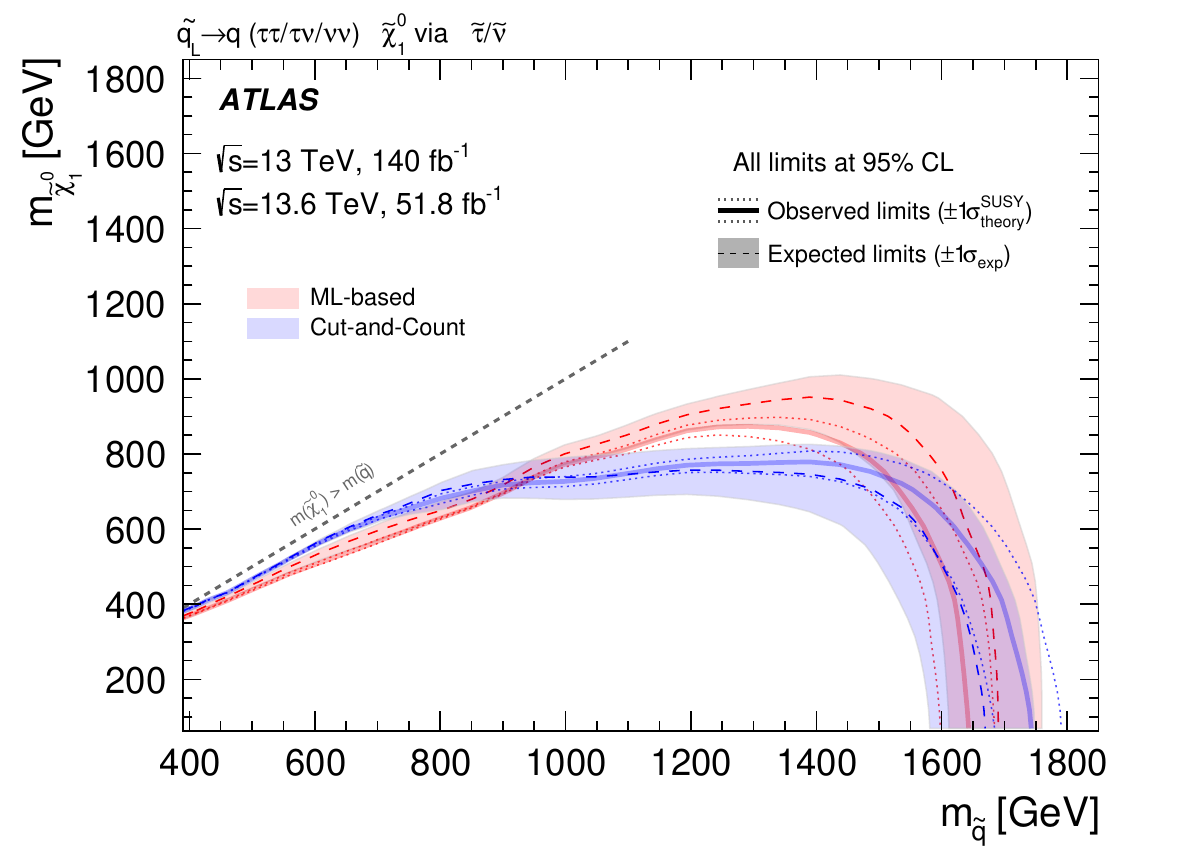}\label{fig:ss_cls}}
\caption{The exclusion contours from the ML-based and cut-and-count approaches, for
\protect\subref{fig:gg_cls} gluino pair production and
\protect\subref{fig:ss_cls} squark pair production.
The results are obtained using the \CLs prescription. The gluino limits are overlayed with the previous, partial \RunTwo, search.
}
\label{fig:sr_results}
\end{figure}

\begin{figure}[htbp!]
\centering

\subfloat[Gluino models, ML]{\includegraphics[width = 0.40\textwidth]{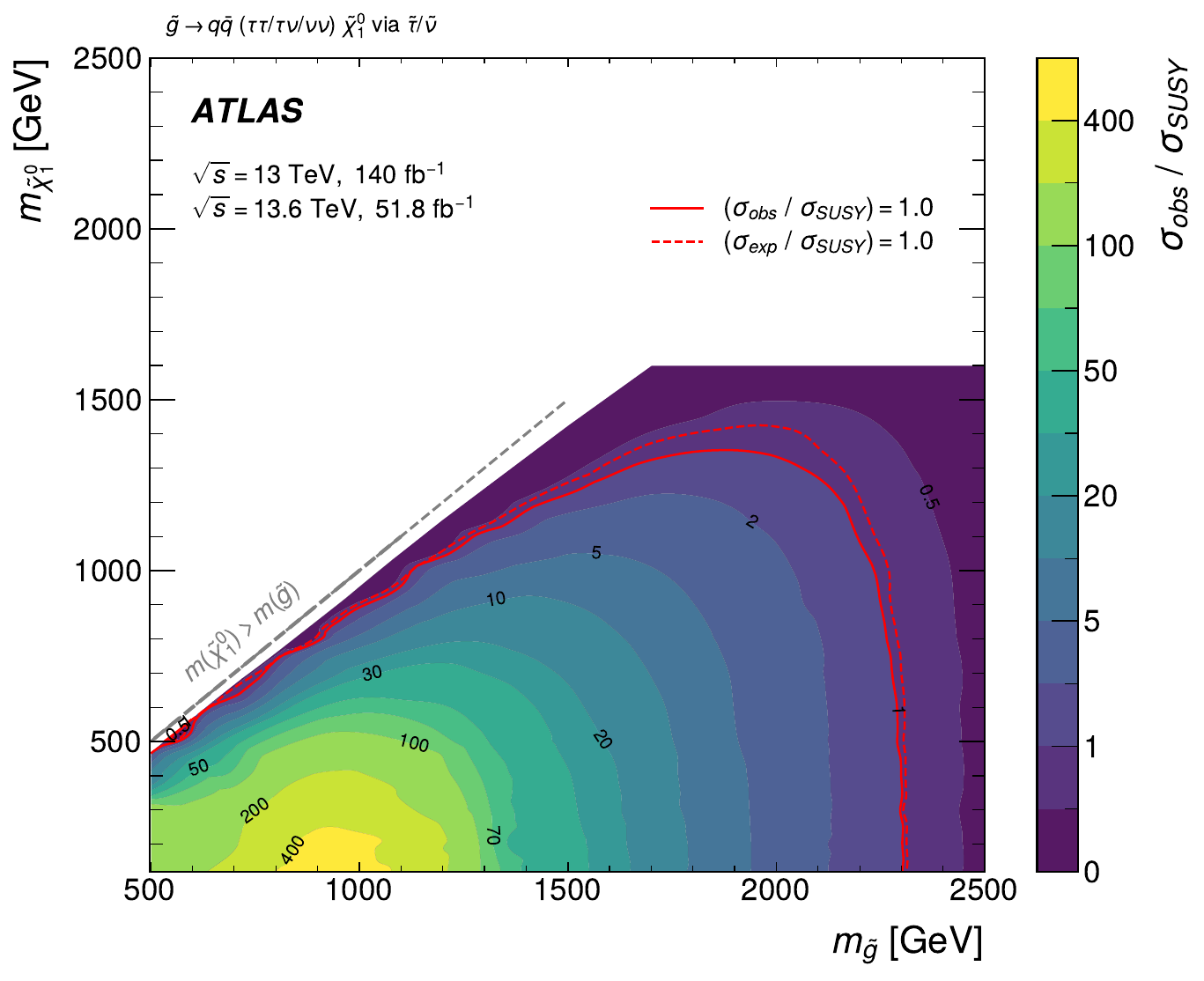}}
\subfloat[Squark models, ML]{\includegraphics[width = 0.40\textwidth]{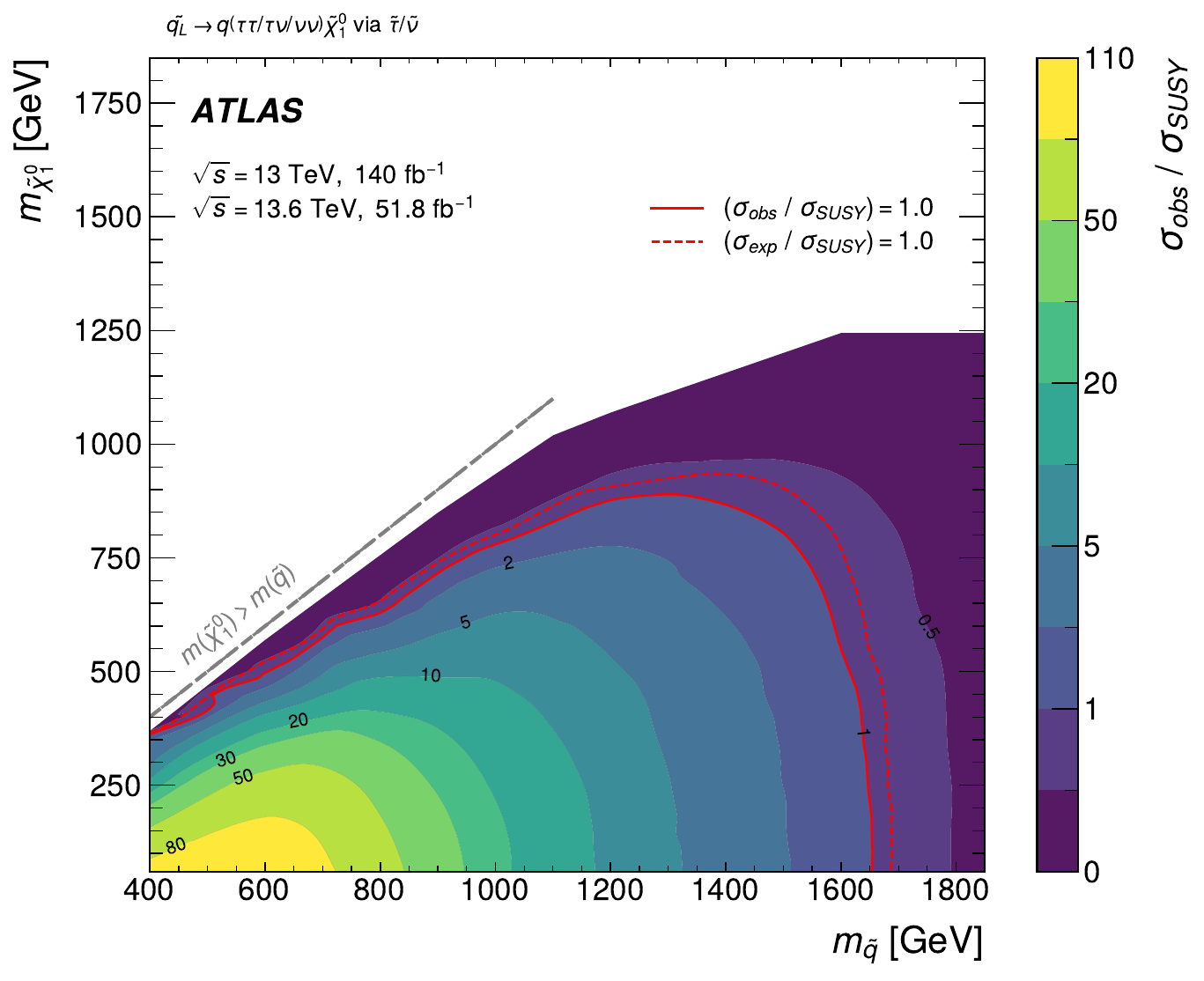}}\\
\subfloat[Gluino models, cut-and-count]{\includegraphics[width = 0.40\textwidth]{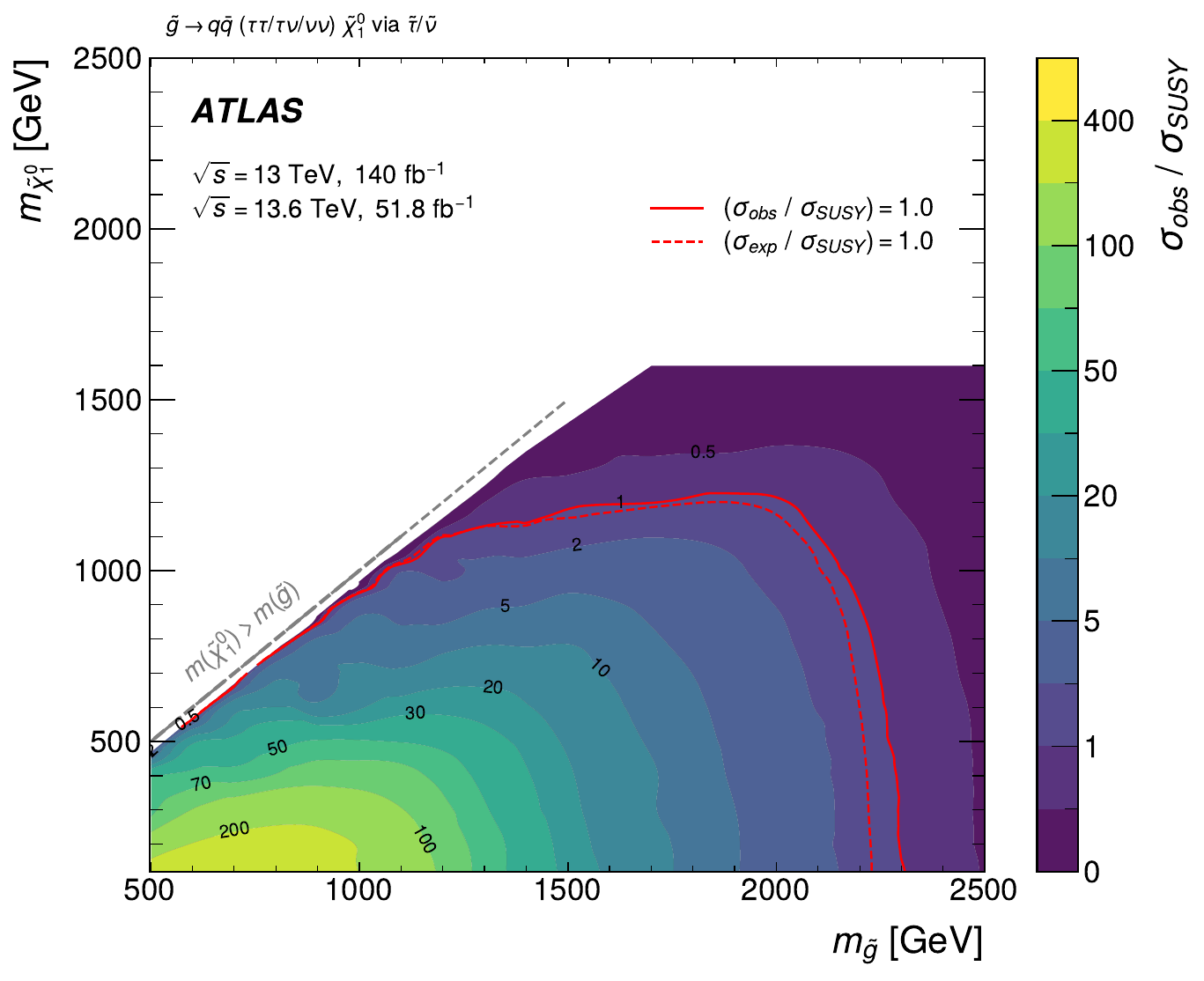}}
\subfloat[Squark models, cut-and-count]{\includegraphics[width = 0.40\textwidth]{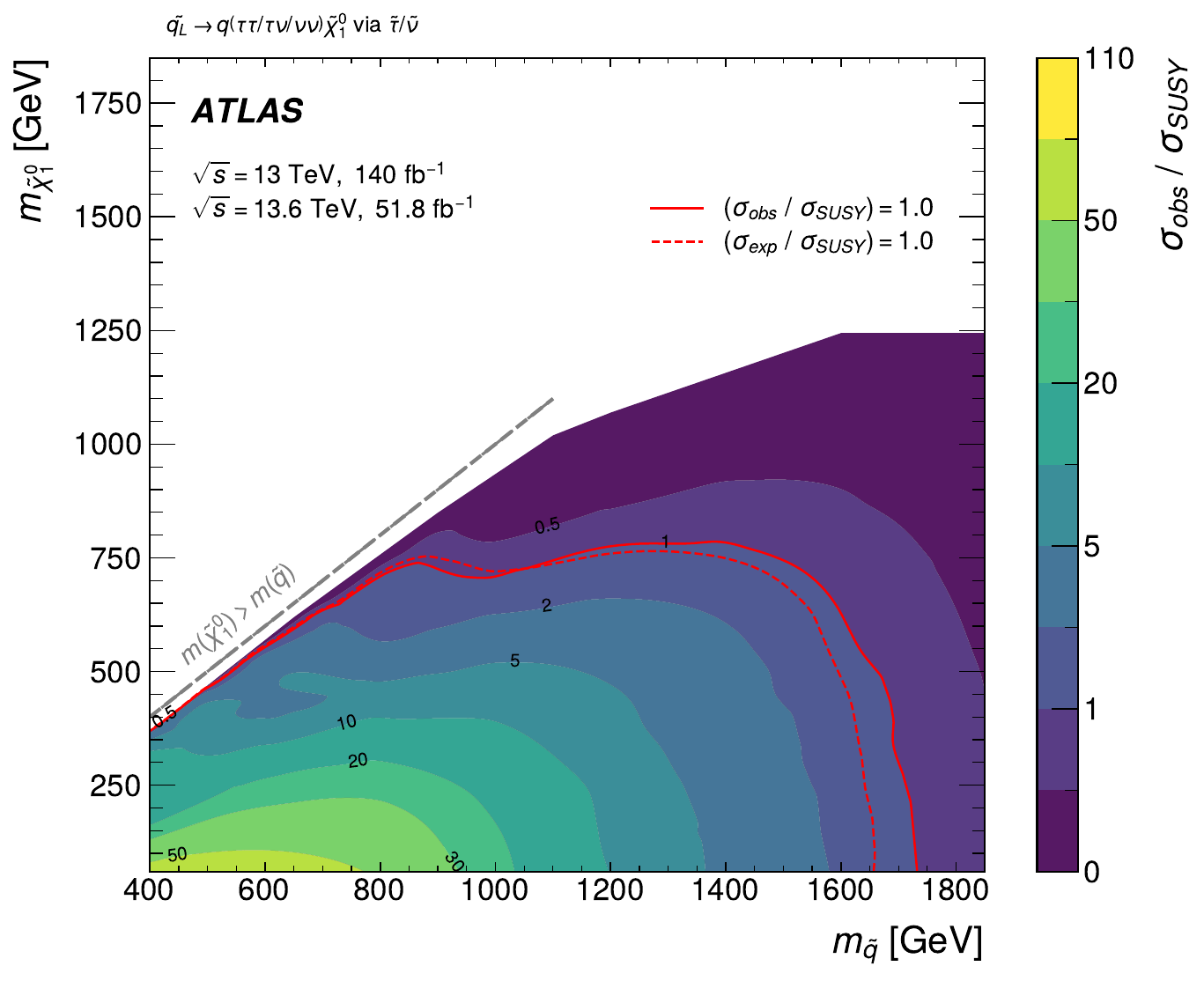}}
\caption{Upper limit on the observed cross-section normalised to the theory-predicted SUSY cross-section, obtained with the ML-based (top row) and cut-and-count (bottom row) approaches.}
\label{fig:ul_comb_ml}
\end{figure}

An exclusion fit is performed for every simplified SUSY model considered by the analysis. All of the relevant SRs are included in the fit together with the CRs. In addition to the NFs, the signal strength parameter $\mu$ is included as the parameter of interest. A comparison of the background-only ($\mu=0$) and signal-plus-background ($\mu=1$) hypothesis tests is performed. A common value of $\mu$ is used for \RunTwo and \RunThr data.
The signal model most compatible with the observed data is GG $1300$ $1200$ in the ML-based SRs, with local $p$-value of $0.00058$ ($Z=3.2$).
Following the convention adopted for ATLAS SUSY searches, the SUSY cross-section uncertainty due to missing higher-order corrections is not included in the fit directly. Instead, the effects of raising and lowering the nominal cross-section by its theoretical uncertainty are shown. The results are presented in the two-dimensional parameter spaces, as exclusion contours at $95\%$ confidence level (CL) following the \CLs prescription, and as upper limits on $\sigma_\text{obs}/\sigma_\text{SUSY}$. The ML-based and cut-and-count analysis results are shown in Figure~\ref{fig:sr_results} for both gluino and squark pair production. The ML approach has better sensitivity in the high-\susymass and high-$m(\ninoone)$ parts of the phase space, excluding gluino masses up to $2.25$~\TeV and $m(\ninoone)$ up to $1.35$~\TeV. For squark models, $m(\ninoone)$ values up to $0.85$~\TeV are excluded. The cut-and-count approach has stronger sensitivity towards the compressed, low-\masssplit region of the parameter space because of the dedicated CompSR regions, excluding models with $\Delta m(\tilde{g}, \tilde\chi^0_1)>150$~\GeV and $\Delta m(\tilde{q}, \tilde\chi^0_1)>130$~\GeV. The observed limits from the cut-and-count analysis also reach gluino masses of $2.3$~\TeV and squark masses of $1.7$~\TeV for the $m(\ninoone)=0.1$~\TeV models.
A comparison of the presented results with the previous searches targeting similar final states with data collected in 2015--2016 is shown in Figure~\ref{fig:gg_cls}. The improvement is mainly due to the more sophisticated $\tau$-lepton reconstruction and identification techniques, the larger data sample, and the use of ML techniques. Upper limits on the signal cross-section are shown in Figure~\ref{fig:ul_comb_ml} for the ML-based and cut-and-count approaches. The upper limits are normalised to the theory SUSY cross-section for the corresponding \susymass pair production. Compared to \RunTwo only, the combined \RunTwo and \RunThr fits in the ML-based (cut-and-count) analysis strengthen the upper limits on the observed cross-section by up to $10\%$ ($30\%$) in the high-$m(\ninoone)$ regions, up to $20\%$ ($50\%$) in the low-\masssplit regions, and up to $10\%$ ($10\%$) in the low-$m_\text{LSP}$ region.

Model-independent statements about the presence of new physics are made for several regions. A generic new signal is assumed to be present in one particular region, a signal-plus-background fit is performed over all CRs and the selected region, and the profile-likelihood-ratio test statistic is evaluated. The final bin of every individual SR is used for this procedure (one at a time) together with all the CRs. In the cut-and-count approach, these bins are the tails of kinematic distributions, sensitive to any new physics with $\tau$-leptons and \met. For the ML-based approach the classifier score is intrinsically tuned to be sensitive to a particular class of SUSY models. If the new physics is sufficiently similar to the models the classifiers are trained on, the sensitivity is maximised.
The limits are set independently for \RunTwo and \RunThr, not assuming any particular cross-section dependence for the new physics.
The results, including the upper limit on the visible cross-section \sigvis at $95\%$ CL, number of signal events given the observed and predicted numbers of background events, confidence level for the background-only hypothesis, and the discovery $p$-value and its associated significance, are presented in Table~\ref{tab:discSR}. Upper limits of $0.02{-}0.31$~fb are placed on the cross-section of a new physics process appearing in these regions.

\FloatBarrier

\begin{table}[t]
\caption{
Upper limits at $95\%$ CL on the visible cross-section \sigvis,
on the number of signal events ($S_\mathrm{obs}^{95}$),
and on the number of signal events given the expected number (and $\pm 1\sigma$ excursions of the expectation) of background events ($S_\mathrm{exp}^{95}$).
The last two columns indicate the \CLb value (i.e.\ the confidence level observed for the background-only hypothesis),
the discovery $p$-value ($p_0$),
and its associated significance $Z$.
}
\label{tab:discSR}
\renewcommand{\arraystretch}{1.2}
\centering
\begin{adjustbox}{width=1\textwidth}
\begin{tabular}{llccccccc}
\toprule{Signal region} & Bin & Total bkg & Data & \sigvis[fb]  &  $S_\mathrm{obs}^{95}$  & $S_\mathrm{exp}^{95}$ & \CLb & $p_0$ ($Z$)  \\
\midrule
\RunTwo cut-and-count  \\
\midrule
CompSR \nolep & $\met > 700$~\GeV  &$\phantom{.}45 \pm 4.3$&    49~~&     0.16\phantom{$0$}&   22.2&   $17.9^{+7.1}_{-5.2}$ &          0.73    &0.26 (0.6) \\
CompSR \onelep & $\met > 700$~\GeV &$4.6 \pm 1.1$&     2&      0.026&   \phantom{$0$}3.7&    $\phantom{0}6.0^{+3.1}_{-1.9}$ &           0.1\phantom{$0$}     &0.50 (0.0) \\
HMSR \nolep   & $\met > 700$~\GeV &$7.9 \pm 1.3$&      4&      0.031&   \phantom{$0$}4.3&    $\phantom{0}7.3^{+3.6}_{-2.3}$ &           0.08    &0.50 (0.0) \\
HMSR \onelep  & $\met > 700$~\GeV &$2.6 \pm 1.0$&     1&      0.024&   \phantom{$0$}3.3&    $\phantom{0}5.1^{+2.7}_{-1.7}$ &           0.14    &0.50 (0.0) \\
TwoTauSR \twotau &  $\mt^{\tau_1} + \mt^{\tau_2} > 600$~\GeV    &$5.0 \pm 0.9$&     4&      0.036&   \phantom{$0$}5.0&      $\phantom{0}6.0^{+3.1}_{-1.9}$ &           0.33    &0.50 (0.0) \\

\midrule
\RunThr cut-and-count  \\
\midrule
CompSR \nolep  & $\met > 700$~\GeV  &$\phantom{.}27 \pm 3.6$&    27~~&     0.29&   14.8&   $14.9^{+6.2}_{-4.3}$ &          0.5\phantom{$0$}     &0.50 (0.0) \\
CompSR \onelep  & $\met > 550$~\GeV &$8.1 \pm 1.6$&     10~~&     0.20&   10.2&   $\phantom{0}7.8^{+3.7}_{-2.4}$ &           0.76    &0.21 (0.8) \\
HMSR \nolep  & $\met > 600$~\GeV    &$7.7 \pm 1.9$&     9&      0.31&   16.0&   $12.6^{+12.4}_{-5.5}$ &         0.63    &0.26 (0.6) \\
HMSR \onelep  & $\met > 550$~\GeV   &$3.9 \pm 1.1$&     5&      0.14&   \phantom{$0$}7.2&    $\phantom{0}5.8^{+3.0}_{-1.8}$ &           0.71    &0.26 (0.6) \\
TwoTauSR \twotau & $\mt^{\tau_1} + \mt^{\tau_2} > 550$~\GeV     &$5.7 \pm 1.5$&     6&      0.14&   \phantom{$0$}7.4&    $\phantom{0}7.0^{+3.4}_{-2.2}$ &           0.55    &0.44 (0.2)  \\
\midrule
\RunTwo ML  \\
\midrule
\nolep SR & MLSR27      &$5.5 \pm 1.3$&     5&      0.07&   10.1&   $11.3^{+6.7}_{-4.2}$ &  0.41    &0.50 (0.0) \\
\onelep SR & MLSR11     &$6.3 \pm 1.9$&     7&      0.15&   20.3&   $18.2^{+6.8}_{-9.1}$ &  0.55    &0.34 (0.4) \\
\twotau SR & MLSR3    &$8.5 \pm 2.2 $&     9&      0.09&   13.3&   $12.4^{+6.9}_{-5.1}$ &  0.53    &0.41 (0.2) \\
\midrule
\RunThr ML  \\
\midrule
\nolep SR  & MLSR13    &$6.3 \pm 1.7$&     7&      0.29&   15.1&   $14.1^{+10.9}_{-6.7}$ & 0.54    &0.35 (0.4) \\
\onelep SR & MLSR5    &$5.7 \pm 1.5$&     6&      0.20&    10.6&   $10.0^{+11.0}_{-4.4}$ & 0.53    &0.43 (0.2) \\
\twotau SR & MLSR1    &$7.2 \pm 1.7$&     8&      0.17&   \phantom{$0$}8.9&    $\phantom{0}7.9^{+4.1}_{-2.6}$ &   0.61    &0.36 (0.4) \\
\bottomrule
\end{tabular}
\end{adjustbox}
\end{table}


\FloatBarrier
\section{Conclusion}
%

A search for pair production of supersymmetric particles with cascade decays in events with hadronically decaying $\tau$-leptons, jets, and missing transverse momentum is presented, performed both with ML-based and cut-and-count approaches. Final states with exactly one $\tau$-lepton and no other leptons, exactly one $\tau$-lepton and at least one other lepton, or at least two $\tau$-leptons are considered. The search uses $\pp$ collisions at $\sqrt s = 13$~\TeV\ and $13.6$~\TeV\ delivered by the Large Hadron Collider to collect data with the ATLAS detector in 2015--2018 and 2022--2023 respectively, corresponding to integrated luminosities of $140$~\ifb and $51.8$~\ifb. The enhanced calibration and identification algorithms, data sample, and analysis techniques improve the sensitivity across the whole range of masses considered compared to the previous searches. No significant deviations from the \SM predictions are observed in either a cut-and-count or ML-based approach. The results are interpreted in the context of simplified supersymmetric models, setting the most stringent limits for these channels. For gluino pair-production models, gluino masses below $2.25$~\GeV\ are excluded, and LSP masses below $1.35$~\TeV\ are excluded for gluino masses around $2$~\TeV. Squark pair-production models with squark masses up to $1.7$~\TeV, and LSP masses up to $0.85$~\TeV\ are excluded.


\FloatBarrier
\section*{Acknowledgements}
%

%

%
%

%
%

We thank CERN for the very successful operation of the LHC and its injectors, as well as the support staff at
CERN and at our institutions worldwide without whom ATLAS could not be operated efficiently.

The crucial computing support from all WLCG partners is acknowledged gratefully, in particular from CERN, the ATLAS Tier-1 facilities at TRIUMF/SFU (Canada), NDGF (Denmark, Norway, Sweden), CC-IN2P3 (France), KIT/GridKA (Germany), INFN-CNAF (Italy), NL-T1 (Netherlands), PIC (Spain), RAL (UK) and BNL (USA), the Tier-2 facilities worldwide and large non-WLCG resource providers. Major contributors of computing resources are listed in Ref.~\cite{ATL-SOFT-PUB-2025-001}.

We gratefully acknowledge the support of ANPCyT, Argentina; YerPhI, Armenia; ARC, Australia; BMWFW and FWF, Austria; ANAS, Azerbaijan; CNPq and FAPESP, Brazil; NSERC, NRC and CFI, Canada; CERN; ANID, Chile; CAS, MOST and NSFC, China; Minciencias, Colombia; MEYS CR, Czech Republic; DNRF and DNSRC, Denmark; IN2P3-CNRS and CEA-DRF/IRFU, France; SRNSFG, Georgia; BMFTR, HGF and MPG, Germany; GSRI, Greece; RGC and Hong Kong SAR, China; ICHEP and Academy of Sciences and Humanities, Israel; INFN, Italy; MEXT and JSPS, Japan; CNRST, Morocco; NWO, Netherlands; RCN, Norway; MNiSW, Poland; FCT, Portugal; MNE/IFA, Romania; MSTDI, Serbia; MSSR, Slovakia; ARIS and MVZI, Slovenia; DSI/NRF, South Africa; MICIU/AEI, Spain; SRC and Wallenberg Foundation, Sweden; SERI, SNSF and Cantons of Bern and Geneva, Switzerland; NSTC, Taipei; TENMAK, T\"urkiye; STFC/UKRI, United Kingdom; DOE and NSF, United States of America.

Individual groups and members have received support from BCKDF, CANARIE, CRC and DRAC, Canada; CERN-CZ, FORTE and PRIMUS, Czech Republic; COST, ERC, ERDF, Horizon 2020, ICSC-NextGenerationEU and Marie Sk{\l}odowska-Curie Actions, European Union; Investissements d'Avenir Labex, Investissements d'Avenir Idex and ANR, France; DFG and AvH Foundation, Germany; Herakleitos, Thales and Aristeia programmes co-financed by EU-ESF and the Greek NSRF, Greece; BSF-NSF and MINERVA, Israel; NCN and NAWA, Poland; La Caixa Banking Foundation, CERCA Programme Generalitat de Catalunya and PROMETEO and GenT Programmes Generalitat Valenciana, Spain; G\"{o}ran Gustafssons Stiftelse, Sweden; The Royal Society and Leverhulme Trust, United Kingdom.

In addition, individual members wish to acknowledge support from CERN: European Organization for Nuclear Research (CERN DOCT); Chile: Agencia Nacional de Investigaci\'on y Desarrollo (FONDECYT 1230812, FONDECYT 1240864, Fondecyt 3240661); China: Chinese Ministry of Science and Technology (MOST-2023YFA1605700, MOST-2023YFA1609300), National Natural Science Foundation of China (NSFC - 12175119, NSFC 12275265); Czech Republic: Czech Science Foundation (GACR - 24-11373S), Ministry of Education Youth and Sports (ERC-CZ-LL2327, FORTE CZ.02.01.01/00/22\_008/0004632), PRIMUS Research Programme (PRIMUS/21/SCI/017); EU: H2020 European Research Council (ERC - 101002463); European Union: European Research Council (BARD No. 101116429, ERC - 948254, ERC 101089007), European Regional Development Fund (SMASH COFUND 101081355, SLO ERDF), Horizon 2020 Framework Programme (MUCCA - CHIST-ERA-19-XAI-00), European Union, Future Artificial Intelligence Research (FAIR-NextGenerationEU PE00000013), Italian Center for High Performance Computing, Big Data and Quantum Computing (ICSC, NextGenerationEU); France: Agence Nationale de la Recherche (ANR-21-CE31-0022, ANR-22-EDIR-0002); Germany: Baden-Württemberg Stiftung (BW Stiftung-Postdoc Eliteprogramme), Deutsche Forschungsgemeinschaft (DFG - 469666862, DFG - CR 312/5-2); China: Research Grants Council (GRF); Italy: Istituto Nazionale di Fisica Nucleare (ICSC, NextGenerationEU), Ministero dell'Università e della Ricerca (NextGenEU 153D23001490006 M4C2.1.1, NextGenEU I53D23000820006 M4C2.1.1, NextGenEU I53D23001490006 M4C2.1.1, SOE2024\_0000023); Japan: Japan Society for the Promotion of Science (JSPS KAKENHI JP22H01227, JSPS KAKENHI JP22H04944, JSPS KAKENHI JP22KK0227, JSPS KAKENHI JP24K23939, JSPS KAKENHI JP25H00650, JSPS KAKENHI JP25H01291, JSPS KAKENHI JP25K01023); Norway: Research Council of Norway (RCN-314472); Poland: Ministry of Science and Higher Education (IDUB AGH, POB8, D4 no 9722), Polish National Science Centre (NCN 2021/42/E/ST2/00350, NCN OPUS 2023/51/B/ST2/02507, NCN OPUS nr 2022/47/B/ST2/03059, NCN UMO-2019/34/E/ST2/00393, UMO-2022/47/O/ST2/00148, UMO-2023/49/B/ST2/04085, UMO-2023/51/B/ST2/00920, UMO-2024/53/N/ST2/00869); Portugal: Foundation for Science and Technology (FCT); Spain: Ministry of Science and Innovation (MCIN \& NextGenEU PCI2022-135018-2, MICIN \& FEDER PID2021-125273NB, RYC2019-028510-I, RYC2020-030254-I, RYC2021-031273-I, RYC2022-038164-I); Sweden: Carl Trygger Foundation (Carl Trygger Foundation CTS 22:2312), Swedish Research Council (Swedish Research Council 2023-04654, VR 2021-03651, VR 2022-03845, VR 2022-04683, VR 2023-03403, VR 2024-05451), Knut and Alice Wallenberg Foundation (KAW 2018.0458, KAW 2022.0358, KAW 2023.0366); Switzerland: Swiss National Science Foundation (SNSF - PCEFP2\_194658); United Kingdom: Leverhulme Trust (Leverhulme Trust RPG-2020-004), Royal Society (NIF-R1-231091); United States of America: U.S. Department of Energy (ECA DE-AC02-76SF00515), Neubauer Family Foundation.

%
%


%
%
%
%
%
%

%
%

%
%
%
\printbibliography
\clearpage
\input{atlas_authlist}
%

%
%

%

%
%
%
%

\end{document}

%% file: atlas_authlist.tex
 
\begin{flushleft}
\hypersetup{urlcolor=black}
{\Large The ATLAS Collaboration}

\bigskip

\AtlasOrcid[0000-0002-6665-4934]{G.~Aad}$^\textrm{\scriptsize 104}$,
\AtlasOrcid[0000-0001-7616-1554]{E.~Aakvaag}$^\textrm{\scriptsize 17}$,
\AtlasOrcid[0000-0002-5888-2734]{B.~Abbott}$^\textrm{\scriptsize 123}$,
\AtlasOrcid[0000-0002-0287-5869]{S.~Abdelhameed}$^\textrm{\scriptsize 119a}$,
\AtlasOrcid[0000-0002-1002-1652]{K.~Abeling}$^\textrm{\scriptsize 55}$,
\AtlasOrcid[0000-0001-5763-2760]{N.J.~Abicht}$^\textrm{\scriptsize 49}$,
\AtlasOrcid[0000-0002-8496-9294]{S.H.~Abidi}$^\textrm{\scriptsize 30}$,
\AtlasOrcid[0009-0003-6578-220X]{M.~Aboelela}$^\textrm{\scriptsize 45}$,
\AtlasOrcid[0000-0002-9987-2292]{A.~Aboulhorma}$^\textrm{\scriptsize 36e}$,
\AtlasOrcid[0000-0001-5329-6640]{H.~Abramowicz}$^\textrm{\scriptsize 157}$
% ...
\bigskip
\\

% ...
$^{17}$Department for Physics and Technology, University of Bergen, Bergen; Norway.\\
% ...
$^{30}$Physics Department, Brookhaven National Laboratory, Upton NY; United States of America.\\
% ...
$^{36}$$^{(a)}$Facult\'e des Sciences Ain Chock, Universit\'e Hassan II de Casablanca;$^{(b)}$Facult\'{e} des Sciences, Universit\'{e} Ibn-Tofail, K\'{e}nitra;$^{(c)}$Facult\'e des Sciences Semlalia, Universit\'e Cadi Ayyad, LPHEA-Marrakech;$^{(d)}$LPMR, Facult\'e des Sciences, Universit\'e Mohamed Premier, Oujda;$^{(e)}$Facult\'e des sciences, Universit\'e Mohammed V, Rabat;$^{(f)}$Institute of Applied Physics, Mohammed VI Polytechnic University, Ben Guerir; Morocco.\\
% ...
$^{45}$Physics Department, Southern Methodist University, Dallas TX; United States of America.\\
% ...
$^{49}$Fakult\"{a}t Physik , Technische Universit{\"a}t Dortmund, Dortmund; Germany.\\
% ...
$^{55}$II. Physikalisches Institut, Georg-August-Universit\"{a}t G\"ottingen, G\"ottingen; Germany.\\
% ...
$^{104}$CPPM, Aix-Marseille Universit\'e, CNRS/IN2P3, Marseille; France.\\
% ...
$^{119}$$^{(a)}$New York University Abu Dhabi, Abu Dhabi;$^{(b)}$United Arab Emirates University, Al Ain; United Arab Emirates.\\
% ...
$^{123}$Homer L. Dodge Department of Physics and Astronomy, University of Oklahoma, Norman OK; United States of America.\\
% ...
$^{157}$Raymond and Beverly Sackler School of Physics and Astronomy, Tel Aviv University, Tel Aviv; Israel.\\
% ...
$^{a}$ Also at Affiliated with an institute formerly covered by a cooperation agreement with CERN.\\
$^{b}$ Also at An-Najah National University, Nablus; Palestine.\\
$^{c}$ Also at Borough of Manhattan Community College, City University of New York, New York NY; United States of America.\\
$^{d}$ Also at Center for Interdisciplinary Research and Innovation (CIRI-AUTH), Thessaloniki; Greece.\\
$^{e}$ Also at Centre of Physics of the Universities of Minho and Porto (CF-UM-UP); Portugal.\\
$^{f}$ Also at CERN, Geneva; Switzerland.\\
% ...

\end{flushleft}
